\documentclass[aps,prd,longbibliography,reprint,twocolumn,amsmath,amssymb,amsfonts,showpacs,footnote,superscriptaddress]{revtex4-2}

\usepackage[T1]{fontenc}
\usepackage{lmodern}
\usepackage{amsmath,amssymb,amsfonts,mathrsfs}
\usepackage{bm}
\usepackage{mathtools}
\usepackage{physics}
\usepackage{hyperref}
\usepackage{xcolor}
\usepackage[T1]{fontenc}
\usepackage{booktabs}
\usepackage{siunitx}
\usepackage{url}

\usepackage{orcidlink}

\definecolor{navyblue}{rgb}{0.0, 0.0, 0.5}
\definecolor{ferrarired}{rgb}{1.0, 0.11, 0.0}
\definecolor{persianblue}{rgb}{0.11, 0.22, 0.73}
\usepackage{float}

\hypersetup{
	colorlinks=true,
	citecolor=blue,%
	linkcolor=ferrarired,%
	urlcolor=persianblue,%
	filecolor=blue,
	linktoc=page
}

\newcommand{\W}[2]{\mathcal{W}\!\left[#1,#2\right]}

\begin{document}


	\title{Ringdown and echoes from compact objects: \\ Debye series and Debye quasinormal modes}

	\author{Mohamed \surname{Ould~El~Hadj}\,\orcidlink{0000-0002-8558-7992}}
	\email{med.ouldelhadj@gmail.com}
	\affiliation{No-affiliation}

	\author{Sam R. \surname{Dolan}\,\orcidlink{0000-0002-4672-6523}}
	\email{s.dolan@sheffield.ac.uk}
	\affiliation{Consortium for Fundamental Physics, School of Mathematical and Physical
		Sciences, University of Sheffield, Hicks Building, Hounsfield Road, Sheffield S3
		7RH, United Kingdom}
	
	\date{\today}

\begin{abstract}   	
We introduce a new series decomposition of the waveform  constructed in the spirit of Debye expansions in scattering theory, and we use this to analyse the time-domain response of compact, horizonless bodies to scalar-field perturbations on curved spacetimes. The Debye decomposition  separates out direct exterior propagation, surface reflection, and successive transmissions through the interior of a compact body, and it provides an intuitive interpretation of the waveform in terms of geodesic trajectories. By analysing the quasinormal-mode (QNM) content of individual Debye terms, we set out a Debye-QNM description that is complementary to the standard QNM description. 
With this framework, we examine a scalar field propagating on two illustrative `Schwarzschild star' compact-body spacetimes: a neutron-star-like model \(R>3M\) and an ultracompact object \(R<3M\). We show that the closed-form Debye reconstruction matches well with the exact waveform, and that it converges even at early times, giving an accurate description of all waveform features including the prompt response. In the neutron-star case, the low-order Debye terms mainly describe the ringdown and a non-modal component associated with the sub-threshold branch cut. In the ultracompact case, the Debye series organizes the waveform into a prompt/ringdown contribution followed by a succession of echo-like wavepackets. The new Debye-QNM expansion and the standard QNM expansion have complementary spectral interpretations: the former identifies modes in individual propagation channels, whereas the latter describes collective resonances that are resummations of the former. This distinction clarifies how echo-like structures emerge from repeated interior propagation, and how pole and branch-cut contributions enter the time-domain signal.
\end{abstract}

	\maketitle
	
	\tableofcontents

\section{Introduction}
\label{sec:introduction}

The time-domain response of compact objects provides a direct probe of wave propagation in strong gravitational fields. In the era of gravitational-wave astronomy and horizon-scale imaging, understanding how the internal structure of horizonless compact objects may imprint itself on observable signals has become an important theoretical problem~\cite{Abbott:2016blz,EventHorizonTelescope:2019kwo,EventHorizonTelescope:2022urf,Cardoso:2019rvt}. 
For black holes, the post-merger signal is commonly interpreted in terms of quasinormal ringing, governed by the poles of the retarded Green function and by boundary conditions that are purely ingoing at the horizon and purely outgoing at spatial infinity~\cite{Nollert:1999ji,Kokkotas:1999bd,Berti:2009kk,Konoplya:2011qq,Berti:2016lat,Baibhav:2023clw,Berti:2025hly}. 
The extraction and interpretation of ringdown modes from time-domain signals has also motivated detailed studies of overtones, nonlinearities, and the regime of validity of linear perturbation theory~\cite{Buonanno:2006ui,London:2014cma,Giesler:2019uxc,Baibhav:2023clw,Giesler:2024hcr}.

For horizonless compact objects, the situation is richer. The absence of an event horizon allows part of the radiation to be reflected at the surface, transmitted through the interior, and repeatedly reprocessed before being emitted back to infinity. The resulting waveform may therefore contain, in addition to the usual ringdown, delayed echo-like pulses and nonmodal contributions associated with branch cuts and low-frequency backscattering. The role of branch cuts in Green-function reconstructions is well established in black-hole perturbation theory~\cite{Price:1972pwII,Price:1972pwI,Leaver:1986gd,Ching:1995tj,Casals:2011aa,Casals:2012tb,Casals:2012ng}. Echo-like responses, on the other hand, have been extensively discussed as possible signatures of ultracompact and exotic compact objects, and as a way of testing the presence or absence of an event horizon~\cite{Cardoso:2016rao,Cardoso:2016oxy,Mark:2017dnq,Cardoso:2017cqb,Maggio:2017ivp,Bueno:2017hyj,Conklin:2017lwb,Wang:2018gin,Testa:2018bzd,Maggio:2019zyv,Hui:2019aox,Cardoso:2019rvt,Maggio:2021uge,Rosato:2025byu,Rosato:2025lxb}. More broadly, echo-like signals may also arise from double-barrier effective potentials in modified-gravity settings, for example in black-hole spacetimes in massive gravity~\cite{Dong:2020odp}. A decade ago, a claim was made of tentative evidence for echoes in the first gravitational-wave data sets from LIGO \cite{Abedi:2016hgu}; but a reanalysis found a reduced significance consistent with noise \cite{Westerweck:2017hus}. Within the LVK Collaboration, systematic searches for echoes have been conducted from GWTC-2 onwards. No statistically significant evidence for echoes was found in the GWTC-2 and GWTC-3 data sets \cite{Gupta:2025utd}, and the recent analysis of GWTC-4 concludes \cite{LIGOScientific:2026wpt}: ``We find no evidence for post-merger echoes in the events that were analyzed.''

In the simplest picture, echoes arise from radiation temporarily trapped between an exterior curvature barrier and an inner reflecting region. For a regular compact star, however, the inner region is not merely a boundary. Waves can penetrate into the stellar interior, propagate through it, and return to the exterior after one or several internal traversals. This makes the interpretation of the signal more subtle: the observed response is not only determined by the global quasinormal-mode spectrum, but also by the propagation history of the waves inside and outside the compact object. Recent work has further emphasized that time-domain reconstructions generally require both modal and nonmodal sectors, including branch-cut contributions~\cite{Su:2026fvj,Ma:2026qbq}. In the standard Schwarzschild problem, these nonmodal terms are usually associated with the branch cut of the Green function conventionally placed along the negative imaginary frequency axis, which gives rise to the late-time tail. In the present compact-object problem, however, the interior wavenumber introduces an additional square-root branch structure. This gives rise to a sub-threshold cut on the real-frequency axis, \(0<\omega<\omega_c\), which is distinct from the usual Schwarzschild branch cut and plays a central role in the Debye reconstruction below.

This point also clarifies the relation with effective-reflectivity echo models. A common phenomenological strategy is to replace the near-surface or near-horizon structure by an effective boundary condition, or equivalently by a frequency-dependent reflectivity \(K(\omega)\), located at some effective tortoise-coordinate position \(x_0\)~\cite{Cardoso:2017cqb,Mark:2017dnq,Wang:2018gin,Testa:2018bzd,Maggio:2019zyv}. Such reduced descriptions are useful for capturing the existence, approximate spacing, and detectability of echo trains. Related approaches have also connected echoes to greybody factors, reflectionless scattering modes, and excitation factors of horizonless compact objects~\cite{Rosato:2025byu,Rosato:2025lxb}. In the present work, however, the stellar interior is not replaced by an effective boundary. Instead, the wave equation is solved through the regular interior, and the internal response is kept explicitly through the frequency-dependent quantity \(\xi_\ell(\omega)=c_\ell^{\rm out}(\omega)/c_\ell^{\rm in}(\omega)\), namely the interior phase factor. The Debye construction may therefore be viewed as a resolved version of an effective-reflectivity description: it retains the propagation through the compact-object interior and exposes the spectral structure associated with the individual propagation channels, in particular the structure controlled by the zeros of \(c_\ell^{\rm in}(\omega)\).

A promising way to disentangle these propagation histories is provided by Debye-type expansions. In ordinary scattering theory, Debye series reorganize a wave amplitude into contributions associated with successive reflections and transmissions~\cite{Nussenzveig:2006}. In the compact-object problem, an analogous construction can be used to decompose the frequency-domain response into terms with a clear trajectory interpretation. In a previous work~\cite{OuldElHadj:2026ewh}, one of us developed such a Debye-series description for the scattering matrix of compact, horizonless objects and studied the associated Regge--Debye poles in the complex angular-momentum plane. The purpose of the present work is to extend this approach to the time-domain response to a scalar-field perturbation.

We consider a massless scalar field propagating on the `Schwarzschild star' spacetime: a static, spherically symmetric, uniform-density compact object with a Schwarzschild exterior. The excitation is described as an initial-value problem with Gaussian Cauchy data localized outside the compact object. Starting from the frequency-domain Green-function representation of the waveform, we rewrite the response kernel in a cavity form and expand the corresponding Fabry--P\'erot-like denominator as a Debye series. This separates the response into a leading exterior contribution and a sequence of terms associated with one or more interior traversals before re-emission to infinity. The resulting decomposition provides a natural framework to identify the parts of the signal associated with direct exterior propagation, surface reflection, propagation through the interior, and branch-cut contributions.

A central point of this work is the distinction between ordinary quasinormal modes (QNMs) and Debye quasinormal modes (D-QNMs). The ordinary QNMs are the poles of the full, resummed response. By contrast, the Debye QNMs, or D-QNMs, are associated with the singularities of the individual Debye building blocks. These spectra are related, but they do not coincide term by term. The Debye representation therefore does not merely provide another way of reconstructing the waveform; it gives access to the modal and nonmodal content of each propagation channel separately.

This distinction is especially important for ultracompact objects. In the ordinary QNM reconstruction, the late-time echo train is typically controlled by long-lived trapped modes of the complete response. In the Debye representation, however, a fixed Debye order corresponds to a definite propagation channel involving a finite number of interior traversals. The dominant D-QNM contribution to a given echo packet need not come from the same spectral family that dominates the QNM reconstruction of the fully resummed signal. This is not a contradiction, but rather reflects the reorganization of the same physical response: ordinary QNMs describe collective resonances of the complete compact-object system, whereas D-QNMs resolve the spectral content of individual propagation paths.

We apply the formalism to two representative configurations. The first one, \(R=6M\), is a neutron-star-like model without long-lived trapped modes. The second one, \(R=2.26M\), is an ultracompact model close to the Buchdahl limit. For both configurations, we compute the relevant QNM and D-QNM spectra, construct the Debye decomposition of the waveform, and analyze the role of the sub-threshold branch cut generated by the interior wavenumber. We show that the two configurations display qualitatively different behavior. In the \(R=6M\) case, the low-order Debye terms contribute mainly to the ringdown and to a nonmodal component controlled by the branch cut, while no clear sequence of well-separated echoes is produced. In the ultracompact case, the Debye series organizes the response into a prompt/ringdown contribution followed by successive echo-like packets, whose timing is governed by propagation through the stellar interior.

The paper is organized as follows. In Sec.~\ref{sec:sec_1}, we introduce the compact-object model, the scalar perturbation equation, and the frequency-domain construction of the waveform. We also review the ordinary QNM spectrum and its contribution to the time-domain response. In Sec.~\ref{sec:sec_2}, we derive the Debye-series representation of the response kernel and identify the D-QNM and branch-cut contributions of each Debye order. In Sec.~\ref{sec:sec_3}, we present the numerical results for the two configurations \(R=6M\) and \(R=2.26M\), comparing the full waveform, the Debye reconstruction, the ordinary QNM content, and the D-QNM-plus-cut decomposition. The technical treatment of the sub-threshold cut, including the Cauchy principal-value and Hadamard finite-part prescriptions, is given in Appendix~\ref{app:cut_regularization}. Appendix~\ref{app:effective_reflectivity} discusses the relation between the Debye construction and effective-reflectivity echo models. Finally, Sec.~\ref{sec:conclusion} summarizes our results and discusses possible extensions.

Throughout this article, we adopt natural units such that \(\hbar=c=G=1\). We also assume a harmonic time dependence of the form \(e^{-i\omega t}\) for the perturbative field.

\section{The initial-value problem and the standard QNM expansion}
\label{sec:sec_1}

\subsection{Theoretical considerations}
\label{sec:sec_1_1}

\subsubsection{Construction of the waveform}
\label{sec:sec_1_1_1}

We consider a massless scalar field $\Phi$ propagating on a static, spherically symmetric compact-object spacetime with a regular center at $r=0$ and a Schwarzschild exterior for $r>R$. In Schwarzschild-like coordinates $\{t,r,\theta,\phi\}$, the line element is
\begin{equation}
	\dd s^2=-F(r)\,\dd t^2+H(r)^{-1}\,\dd r^2+r^2\,\dd\sigma^2,
	\label{eq:metric_waveform}
\end{equation}
where $\dd\sigma^2=\dd\theta^2+\sin^2\theta\,\dd\phi^2$, and in the exterior region
\begin{equation}
	F(r)=H(r)=1-\frac{2M}{r},
	\qquad r>R.
	\label{eq:schw_ext_waveform}
\end{equation}

The interior functions $F(r)$ and $H(r)$ are specified by a stellar model. In this work we adopt, as in Ref.~\cite{OuldElHadj:2019kji}, the incompressible perfect-fluid (constant-density) model~\cite{Shapiro1983}, for which the interior metric functions are
\begin{subequations}\label{Interior_Solution}
	\begin{align}\label{Interior_Solution_f}
		F(r) &= \frac{1}{4}\left(1-\frac{2 M r^2}{R^3}\right)+\frac{9}{4}\left(1-\frac{2M}{R}\right) \nonumber \\
		&\quad -\frac{3}{2} \sqrt{\left(1-\frac{2M}{R}\right)\left(1-\frac{2 M r^2}{R^3}\right)}, \\
		H(r) &= 1-\frac{2 M r^2}{R^3}.
		\label{Interior_Solution_h}
	\end{align}
\end{subequations}

The corresponding partial amplitude $\phi_\ell(t,r)$ satisfies
\begin{equation}
	\left[
	-\frac{\partial^2}{\partial t^2}
	+\frac{\partial^2}{\partial r_\ast^2}
	- V_\ell(r)
	\right]\phi_\ell(t,r)=0,
	\label{eq:RW_time_compact_waveform}
\end{equation}
where the tortoise coordinate is defined by
\begin{equation}
	\frac{\dd r_\ast}{\dd r}=\frac{1}{\sqrt{F(r)H(r)}},
	\label{eq:tortoise_waveform}
\end{equation}
with the additive constant chosen so that $r_\ast(r)$ is continuous at $r=R$.
For a massless scalar field, the effective potential is
\begin{equation}
	V_{\ell}(r)=F(r)\left[\frac{\ell(\ell+1)}{r^{2}}
	+\frac{H(r)}{2r}\left(\frac{F'(r)}{F(r)}+\frac{H'(r)}{H(r)}\right)\right].
	\label{eq:V_general_scalar_waveform}
\end{equation}
where a prime denotes differentiation with respect to r. In the Schwarzschild exterior, it reduces to the familiar Regge–Wheeler-type potential
\begin{align}
	V_\ell^{\rm (ext)}(r)
		=\left(1-\frac{2M}{r}\right)\left[\frac{\ell(\ell+1)}{r^2}+\frac{2M}{r^3}\right].
		\label{eq:Vext_waveform}
\end{align}
At the surface $r=R$, we impose continuity of the field and of its first $r_\ast$-derivative.

\begin{figure}[!htb]
	\centering
	\includegraphics[scale=0.50]{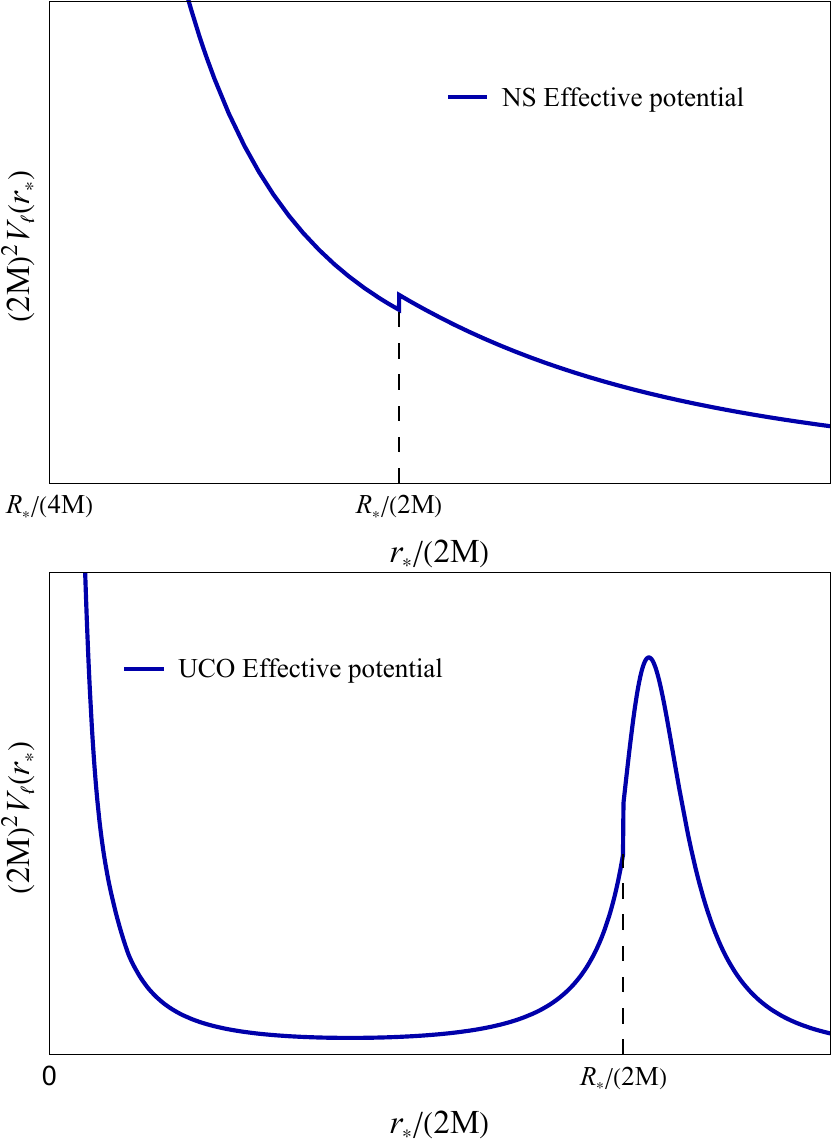}
	\caption{
		Effective scalar potential \(V_\ell\) for a quadrupole \((\ell=2)\) perturbation of a compact body of constant density, shown for two tenuities: \(R/M=6\) and \(R/M=2.26\). 
		The horizontal axis is the tortoise coordinate \(r_\ast/(2M)\), defined in Eq.~\eqref{eq:tortoise_waveform}. 
		For \(R/M=6\), the stellar surface lies outside the ultracompact region, whereas for \(R/M=2.26\) it lies inside the exterior curvature barrier, producing a trapping cavity. 
		The vertical dashed line marks the stellar surface, where the effective potential is discontinuous.}
	\label{fig:Effective_Potential_NS_UCO}
\end{figure}
Figure~\ref{fig:Effective_Potential_NS_UCO} illustrates the structure of this effective potential as a function of the tortoise coordinate for two representative configurations. In the neutron-star-like case, the stellar surface lies outside the region where a trapping cavity can form. By contrast, in the ultracompact case, the surface is located inside the exterior curvature barrier, producing a cavity between the surface and the barrier. The figure also displays the jump of the effective potential at the stellar surface, which follows from the matching of the interior and exterior geometries.

We now describe the perturbation by an initial-value problem with Gaussian initial data. More precisely, we assume that the partial amplitude $\phi_\ell(t, r)$ solution of \eqref{eq:RW_time_compact_waveform} is given, at $t = 0$, by
\begin{equation}
	\phi_{\ell}(t = 0,r)=\phi_{0}\,
	\exp\!\left[-\frac{a^2}{(2M)^2}\bigl[r_\ast(r)-r_\ast(r_0)\bigr]^2\right].
	\label{eq:Cauchy_data_waveform}
\end{equation}
and satisfies $\partial_t\phi_\ell(t = 0,r)=0$. In what follows, we assume that $r_0>R$ and that the Gaussian is sufficiently narrow so that its support is effectively concentrated in the exterior region.

Let $G_\ell^{\rm ret}(t;r,r')$ denote the retarded Green function associated with Eq.~\eqref{eq:RW_time_compact_waveform}, i.e.
\begin{equation}
	\left[
	-\frac{\partial^2}{\partial t^2}
	+\frac{\partial^2}{\partial r_\ast^2}
	- V_\ell(r)
	\right]G_\ell^{\rm ret}(t;r,r')
	=-\delta(t)\,\delta(r_\ast-r_\ast'),
	\label{eq:Gret_waveform}
\end{equation}
together with the causality condition $	G_\ell^{\rm ret}(t;r,r')=0,\qquad t\le 0$, Green's identity then gives, for $t>0$,
\begin{equation}
	\phi_\ell(t,r)=\int_{r_\ast(0)}^{+\infty}
	\partial_t G_\ell^{\rm ret}(t;r,r')\,\phi_{\ell}(t=0,r')\,\dd r'_\ast.
	\label{eq:TimeEvolution_waveform}
\end{equation}

We now introduce the frequency-domain representation
\begin{equation}
	G_\ell^{\rm ret}(t;r,r')
	=
	-\int_{-\infty+ic}^{+\infty+ic}\frac{\dd\omega}{2\pi}\,
	\frac{\phi^{\rm reg}_{\omega\ell}(r_<)\,\phi^{\rm up}_{\omega\ell}(r_>)}{W_\ell(\omega)}\,
	e^{-i\omega t},
	\label{eq:Gret_omega_waveform}
\end{equation}
where $c>0$, $r_< = \min(r,r')$, $r_> = \max(r,r')$, and $	W_\ell(\omega)$ is the Wronskian with respect to $r_\ast$ of the functions $\phi^{\rm reg}$ and  $\phi^{\rm up}$. Here $\phi^{\rm reg}_{\omega\ell}$ denotes the solution of
\begin{equation}
	\frac{\dd^2\phi_{\omega\ell}}{\dd r_\ast^2}
	+\bigl[\omega^2-V_\ell(r)\bigr]\phi_{\omega\ell}=0
	\label{eq:RW_freq_waveform}
\end{equation}
which is regular at the center,
\begin{equation}
	\phi^{\rm reg}_{\omega\ell}(r)\propto r^{\ell+1},
	\qquad r\to 0,
	\label{eq:bc_reg_center_waveform}
\end{equation}
and is continued to the exterior by $C^1$ matching at $r=R$.
The second solution, $\phi^{\rm up}_{\omega\ell}$, is defined by its purely outgoing behavior at spatial infinity,
\begin{equation}
	\phi^{\rm up}_{\omega\ell}(r)\underset{r_\ast\to+\infty}{\sim}e^{+i\omega r_\ast}.
	\label{eq:bc_up_waveform}
\end{equation}
In the exterior region, the regular solution has the asymptotic decomposition
\begin{equation}
	\phi^{\rm reg}_{\omega\ell}(r)\underset{r_\ast\to+\infty}{\sim}
	A_\ell^{(-)}(\omega)\,e^{-i\omega r_\ast}
	+A_\ell^{(+)}(\omega)\,e^{+i\omega r_\ast},
	\label{eq:asympt_reg_waveform}
\end{equation}
so that
\begin{equation}
	W_\ell(\omega)=2i\omega\,A_\ell^{(-)}(\omega).
	\label{eq:W_Aminus_waveform}
\end{equation}

Substituting Eq.~\eqref{eq:Gret_omega_waveform} into Eq.~\eqref{eq:TimeEvolution_waveform}, differentiating with respect to (t), and using Eq.~\eqref{eq:W_Aminus_waveform}, while also assuming that the initial data are strongly localized near ($r=r_0>R$) and that the observer is located at a larger radius, so that ($r>r_0$) over the effective support of ($\phi_{\ell}(t = 0, r)$), we have ($r_< = r'$) and ($r_> = r$), we obtain
\begin{align}
	\phi_\ell(t,r)
	=&
	\frac{1}{2}\int_{-\infty+ic}^{+\infty+ic}\frac{\dd\omega}{2\pi}\,
	\left(\frac{e^{-i\omega t}}{A_\ell^{(-)}(\omega)}\right)
	\phi^{\rm up}_{\omega\ell}(r)\, \nonumber \\
	&\times \int_{r_\ast(0)}^{+\infty}\dd r'_\ast\,
	\phi_{\ell}(t = 0, r')\,\phi^{\rm reg}_{\omega\ell}(r').
	\label{eq:phi_reduced_frequency_waveform}
\end{align}
For a real potential and real initial data, the integrand has the usual symmetry under $\omega\mapsto-\omega^\ast$, and the waveform can therefore be written as
\begin{multline}
	\phi_\ell(t,r)=
	\frac{1}{2\pi}\,\Re\!\Biggl[
	\int_{0+ic}^{+\infty+ic}\dd\omega\,
	\left(\frac{e^{-i\omega t}}{A_\ell^{(-)}(\omega)}\right)
	\phi^{\rm up}_{\omega\ell}(r)
	\\
	\times
	\int_{r_\ast(0)}^{+\infty}\dd r'_\ast\,
	\phi_{\ell}(t = 0 ,r')\,\phi^{\rm reg}_{\omega\ell}(r')
	\Biggr].
	\label{eq:response_partial_waveform}
\end{multline}

Using Eq.~\eqref{eq:bc_up_waveform}, the waveform measured at infinity takes the form
\begin{multline}
	\phi_\ell(t,r)=
	\frac{1}{2\pi}\,\Re\!\Biggl[
	\int_{0+ic}^{+\infty+ic}\dd\omega\,
	\left(\frac{e^{-i\omega (t-r_\ast)}}{A_\ell^{(-)}(\omega)}\right)
	\\
	\times
	\int_{r_\ast(0)}^{+\infty}\dd r'_\ast\,
	\phi_{\ell}(t=0,r')\,\phi^{\rm reg}_{\omega\ell}(r')
	\Biggr].
	\label{eq:response_infinity_waveform}
\end{multline}
This is the expression that will be used to reconstruct numerically the waveform seen by an observer at infinity.

\subsubsection{Quasinormal mode spectrum}
\label{sec:sec_1_1_2}

The quasinormal frequencies are the zeros of $A_\ell^{(-)}(\omega)$, or equivalently of $W_\ell(\omega)$ for each integer multipole index $\ell \in \mathbb{N}$, with $\omega_{\ell n} \in \mathbb{C}$. For compact objects, the spectrum may in general split into several distinct families. It is therefore convenient to denote the QNM frequencies by $\omega_{\ell n}^{(a)}$, where $a$ labels the family and $n=0,1,2,\dots$ orders the modes within that family by increasing damping. 

Quasinormal oscillations of spherically symmetric compact objects have been extensively investigated in the literature; see, for example, Refs.~\cite{Kokkotas:1986gd,Kokkotas:1992ka,Andersson:1997eq,Kokkotas:1999bd,Leins:1993zz}. Newly born neutron stars, produced in the aftermath of core-collapse supernovae, are expected to undergo strong pulsations with substantial initial energy, thereby emitting gravitational radiation. The classification of fluid oscillation modes in relativistic stars was first introduced by Thorne and Campolattaro in 1967 \cite{thorne1967non}, following the Newtonian description of stellar pulsations but incorporating gravitational-wave damping. This topic was revisited about twenty years later \cite{Detweiler:1985zz,Kokkotas:1986gd}, and Kokkotas and Schutz \cite{Kokkotas:1992ka} identified an additional class of oscillations, now known as $w$-modes. These modes involve little coupling to the stellar fluid and, in the axial sector, no fluid motion whatsoever. They are strongly damped and are best interpreted as oscillations of the perturbed spacetime itself. A broad overview of gravitational quasinormal modes of relativistic stars and black holes may be found in Ref.~\cite{Kokkotas:1999bd}.

The family of $w$-modes can be separated into three main branches:
\begin{enumerate}
	\item \textit{Curvature modes}: the usual $w$-modes, present in all relativistic stars. Their damping becomes stronger as the stellar compactness decreases, i.e.~as $R/M$ increases and Im$\{\omega\}$ grows.
	\item \textit{Interface modes} (or $\omega_{\text{II}}$-modes \cite{Leins:1993zz}): these are distinguished by extremely rapid decay, corresponding to a large negative imaginary part of $\omega_{\ell n}$. They resemble the modes arising in the scattering of acoustic waves by a hard sphere.
	\item \textit{Trapped modes} \cite{Chandrasekhar449}: these may occur when the effective radial potential develops a well or cavity, as happens for ultra-compact objects with $R/M<3$. As the potential well becomes deeper, the number of trapped modes increases while their damping rate decreases.
\end{enumerate}

\subsubsection{Extraction of the QNM contribution}
\label{sec:sec_1_1_3}

The contour of integration in Eq.~\eqref{eq:response_infinity_waveform} may be deformed in order to capture the QNM contribution. By Cauchy's theorem, and neglecting the other contributions arising from the large-$|\omega|$ arcs and from cut terms, one obtains a residue series over the quasinormal frequencies in the lower half-plane. This isolates the compact-object ringing generated by the initial data:
\begin{equation}
	\phi_\ell^{\rm QNM}(t,r)=
	2\,\Re\!\left[
	\sum_{a}\sum_{n}
	-i\,\omega_{\ell n}^{(a)}\,
	\mathcal{C}_{\ell n}^{(a)}\,
	e^{-i\omega_{\ell n}^{(a)}(t-r_\ast)}
	\right].
	\label{eq:TimeEvolution_QNM_infinity_waveform}
\end{equation}
The corresponding intrinsic excitation factor is
\begin{equation}
	\mathcal{B}_{\ell n}^{(a)}
	=
	\left[
	\frac{1}{2\omega}\,
	\frac{A_\ell^{(+)}(\omega)}
	{\displaystyle \frac{dA_\ell^{(-)}(\omega)}{d\omega}}
	\right]_{\omega=\omega_{\ell n}^{(a)}},
	\label{eq:ExcitationFactor_waveform}
\end{equation}
and the excitation coefficient, which incorporates the chosen initial data, is
\begin{equation}
	\mathcal{C}_{\ell n}^{(a)}
	=
	\mathcal{B}_{\ell n}^{(a)}
	\int_{r_\ast(0)}^{+\infty}\dd r'_\ast\,
	\frac{\phi_{\ell}(t = 0, r')\, \phi^{\rm reg}_{\omega_{\ell n}^{(a)}\ell}(r')}{A_\ell^{(+)}\!\left(\omega_{\ell n}^{(a)}\right)}.
	\label{eq:ExcitationCoefficient_waveform}
\end{equation}

For further details concerning the excitation factors (intrinsic quantities) and the excitation coefficients (extrinsic quantities), we refer to Refs.~\cite{Decanini:2014bwa,Decanini:2014kha,Berti:2006wq}. As in the black-hole problem, the representation \eqref{eq:TimeEvolution_QNM_infinity_waveform} is not expected to approximate the full waveform at arbitrarily early times. Indeed, a backward extrapolation of the damped QNM contribution \(\phi_\ell^{\rm QNM}(t,r)\) is exponentially divergent and cannot provide a physically meaningful description of the prompt response. The QNM waveform must therefore be used only after a physically motivated onset time. In the applications below, this onset is identified with the appropriate ringdown time \(t_{\rm ring}\), estimated from propagation considerations involving the support of the initial data, the relevant exterior scattering region, and the observer. The corresponding estimates are discussed in Sec.~\ref{sec:sec_1_2_3}. 

A complementary finite-time viewpoint has recently been emphasized in the Green-function literature~\cite{Su:2026fvj,Chavda:2024awq}. In this formulation, the effective QNM amplitude, after the modal exponential has been factored out, can depend on time because the portion of the initial data that has causally interacted with the scattering region changes with time. The standard residue construction used here instead gives the time-independent excitation coefficients in Eq.~\eqref{eq:ExcitationCoefficient_waveform}. For localized initial data, these effective finite-time amplitudes are expected to approach the constant residue coefficients once the relevant part of the initial packet has interacted with the scattering region. In the comparisons below, the QNM reconstructions are therefore used in the ringdown and echo windows, not as a description of the prompt causal transition.

\subsection{Results and discussions}
\label{sec:sec_1_2}

\subsubsection{Numerical methods}
\label{sec:sec_1_2_1}

The quasinormal frequencies displayed in this section are computed with the same numerical strategy as that used in Ref.~\cite{OuldElHadj:2019kji} for the determination of the Regge poles of \(A_\ell^{(-)}(\omega)\). In the present problem, the roles of the variables are exchanged: instead of fixing the frequency \(\omega\) and solving for complex angular momenta, we fix the multipolar index \(\ell\), here \(\ell=2\), and determine the zeros of \(A_\ell^{(-)}(\omega)\) in the complex-frequency plane. Since the method has already been described in detail in Ref.~\cite{OuldElHadj:2019kji}, we do not repeat its technical implementation here.

The time-domain waveforms, Eq.~\eqref{eq:response_infinity_waveform}, and the corresponding QNM reconstructions, Eq.~\eqref{eq:TimeEvolution_QNM_infinity_waveform}, are obtained from numerical solutions of the radial equation \eqref{eq:RW_freq_waveform}. The regular interior solution is initialized near the center by means of a Frobenius expansion implementing the boundary condition \eqref{eq:bc_reg_center_waveform}, as in Refs.~\cite{OuldElHadj:2019kji,OuldElHadj:2026ewh}. The solution is then integrated up to the stellar surface and matched at \(r=R\) to the exterior Regge--Wheeler solutions.

In the exterior region, the radial equation is integrated and the numerical solution is compared with the ingoing and outgoing asymptotic expansions at spatial infinity, Eq.~\eqref{eq:asympt_reg_waveform}. This matching gives the scattering coefficient \(A_\ell^{(-)}(\omega)\). The same numerical ingredients are used to evaluate the intrinsic excitation factors \eqref{eq:ExcitationFactor_waveform} and the excitation coefficients \eqref{eq:ExcitationCoefficient_waveform}. We follow the implementation of Refs.~\cite{Decanini:2014bwa,Decanini:2016ifm}, adapted here to compact-object backgrounds. The modal reconstructions shown below are finite sums over the relevant QNM families; the number of modes retained in each case is specified in the corresponding figure captions.

For the time-domain waveforms obtained from the real-frequency inverse transform, the response was sampled on the finite frequency window \(-6\leq 2M\omega\leq 6\), with a standard frequency resolution \(\delta(2M\omega)=10^{-4}\).

The ultracompact configuration requires particular care because the least-damped trapped modes lie very close to the real-frequency axis. For this case, we therefore repeated the waveform calculation using the finer spacing
$
\delta(2M\omega)=10^{-5}.
$
For the fundamental trapped mode listed in Table~\ref{tab:qnm_modes_UCO},
$
|{\rm Im}(2M\omega_{\ell n}^{(T)})|\simeq 0.84\times10^{-4}.
$
The refined spacing is thus smaller than this characteristic damping scale by a factor of approximately \(8.4\), providing a substantially finer sampling of the narrowest trapped-mode structure.

To quantify the sensitivity of the time-domain waveform to the frequency sampling, we also computed the same ultracompact waveform with the standard spacing \(\delta(2M\omega)=10^{-4}\). Taking the \(\delta(2M\omega)=10^{-5}\) result as the reference, we define the relative maximum difference by
\begin{equation}
L_\infty=
\frac{
	\max_t|\phi_{10^{-4}}(t)-\phi_{10^{-5}}(t)|
}{
	\max_t|\phi_{10^{-5}}(t)|
},
\end{equation}
and the relative \(L_2\) difference by
\begin{equation}
L_2=
\left[
\frac{
	\sum_t|\phi_{10^{-4}}(t)-\phi_{10^{-5}}(t)|^2
}{
	\sum_t|\phi_{10^{-5}}(t)|^2
}
\right]^{1/2}.
\end{equation}
We obtain $L_\infty=1.20\times10^{-4}$ and $L_2=6.23\times10^{-4}$,
with a maximum absolute difference of $3.93\times10^{-5}$. Thus, changing the frequency resolution by one order of magnitude changes the reconstructed time-domain waveform by less than \(10^{-3}\) according to both relative measures.

As an additional a posteriori check, we analysed the Fourier spectrum of the late-time ultracompact waveform obtained with the standard spacing \(\delta(2M\omega)=10^{-4}\). Distinct spectral peaks occur at frequencies coinciding with the real parts of the low-lying trapped-mode QNM frequencies listed in Table~\ref{tab:qnm_modes_UCO}. This shows that the long-lived trapped-mode frequencies are already present in the waveform obtained with the standard frequency sampling.

\subsubsection{Quasinormal mode spectrum}
\label{sec:sec_1_2_2}

\begin{figure}[!htb]
	\centering
	\includegraphics[scale=0.50]{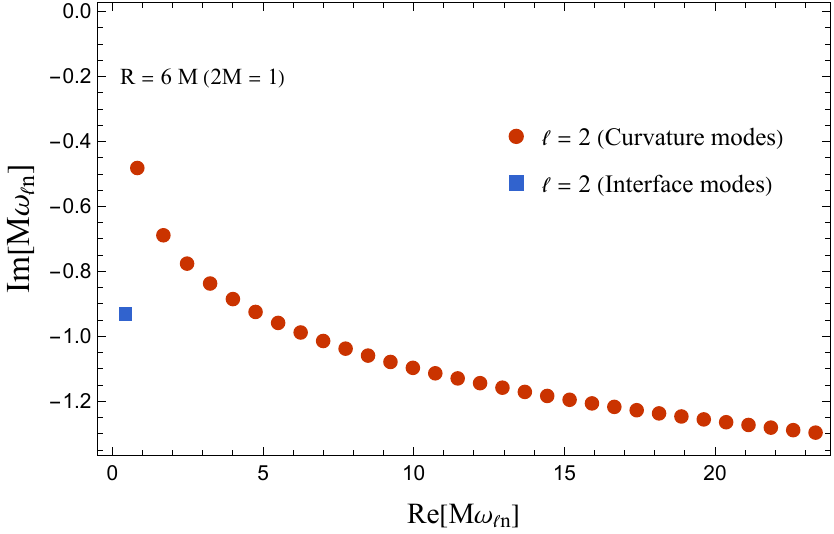}
	\caption{ Quasinormal-frequency spectrum for the \(\ell=2\) perturbation of a neutron-star-like compact object with radius \(R=6M\), in units where \(2M=1\). 
		The red circles denote the curvature-mode family, while the blue squares denote the interface-mode family.}
	\label{fig:QNFs_NS_R6M_l_2}
\end{figure}

\begin{figure}[!htb]
	\centering
	\includegraphics[scale=0.50]{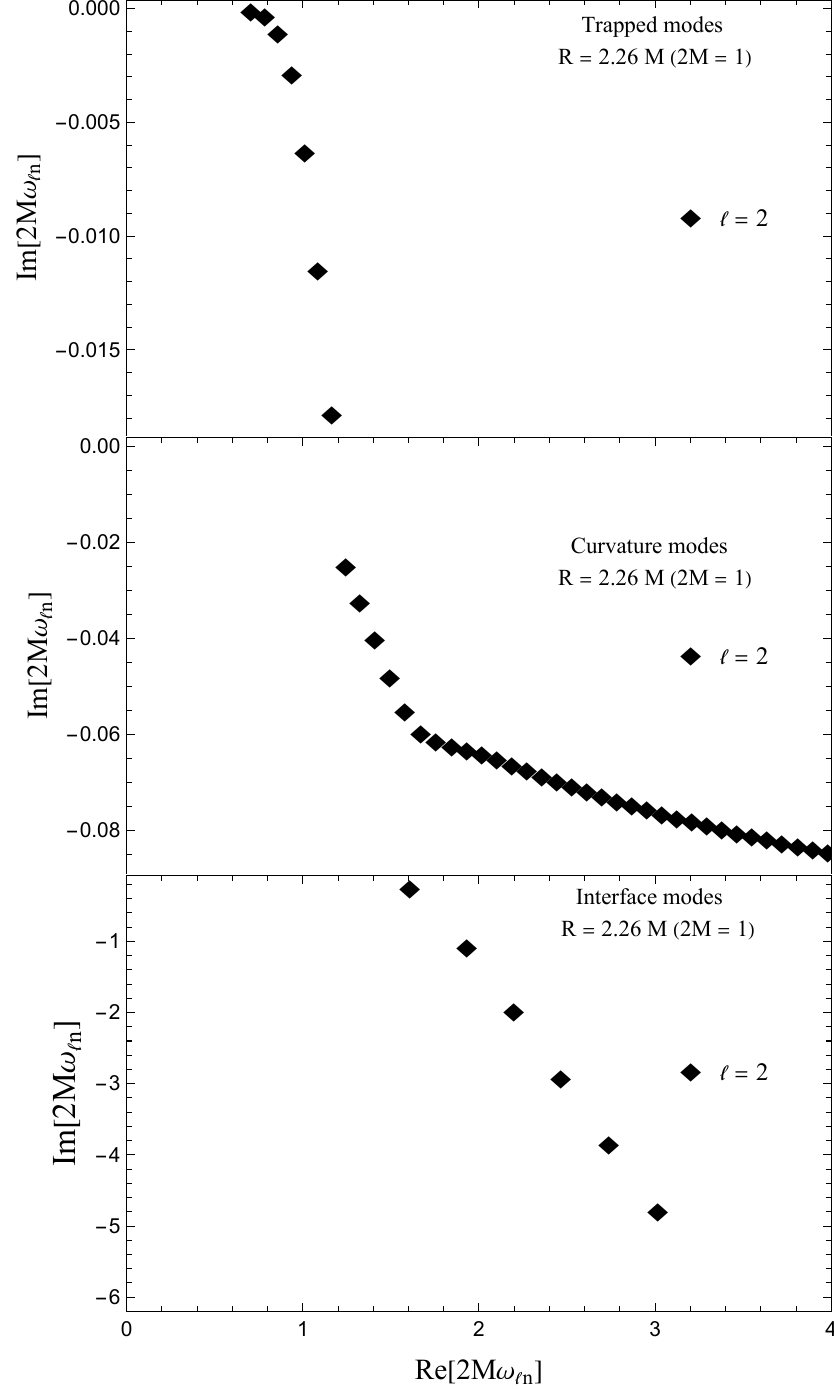}
		\caption{Quasinormal-mode spectrum for the $\ell=2$ perturbation of an ultracompact object with radius $R=2.26M$ (in units where $2M=1$). The top, middle, and bottom panels display respectively the trapped, curvature, and interface-mode families.}
	\label{fig:QNFs_UCO_R226M_l_2}
\end{figure}

In this section, we present numerical results for the quasinormal-mode spectrum of the massless scalar field in two representative configurations: (i) a neutron-star-like configuration with $R=6M$, and (ii) an ultracompact object with $R=2.26M$, close to the Buchdahl limit. In both cases, we focus on the $\ell=2$ perturbation. The corresponding spectra are displayed in Figs.~\ref{fig:QNFs_NS_R6M_l_2} and \ref{fig:QNFs_UCO_R226M_l_2}, while the lowest quasinormal frequencies are listed in Tables~\ref{tab:qnm_modes_NS} and \ref{tab:qnm_modes_UCO}.

For the $R=6M$ model, the spectrum contains two distinct families, namely the curvature modes and a low-lying interface branch, as shown in Fig.~\ref{fig:QNFs_NS_R6M_l_2}. No trapped modes are found in this case, consistently with the absence of a sufficiently deep cavity in the corresponding effective potential.

The situation changes qualitatively for the ultracompact configuration. As shown in Fig.~\ref{fig:QNFs_UCO_R226M_l_2}, the spectrum separates into three distinct families, namely trapped modes, curvature modes, and interface modes. The trapped branch is characterized by very small imaginary parts, indicating long-lived oscillations confined inside the effective potential cavity. By contrast, the curvature modes are more strongly damped and follow the usual photon-sphere-type branch, while the interface modes are the most strongly damped and are associated with the stellar surface.

\begin{table}[t]
	\centering
	\caption{The lowest quasinormal frequencies for the massless scalar field. The radius of the compact body is $R=6M$.}
	\label{tab:qnm_modes_NS}
	\small
	\setlength{\tabcolsep}{6pt}
	\renewcommand{\arraystretch}{1.15}
	\begin{tabular}{c c l l l}
		\hline\hline
		$\ell$ & $n$ & $2M\omega_{\ell n}^{(C)}$ & $2M\omega_{\ell n}^{(I)}$ \\
		\hline
		2 & 0  & $0.825581 - 0.481490i$ & $0.434659 - 0.931937i$ \\
		& 1  & $1.691780 - 0.688623i$ & \\
		& 2  & $2.477731 - 0.775894i$ & \\
		& 3  & $3.242721 - 0.837147i$ & \\
		& 4  & $3.999125 - 0.885099i$ & \\
		& 5  & $4.751044 - 0.924692i$ & \\
		& 6  & $5.500284 - 0.958478i$ & \\
		& 7  & $6.247781 - 0.987972i$ & \\
		& 8  & $6.994073 - 1.014156i$ & \\
		& 9  & $7.739495 - 1.037705i$ & \\
		& 10 & $8.484266 - 1.059105i$ & \\
		\hline\hline
	\end{tabular}
\end{table}

\begin{table*}[t]
	\centering
	\caption{The lowest quasinormal frequencies for the massless scalar field. The radius of the compact body is $R=2.26M$.}
	\label{tab:qnm_modes_UCO}
	\small
	\setlength{\tabcolsep}{6pt}
	\renewcommand{\arraystretch}{1.15}
	\begin{tabular}{c c l l l}
		\hline\hline
			$\ell$ & $n$ & $2M\omega_{\ell n}^{(T)}$ & $2M\omega_{\ell n}^{(C)}$ & $2M\omega_{\ell n}^{(I)}$ \\
			\hline
			2 & 0  & $0.709336 - 0.000084i$ & $1.246783 - 0.024798i$ & $1.611191 - 0.240750i$ \\
			& 1  & $0.785701 - 0.000320i$ & $1.327814 - 0.032206i$ & $1.930829 - 1.079527i$ \\
			& 2  & $0.861624 - 0.001040i$ & $1.410388 - 0.039952i$ & $2.197702 - 1.980245i$ \\
			& 3  & $0.937107 - 0.002836i$ & $1.494895 - 0.047868i$ & $2.465613 - 2.909475i$ \\
			& 4  & $1.012678 - 0.006307i$ & $1.581970 - 0.055063i$ & $2.740755 - 3.846993i$ \\
			& 5  & $1.089206 - 0.011474i$ & $1.670916 - 0.059544i$ & $3.015896 - 4.784510i$ \\
			& 6  & $1.167222 - 0.017807i$ & $1.759103 - 0.061298i$ & \\
			& 7  &                        & $1.845853 - 0.062184i$ & \\
			& 8  &                        & $1.931760 - 0.063039i$ & \\
			& 9  &                        & $2.017270 - 0.064009i$ & \\
			& 10 &                        & $2.102608 - 0.065076i$ & \\
		\hline\hline
	\end{tabular}
\end{table*}

\subsubsection{Interpreting the response: Characteristic timescales}
\label{sec:sec_1_2_3}

To aid the interpretation of the time-domain signals (presented in the next section), it is useful to first introduce a set of characteristic propagation times. We first write explicitly the tortoise coordinate associated with Eq.~\eqref{eq:tortoise_waveform}. In the exterior Schwarzschild region, \(r\geq R\),
\begin{equation}
	r_*(r)=r+2M\ln\left(\frac{r}{2M}-1\right),
	\label{eq:Ext_tortoise}
\end{equation}
whereas in the interior region, \(0\leq r\leq R\),
\begin{equation}
	r_*(r)=r_*(R)-\int_r^R \frac{d\rho}{\sqrt{F(\rho)H(\rho)}}.
	\label{eq:Int_tortoise}
\end{equation}
The effective potential as a function of $r_*$ is shown in Fig.~\ref{fig:Effective_Potential_NS_UCO}.

We define the interior optical length
\begin{equation}
	L_{\rm int}\equiv \int_0^R \frac{d\rho}{\sqrt{F(\rho)H(\rho)}} ,
	\label{eq:Int_optical_length}
\end{equation}
so that
\begin{equation}
	r_*(0)=r_*(R)-L_{\rm int}.
	\label{eq:Center_coord}
\end{equation}
This length measures the propagation delay across the stellar interior and will be used below to identify the expected echo windows. 

The waveforms considered below are generated by Gaussian Cauchy data localized outside the compact object. Let \(r_0\) denote the center of the initial Gaussian packet and \(r_{\rm obs}\) the position of the observer. The arrival time of the direct signal is estimated by
\begin{equation}
	t_{\rm dir}\simeq r_*(r_{\rm obs})-r_*(r_0).
	\label{eq:dir_time}
\end{equation}
For an ultracompact configuration, \(R<3M\), the first ringdown pulse is naturally associated with propagation to the exterior curvature barrier. In the geometric-optics estimate used below, this region is identified with the photon-sphere radius \(r=3M\). Its characteristic onset is therefore estimated by
\begin{equation}
	t_{\rm ring}\simeq r_*(r_0)+r_*(r_{\rm obs})-2r_*(3M).
	\label{eq:ring_time_uco}
\end{equation}
The first echo is instead controlled by the optical length of the interior and is estimated as
\begin{align}
	t_{{\rm echo},1}
	&\simeq r_*(r_0)+r_*(r_{\rm obs})-2r_*(0)
	\nonumber\\
	&=r_*(r_0)+r_*(r_{\rm obs})-2r_*(R)+2L_{\rm int}.
	\label{eq:echo1_time}
\end{align}
The delay between successive echoes is approximated by the round-trip time across the effective cavity,
\begin{align}
	\Delta t_{\rm echo}
	&\simeq 2\bigl[r_*(3M)-r_*(0)\bigr]
	\nonumber\\
	&=2\bigl[r_*(3M)-r_*(R)+L_{\rm int}\bigr].
	\label{eq:echo_delay}
\end{align}
It is useful to compare Eq.~\eqref{eq:echo_delay} with the echo-delay estimates commonly used in phenomenological models of ultracompact objects (see for exemple Ref.~\cite{Cardoso:2017cqb}). In models where the stellar interior is replaced by an effective reflecting surface located at a tortoise coordinate \(x_0\), the echo delay is usually estimated as
\begin{equation}
	\Delta t_{\rm echo}^{\rm eff}
	\simeq
	2\bigl[r_*(3M)-x_0\bigr].
	\label{eq:echo_delay_effective}
\end{equation}
If the reflecting surface is placed at \(R=2M(1+\epsilon)\), then \(x_0=r_*(R)\simeq 2M[1+\ln\epsilon]\) for \(\epsilon\ll1\), and the dominant contribution becomes
\begin{equation}
	\Delta t_{\rm echo}^{\rm eff}
	\sim
	4M|\ln\epsilon| .
	\label{eq:echo_delay_log}
\end{equation}
In the present stellar model, however, the interior is not replaced by an effective boundary. The wave propagates through the regular interior, and the corresponding return point is \(r_*(0)=r_*(R)-L_{\rm int}\). The delay in Eq.~\eqref{eq:echo_delay} therefore contains, in addition to the exterior tortoise-coordinate contribution, the interior optical length \(L_{\rm int}\), which encodes the propagation time through the compact-object interior.

Thus the characteristic arrival time of the \(n\)-th echo is
\begin{equation}
	t_{{\rm echo},n}
	\simeq
	t_{{\rm echo},1}+(n-1)\Delta t_{\rm echo},
	\qquad n=1,2,3,\dots .
	\label{eq:echo_n_time}
\end{equation}
Since the Gaussian initial data are not compactly supported, these estimates should be understood as characteristic time windows rather than sharp causal thresholds.

In order to compare the ultracompact configuration \(R=2.26M\) with the neutron-star-like configuration \(R=6M\), we fix the source position by its tortoise-coordinate distance from the stellar surface rather than by its areal radius. Indeed, the initial data are prescribed as functions of \(r_*(r)-r_*(r_0)\), so the relevant measure of the source location is \(r_*(r_0)-r_*(R)\). We therefore choose \(r_0\) in the \(R=6M\) case so that this quantity is the same as in the \(R=2.26M\) case. With \(r_0=10M\) for \(R=2.26M\), this gives \(r_0\simeq 17.84M\) for \(R=6M\). This choice places the initial packet at the same optical distance from the surface in both spacetimes.

For the neutron-star-like case, \(R>3M\), the stellar surface lies outside the photon-sphere region, and there is no exterior cavity bounded by the surface and the curvature barrier. It is then more appropriate to introduce two surface/interior propagation times. The shortest reflected trajectory gives
\begin{equation}
	t_{\rm surf}
	\simeq
	r_*(r_0)+r_*(r_{\rm obs})-2r_*(R),
	\label{eq:surf_time}
\end{equation}
which corresponds to propagation from the source to the stellar surface and back to the observer. A complete traversal of the interior gives
\begin{align}
	t_{\rm int}^{(1)}
	&\simeq
	r_*(r_0)+r_*(r_{\rm obs})-2r_*(0)
	\nonumber\\
	&=
	r_*(r_0)+r_*(r_{\rm obs})-2r_*(R)+2L_{\rm int}.
	\label{eq:int_time_1}
\end{align}
If needed, successive interior round trips would be characterized by
\begin{equation}
	t_{\rm int}^{(n)}
	\simeq
	t_{\rm int}^{(1)}+(n-1)\,2L_{\rm int},
	\qquad n=1,2,3,\dots .
	\label{eq:int_time_n}
\end{equation}

Finally, for the \(R=6M\) configuration, the onset of the observed ringdown may be parametrized by an effective penetration depth \(r_*^{\rm pen}\), defined through
\begin{equation}
	t_{\rm ring}
	\simeq
	r_*(r_0)+r_*(r_{\rm obs})-2r_*^{\rm pen},
	\label{eq:ring_time_pen}
\end{equation}
with
\begin{equation}
	r_*(0)<r_*^{\rm pen}<r_*(R).
\end{equation}
This effective quantity should not be interpreted as a sharp reflection point. Rather, it provides a diagnostic of how deeply the wave probes the stellar interior before being reradiated toward the observer. The propagation times defined in Eqs.~\eqref{eq:dir_time}--\eqref{eq:ring_time_pen} will be used below to interpret the ringdown and echo windows in the numerical waveforms.

\subsubsection{Numerical results: Waveforms and echoes}
\label{sec:sec_1_2_4}

In this section, we present numerical results for the time-domain waveform and its quasinormal-mode reconstruction in the same two representative configurations considered previously: (i) a neutron-star-like object with \(R=6M\), and (ii) an ultracompact object with \(R=2.26M\). In both cases, we focus on the \(\ell=2\) perturbation and compare the full waveform with partial or complete QNM reconstructions built from the different modal families identified above.

\begin{figure}[!htb]
	\centering
	\includegraphics[scale=0.55]{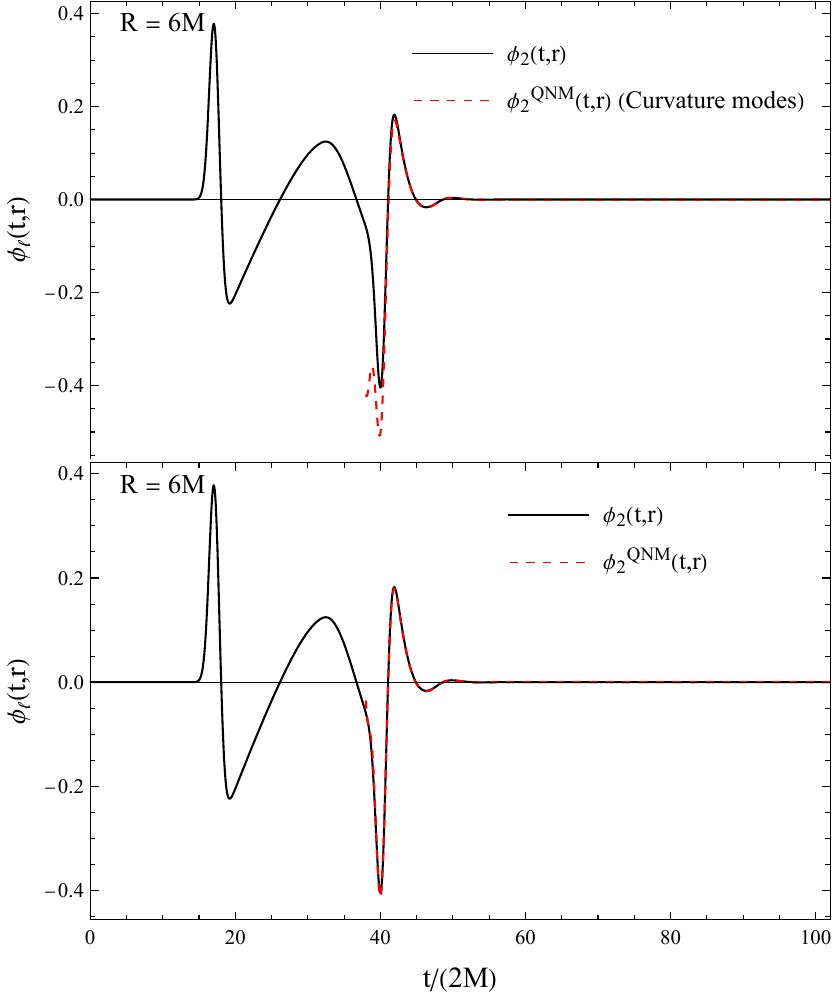}
	\caption{Time-domain waveform \(\phi(t,r)\) (black solid line) and quasinormal-mode reconstruction for the neutron-star-like case \(R=6M\). The observer is located at \(r_{\rm obs}=50M\). The Gaussian initial data in Eq.~\eqref{eq:Cauchy_data_waveform} are specified by \(\phi_0=1\), \(a=1\), and \(r_0=17.84M\). Top panel: contribution of the curvature branch only, summed over \(n=0,\ldots,4\). Bottom panel: contribution obtained after adding the fundamental interface mode \(n=0\). While the curvature modes dominate the late-time response, the interface mode provides a correction at the beginning of the ringdown. In contrast with the ultracompact case, no distinct echo-like structure is observed, in agreement with the absence of trapped modes and with the characteristic propagation times \(t_{\rm surf}\) and \(t_{\rm int}^{(1)}\) introduced in Eqs.~\eqref{eq:surf_time} and \eqref{eq:int_time_1}.}
	\label{fig:Waveform_vs_QNM_Echoes_NS}
\end{figure}

We begin with the neutron-star-like configuration. As shown in Fig.~\ref{fig:Waveform_vs_QNM_Echoes_NS}, the waveform is primarily described by the curvature branch, whose contribution already captures the dominant late-time decay. The addition of the fundamental interface mode improves the agreement at the beginning of the ringdown, but does not qualitatively modify the overall structure of the signal. In contrast with the ultracompact case discussed below, no clear sequence of echo-like pulses is observed. This is consistent with the absence of trapped modes in the \(R=6M\) spectrum and with the fact that the effective potential does not support a cavity capable of producing long-lived repeated internal reflections.

This interpretation is reinforced by the characteristic times introduced in Sec.~\ref{sec:sec_1_2_1}. In the neutron-star-like case, the shortest reflected timescale is \(t_{\rm surf}\), Eq.~\eqref{eq:surf_time}, whereas a full traversal of the stellar interior would occur on the timescale \(t_{\rm int}^{(1)}\), Eq.~\eqref{eq:int_time_1}. The observed onset of the ringdown lies between these two scales and may be parametrized by the effective penetration depth \(r_*^{\rm pen}\), Eq.~\eqref{eq:ring_time_pen}, indicating that the wave probes the surface and a finite outer portion of the stellar interior before being reradiated.

\begin{figure}[!htb]
	\centering
	\includegraphics[scale=0.55]{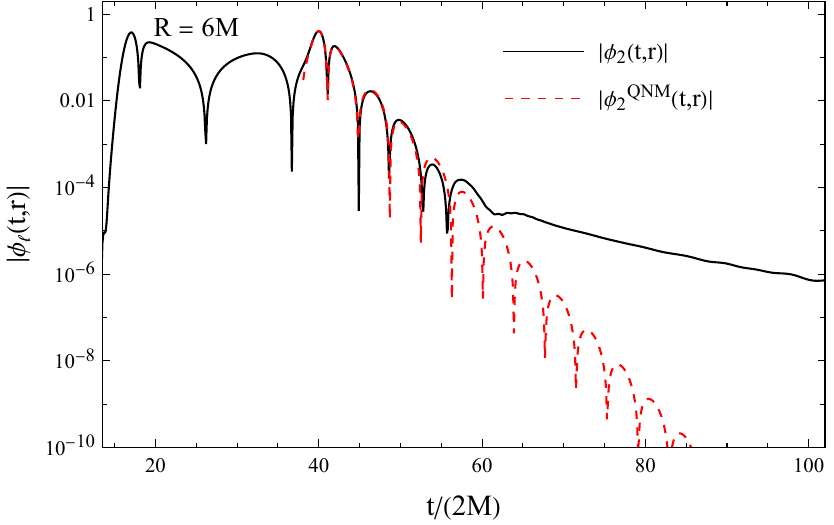}
	\caption{Logarithmic representation of the full waveform \(|\phi(t,r)|\) (black solid line) and of its QNM reconstruction for the neutron-star-like case \(R=6M\). The observer is located at \(r_{\rm obs}=50M\). The Gaussian initial data in Eq.~\eqref{eq:Cauchy_data_waveform} are specified by \(\phi_0=1\), \(a=1\), and \(r_0=17.84M\). The logarithmic scale makes it possible to test whether weak secondary pulses appear at late times. No clear echo-like structure is observed. In particular, the first interior-return timescale \(t_{\rm int}^{(1)}\) [Eq.~\eqref{eq:int_time_1}] falls essentially within the onset of the main ringdown, so that any weak reflected signal remains buried under the dominant curvature-mode response.}
	\label{fig:Waveform_vs_QNM_Echoes_NS_Log}
\end{figure}

This conclusion is further supported by the logarithmic representation shown in Fig.~\ref{fig:Waveform_vs_QNM_Echoes_NS_Log}. In the \(R=6M\) case, the first interior-return timescale \(t_{\rm int}^{(1)}\) lies essentially within the onset of the main ringdown, so that any weak reflected signal associated with it is expected to remain buried under the dominant curvature-mode response. At later times, the logarithmic plot does not reveal any distinct secondary pulses or any clear excess above the late-time tail. Therefore, if internal reflections are present in this configuration, they remain too weak to generate separately identifiable echo-like features in the waveform.

\begin{figure}[!htb]
	\centering
	\includegraphics[scale=0.55]{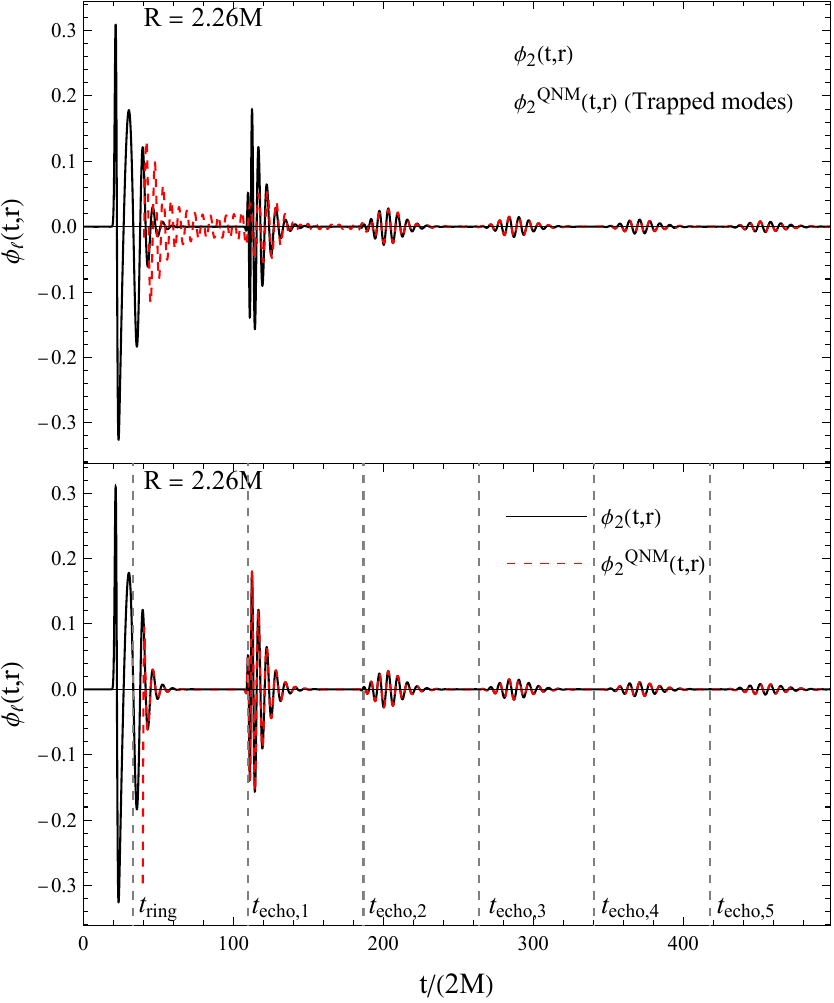}
	\caption{Comparison between the full time-domain waveform \(\phi(t,r)\) (black solid line) and its quasinormal-mode reconstruction for the ultracompact configuration \(R=2.26M\). The observer is located at \(r_{\rm obs}=50M\). The Gaussian initial data in Eq.~\eqref{eq:Cauchy_data_waveform} are specified by \(\phi_0=1\), \(a=1\), and \(r_0=10M\). In the top panel, the QNM contribution is built from the trapped branch only, with \(n=0,\ldots,6\). In the bottom panel, the reconstruction includes the trapped branch with \(n=0,\ldots,6\), the curvature branch with \(n=0,\ldots,64\), and the interface branch with \(n=0,1\). The vertical grey dashed lines mark the characteristic timescales estimated from Eqs.~\eqref{eq:ring_time_uco}--\eqref{eq:echo_n_time}: \(t_{\rm ring}\simeq 33\), \(t_{\rm echo,1}\simeq 110\), \(t_{\rm echo,2}\simeq 187\), \(t_{\rm echo,3}\simeq 264\), \(t_{\rm echo,4}\simeq 341\), and \(t_{\rm echo,5}\simeq 418\), expressed in the observer-time variable \(t/(2M)\). In particular, \(t_{\rm ring}\) marks the characteristic onset of the ringdown window, not a sharp pointwise threshold for the finite truncated QNM sum. The trapped modes reproduce the late-time sequence of echo-like pulses, whereas an accurate description of the initial ringdown near \(t_{\rm ring}\) requires the inclusion of the curvature and interface branches.}
	\label{fig:Waveform_vs_QNM_Echoes_UCO_1}
\end{figure}

We now turn to the ultracompact configuration. Figure~\ref{fig:Waveform_vs_QNM_Echoes_UCO_1} illustrates the time-domain response and its QNM reconstruction for \(R=2.26M\). In the top panel, the reconstruction is built from the trapped branch only. These weakly damped modes already reproduce the late-time sequence of echo-like pulses rather well, showing that the trapped family provides the dominant contribution to the long-lived part of the signal. This is consistent with the geometric estimates introduced in Sec.~\ref{sec:sec_1_2_1}: the first oscillatory burst is expected near the ringdown timescale \(t_{\rm ring}\), Eq.~\eqref{eq:ring_time_uco}, while the subsequent echo-like pulses are expected in the time windows centered around \(t_{{\rm echo},n}\), Eq.~\eqref{eq:echo_n_time}, with separation \(\Delta t_{\rm echo}\), Eq.~\eqref{eq:echo_delay}. The trapped modes account very well for these later packets, but they do not accurately describe the early part of the waveform. As shown in the bottom panel, once the curvature and interface branches are added, the agreement with the full waveform is significantly improved during the initial ringdown. This confirms that, in the ultracompact case, the late-time response is mainly governed by the trapped modes, whereas the early-time signal also receives important contributions from the curvature and interface families.

\section{The Debye expansion and Debye-QNMs}
\label{sec:sec_2}

In this section, we derive a Debye-type decomposition of the frequency-domain response associated with the initial-value problem introduced in Sec.~\ref{sec:sec_1}. 
Our aim is to rewrite the frequency-domain kernel, from which the waveform is reconstructed according to Eq.~\eqref{eq:response_infinity_waveform}, in a form that separates the different physical propagation channels: direct exterior propagation, direct reflection at the stellar surface, and contributions involving one or more interior traversals before re-emission to infinity. 
To this end, we combine the exterior basis with an interior basis normalized at the surface, and then expand the resulting Fabry--P\'erot-like denominator as a geometric series.

\subsection{Debye representation of the frequency-domain response}
\label{sec:sec_2_1}

We now recall the main ingredients needed to construct the Debye-type decomposition of the frequency-domain response. 
The detailed scattering construction has been presented in Ref.~\cite{OuldElHadj:2026ewh}; here we only introduce the notation required for the waveform problem.

In the exterior Schwarzschild region, $r>R$, we use the two standard solutions 
$f^{\rm dn}_{\omega\ell}$ and $f^{\rm up}_{\omega\ell}$, defined by their asymptotic behaviour at spatial infinity,
\begin{equation}
	\left\{
	\begin{aligned}
		f^{\rm dn}_{\omega\ell}(r) &\underset{r_\ast\to+\infty}{\sim} e^{-i\omega r_\ast},\\
		f^{\rm up}_{\omega\ell}(r) &\underset{r_\ast\to+\infty}{\sim} e^{+i\omega r_\ast},
	\end{aligned}
	\right.
	\label{eq:debye_response_jost}
\end{equation}
with Wronskian
\begin{equation}
	\W{f^{\rm dn}_{\omega\ell}}{f^{\rm up}_{\omega\ell}}=2i\omega .
	\label{eq:WJost_Debye}
\end{equation}
Throughout this work, the Wronskian is taken with respect to the tortoise coordinate,
\(
\W{f}{g}=f\,\partial_{r_\ast}g-g\,\partial_{r_\ast}f .
\)
The superscript ``dn'' stands for down-going. It replaces the more standard ``in'' label in order to avoid confusion with the interior solution \(u_\ell^{\rm in}\).

Inside the compact object we introduce a local basis at the surface,
\(\{u^{\rm in}_{\omega\ell},u^{\rm out}_{\omega\ell}\}\), defined by
\begin{equation}
	\left\{
	\begin{aligned}
		u^{\rm in}_{\omega\ell}(R_\ast)&=1,
		&\qquad
		\left.\partial_{r_\ast}u^{\rm in}_{\omega\ell}\right|_{R_\ast}&=-ik_{\rm int},\\
		u^{\rm out}_{\omega\ell}(R_\ast)&=1,
		&\qquad
		\left.\partial_{r_\ast}u^{\rm out}_{\omega\ell}\right|_{R_\ast}&=+ik_{\rm int},
	\end{aligned}
	\right.
	\label{eq:debye_response_uin_uout}
\end{equation}
where
\begin{equation}
	k_{\rm int}(\omega,\ell)=\sqrt{\omega^2-V_\ell(R^-)} .
	\label{eq:kint_def_Debye}
\end{equation}
This normalization gives
\begin{equation}
	\W{u^{\rm in}_{\omega\ell}}{u^{\rm out}_{\omega\ell}}=2ik_{\rm int}.
	\label{eq:Wuin_uout_Debye}
\end{equation}

The surface matching of the field and of its first $r_\ast$-derivative relates the exterior and interior bases. In particular, one may write
\begin{equation}
	u^{\rm in}_{\omega\ell}
	=
	\alpha^{\rm in}_\ell(\omega) f^{\rm dn}_{\omega\ell}
	+
	\beta^{\rm out}_\ell(\omega) f^{\rm up}_{\omega\ell},
	\label{eq:uin_expand_Debye}
\end{equation}
and
\begin{equation}
	f^{\rm up}_{\omega\ell}
	=
	\gamma^{\rm out}_\ell(\omega) u^{\rm out}_{\omega\ell}
	+
	\delta^{\rm in}_\ell(\omega) u^{\rm in}_{\omega\ell}.
	\label{eq:fup_expand_Debye}
\end{equation}
The coefficients appearing in these connection formulae are obtained from Wronskians evaluated at the surface
\begin{subequations}
	\label{eq:connection_coeffs_Debye}
	\begin{align}
		\alpha^{\rm in}_\ell(\omega)
		&=
		\frac{
			\W{u^{\rm in}_{\omega\ell}}{f^{\rm up}_{\omega\ell}}
		}{
			\W{f^{\rm dn}_{\omega\ell}}{f^{\rm up}_{\omega\ell}}
		}\Bigg|_{R_\ast},
		\label{eq:connection_coeffs_Debye_alpha}
		 \\
		\beta^{\rm out}_\ell(\omega)
		&=
		\frac{
			\W{f^{\rm dn}_{\omega\ell}}{u^{\rm in}_{\omega\ell}}
		}{
			\W{f^{\rm dn}_{\omega\ell}}{f^{\rm up}_{\omega\ell}}
		}\Bigg|_{R_\ast},
		\\
		\gamma^{\rm out}_\ell(\omega)
		&=
		\frac{
			\W{f^{\rm up}_{\omega\ell}}{u^{\rm in}_{\omega\ell}}
		}{
			\W{u^{\rm out}_{\omega\ell}}{u^{\rm in}_{\omega\ell}}
		}\Bigg|_{R_\ast},
		\\
		\delta^{\rm in}_\ell(\omega)
		&=
		\frac{
			\W{u^{\rm out}_{\omega\ell}}{f^{\rm up}_{\omega\ell}}
		}{
			\W{u^{\rm out}_{\omega\ell}}{u^{\rm in}_{\omega\ell}}
		}\Bigg|_{R_\ast}.
	\end{align}
\end{subequations}

Let $\phi^{\rm reg}_{\omega\ell}$ denote the solution regular at the center. In the interior basis it can be written as
\begin{equation}
	\phi^{\rm reg}_{\omega\ell}
	=
	c^{\rm in}_\ell(\omega) u^{\rm in}_{\omega\ell}
	+
	c^{\rm out}_\ell(\omega) u^{\rm out}_{\omega\ell}, \qquad 0\le r\le R.
	\label{eq:phireg_expand_u_Debye}
\end{equation}
where
\begin{subequations}
	\label{eq:cin_cout_Debye}
	\begin{align}
		c^{\rm in}_\ell(\omega)
		&=
		\frac{
			\W{\phi^{\rm reg}_{\omega\ell}}{u^{\rm out}_{\omega\ell}}
		}{
			\W{u^{\rm in}_{\omega\ell}}{u^{\rm out}_{\omega\ell}}
		}\Bigg|_{R_\ast},
		\label{eq:cin_cout_Debye_cin}
		\\
		c^{\rm out}_\ell(\omega)
		&=
		\frac{
			\W{u^{\rm in}_{\omega\ell}}{\phi^{\rm reg}_{\omega\ell}}
		}{
			\W{u^{\rm in}_{\omega\ell}}{u^{\rm out}_{\omega\ell}}
		}\Bigg|_{R_\ast}.
	\end{align}
\end{subequations}
It is convenient to introduce the interior phase factor
\begin{equation}
	\xi_\ell(\omega)\equiv \frac{c^{\rm out}_\ell(\omega)}{c^{\rm in}_\ell(\omega)},
	\label{eq:debye_response_xi}
\end{equation}
so that
\begin{equation}
	\phi^{\rm reg}_{\omega\ell}(r)
	=
	c^{\rm in}_\ell(\omega)
	\Bigl[
	u^{\rm in}_{\omega\ell}(r)+\xi_\ell(\omega)\,u^{\rm out}_{\omega\ell}(r)
	\Bigr],
	\quad 0\le r\le R.
	\label{eq:debye_response_phireg_int_xi}
\end{equation}

Using Eqs.~\eqref{eq:uin_expand_Debye} and \eqref{eq:fup_expand_Debye}, the regular solution Eq.~\eqref{eq:debye_response_phireg_int_xi} can be rewritten 
\begin{equation}
	\begin{split}
		&\phi^{\rm reg}_{\omega\ell}(r)
		= \alpha^{\rm in}_\ell\,c^{\rm in}_\ell\Bigl(1-\frac{\delta^{\rm in}_\ell}{\gamma^{\rm out}_\ell}\,\xi_\ell\Bigr)\,f^{\rm dn}_{\omega\ell}(r) \\
		& + \Big(\beta^{\rm out}_\ell\,c^{\rm in}_\ell\Bigl(1-\frac{\delta^{\rm in}_\ell}{\gamma^{\rm out}_\ell}\,\xi_\ell\Bigr)
		+ \frac{1}{\gamma^{\rm out}_\ell}\,c^{\rm out}_\ell\Big)\,f^{\rm up}_{\omega\ell}(r).
	\end{split}
	\label{eq:phireg_expand_ext_repeat_bis}
\end{equation}
and comparing with the far-field decomposition
\begin{equation}
	\phi^{\rm reg}_{\omega\ell}(r)
	= A^{(-)}_\ell(\omega)\,f^{\rm dn}_{\omega\ell}(r) + A^{(+)}_\ell(\omega)\,f^{\rm up}_{\omega\ell}(r), \quad r\geq R.
	\label{eq:phireg_expand_ext_repeat}
\end{equation}
we obtain
\begin{equation}
	A^{(-)}_\ell(\omega)
	=
	\alpha^{\rm in}_\ell c^{\rm in}_\ell
	\left(
	1-\frac{\delta^{\rm in}_\ell}{\gamma^{\rm out}_\ell}\xi_\ell
	\right),
	\label{eq:Aminus_Debye_reduced}
\end{equation}
and
\begin{equation}
	A^{(+)}_\ell(\omega)
	=
	\beta^{\rm out}_\ell c^{\rm in}_\ell
	\left(
	1-\frac{\delta^{\rm in}_\ell}{\gamma^{\rm out}_\ell}\xi_\ell
	\right)
	+
	\frac{c^{\rm out}_\ell}{\gamma^{\rm out}_\ell}.
	\label{eq:Aplus_Debye_reduced}
\end{equation}
The ratio $A^{(+)}_\ell/A^{(-)}_\ell$ therefore admits an exact cavity-like form,
\begin{equation}
	\frac{A^{(+)}_\ell(\omega)}{A^{(-)}_\ell(\omega)}
	=
	R_{22,\ell}
	+
	\frac{
		T_{12,\ell}T_{21,\ell}\xi_\ell
	}{
		1-R_{11,\ell}\xi_\ell
	},
	\label{eq:Rtot_cavity_Debye}
\end{equation}
where
\begin{subequations}\label{eq:debye_response_RijTij}
	\begin{align}
		R_{22,\ell}(\omega)&\equiv \frac{\beta^{\rm out}_\ell(\omega)}{\alpha^{\rm in}_\ell(\omega)},
		\label{eq:debye_response_R22}\\
		T_{21,\ell}(\omega)&\equiv \frac{1}{\alpha^{\rm in}_\ell(\omega)},
		\label{eq:debye_response_T21}\\
		T_{12,\ell}(\omega)&\equiv \frac{1}{\gamma^{\rm out}_\ell(\omega)},
		\label{eq:debye_response_T12}\\
		R_{11,\ell}(\omega)&\equiv \frac{\delta^{\rm in}_\ell(\omega)}{\gamma^{\rm out}_\ell(\omega)}.
		\label{eq:debye_response_R11}
	\end{align}
\end{subequations}
Here we introduced the effective reflection and transmission coefficients and the index $1$ refers to the interior region and the index $2$ to the exterior region (see Refs.\cite{Nussenzveig:2006,OuldElHadj:2026ewh}).

We now apply the cavity representation introduced above to the time-domain waveform.  The response at infinity Eq.~\eqref{eq:response_infinity_waveform} can be written in the form
\begin{equation}
	\phi_\ell(t,r)
	=
	\frac{1}{2\pi}\,\Re\!\left[
	\int_{0+ic}^{+\infty+ic} \hspace{-5pt}
	e^{-i\omega (t-r_\ast)}\,
	\mathcal{D}_\ell(\omega)\,
	d\omega
	\right],
	\label{eq:debye_response_phi_inf}
\end{equation}
where the frequency-domain response kernel is
\begin{equation}
	\mathcal{D}_\ell(\omega)
	\equiv
	\frac{1}{A^{(-)}_\ell(\omega)}
	\int_{r_\ast(0)}^{+\infty}
	\phi_{\ell}(t=0,r')\,\phi^{\rm reg}_{\omega\ell}(r')\,dr'_\ast .
	\label{eq:debye_response_D_def}
\end{equation}
In order to express this kernel in terms of the Debye scattering coefficients, we project the initial data onto the interior and exterior basis functions.  We define
\begin{subequations}\label{eq:debye_response_overlap_defs}
	\begin{align}
		Q^{\rm in}_\ell(\omega)
		&\equiv
		\int_{r_\ast(0)}^{R_\ast}
		\phi_{\ell}(t=0,r')\,u^{\rm in}_{\omega\ell}(r')\,dr'_\ast,
		\label{eq:debye_response_Qin}\\
		Q^{\rm out}_\ell(\omega)
		&\equiv
		\int_{r_\ast(0)}^{R_\ast}
		\phi_{\ell}(t=0,r')\,u^{\rm out}_{\omega\ell}(r')\,dr'_\ast,
		\label{eq:debye_response_Qout}\\
		P^{(-)}_\ell(\omega)
		&\equiv
		\int_{R_\ast}^{+\infty}
		\phi_{\ell}(t=0,r')\,f^{\rm dn}_{\omega\ell}(r')\,dr'_\ast,
		\label{eq:debye_response_Pminus}\\
		P^{(+)}_\ell(\omega)
		&\equiv
		\int_{R_\ast}^{+\infty}
		\phi_{\ell}(t=0,r')\,f^{\rm up}_{\omega\ell}(r')\,dr'_\ast .
		\label{eq:debye_response_Pplus}
	\end{align}
\end{subequations}
Using the interior expansion \eqref{eq:phireg_expand_u_Debye} and the exterior expansion \eqref{eq:phireg_expand_ext_repeat}, the integral \eqref{eq:debye_response_D_def} is therefore
\begin{multline}
	\int_{r_\ast(0)}^{+\infty}
	\phi_{\ell}(t=0,r')\,\phi^{\rm reg}_{\omega\ell}(r')\,dr'_\ast
	=\\
	A^{(-)}_\ell(\omega)\,P^{(-)}_\ell(\omega)
	+
	A^{(+)}_\ell(\omega)\,P^{(+)}_\ell(\omega)
	\\
	+
	c^{\rm in}_\ell(\omega)
	\Bigl[
	Q^{\rm in}_\ell(\omega)
	+
	\xi_\ell(\omega)\,Q^{\rm out}_\ell(\omega)
	\Bigr].
	\label{eq:debye_response_overlap_identity}
\end{multline}
Substitution into Eq.~\eqref{eq:debye_response_D_def} gives the exact identity
\begin{multline}
	\mathcal{D}_\ell(\omega)
	=
	P^{(-)}_\ell(\omega)
	+
	\frac{A^{(+)}_\ell(\omega)}{A^{(-)}_\ell(\omega)}\,
	P^{(+)}_\ell(\omega)
	\\
	+
	\frac{c^{\rm in}_\ell(\omega)}{A^{(-)}_\ell(\omega)}
	\Bigl[
	Q^{\rm in}_\ell(\omega)
	+
	\xi_\ell(\omega)\,Q^{\rm out}_\ell(\omega)
	\Bigr].
	\label{eq:debye_response_D_exact_pre}
\end{multline}
Using
\begin{equation}
	\frac{c^{\rm in}_\ell(\omega)}{A^{(-)}_\ell(\omega)}
	=
	\frac{T_{21,\ell}(\omega)}
	{1-R_{11,\ell}(\omega)\,\xi_\ell(\omega)},
	\label{eq:debye_response_cin_over_Aminus}
\end{equation}
together with Eq.~\eqref{eq:Rtot_cavity_Debye}, we obtain the cavity form of the
exact response kernel,
\begin{align}
	&\mathcal{D}_\ell(\omega)
	=
	P^{(-)}_\ell(\omega)
	+
	R_{22,\ell}(\omega)\,P^{(+)}_\ell(\omega)
	\nonumber\\
	&\qquad
	+
	\frac{T_{21,\ell}(\omega)}
	{1-R_{11,\ell}(\omega)\,\xi_\ell(\omega)}
	\biggl[
	\xi_\ell(\omega) T_{12,\ell}(\omega)\,P^{(+)}_\ell(\omega) \nonumber\\
	&\qquad+
	Q^{\rm in}_\ell(\omega)
	+
	\xi_\ell(\omega) Q^{\rm out}_\ell(\omega)
	\biggr].
	\label{eq:debye_response_D_exact}
\end{align}
Equation~\eqref{eq:debye_response_D_exact} is an exact frequency-domain representation of the response kernel in cavity form. It organizes the response into a direct exterior contribution, a component reflected from
the exterior side of the compact object, and a cavity-mediated contribution involving transmission into the interior, repeated internal reflections, and subsequent re-emission to infinity.

\subsection{Debye series}
\label{sec:sec_2_2}

The Debye expansion follows from the geometric expansion of the cavity denominator appearing in Eq.~\eqref{eq:debye_response_D_exact}. More precisely, expanding the factor \((1-R_{11,\ell}\xi_\ell)^{-1}\), we write the response kernel as
\begin{equation}
	\mathcal{D}_\ell(\omega)
	=
	\mathcal{D}^{(0)}_\ell(\omega)
	+
	\sum_{p=1}^{\infty}\mathcal{D}^{(p)}_\ell(\omega),
	\label{eq:debye_response_Dsum}
\end{equation}
where the zeroth-order contribution is
\begin{equation}
	\mathcal{D}^{(0)}_\ell(\omega)
	=
	P^{(-)}_\ell(\omega)
	+
	R_{22,\ell}(\omega)\,P^{(+)}_\ell(\omega),
	\label{eq:debye_response_D0}
\end{equation}
and, for \(p\geq1\),
\begin{align}
	\mathcal{D}^{(p)}_\ell(\omega)
	&=
	T_{21,\ell}(\omega)
	\biggl[
	R_{11,\ell}(\omega)^{p-1}
	\xi_\ell(\omega)^{p-1}
	Q^{\rm in}_\ell(\omega)
	\nonumber\\
	&\quad
	+
	R_{11,\ell}(\omega)^{p-1}
	\xi_\ell(\omega)^p
	Q^{\rm out}_\ell(\omega)
	\nonumber\\
	&\quad
	+
	T_{12,\ell}(\omega)
	R_{11,\ell}(\omega)^{p-1}
	\xi_\ell(\omega)^p
	P^{(+)}_\ell(\omega)
	\biggr].
	\label{eq:debye_response_Dp_exact}
\end{align}
The expansion is first understood in the domain where the geometric series is convergent and then used, by analytic continuation, to identify the separate Debye contributions to the waveform.

The corresponding time-domain decomposition is obtained by substituting Eq.~\eqref{eq:debye_response_Dsum} into Eq.~\eqref{eq:debye_response_phi_inf}. This gives
\begin{equation}
	\phi_\ell(t,r)
	=
	\phi^{(0)}_\ell(t,r)
	+
	\sum_{p=1}^{\infty}\phi^{(p)}_\ell(t,r),
	\label{eq:debye_response_time_sum}
\end{equation}
with
\begin{equation}
	\phi^{(p)}_\ell(t,r)
	=
	\frac{1}{2\pi}\,
	\Re\!\left[
	\int_{0+ic}^{+\infty+ic}
	e^{-i\omega (t-r_\ast)}
	\mathcal{D}^{(p)}_\ell(\omega)
	d\omega
	\right].
	\label{eq:debye_response_time_p}
\end{equation}

This decomposition has a direct propagation interpretation. The term \(\mathcal{D}^{(0)}_\ell\) contains the purely exterior part of the response: the direct exterior contribution \(P^{(-)}_\ell\), together with the component reflected once by the surface, \(R_{22,\ell}P^{(+)}_\ell\). The terms \(\mathcal{D}^{(p)}_\ell\), with \(p\geq1\), contain the contributions for which the wave probes the interior cavity. In Eq.~\eqref{eq:debye_response_Dp_exact}, the factors \(R_{11,\ell}^{p-1}\xi_\ell^{p-1}\) and \(R_{11,\ell}^{p-1}\xi_\ell^{p}\) count the successive internal cycles, while the transmission amplitudes \(T_{21,\ell}\) and \(T_{12,\ell}\) describe the matching between the exterior and interior sectors. In particular, the term proportional to \(P^{(+)}_\ell\) corresponds to an exterior wave transmitted through the surface, propagated through the interior cavity, and re-emitted towards infinity after \(p\) interior traversals and \(p-1\) internal reflections.

For the analysis of the singular structure of each Debye term, it is useful to rewrite Eqs.~\eqref{eq:debye_response_D0} and \eqref{eq:debye_response_Dp_exact} in terms of the surface connection coefficients \(\alpha^{\rm in}_\ell\), \(\beta^{\rm out}_\ell\), \(\gamma^{\rm out}_\ell\), and \(\delta^{\rm in}_\ell\). These coefficients are defined by matching the exterior basis to the interior basis at the stellar surface. With the conventions introduced above, they satisfy
\begin{equation}
	\gamma^{\rm out}_\ell(\omega)
	=
	\frac{\omega}{k_{\rm int}(\omega,\ell)}
	\alpha^{\rm in}_\ell(\omega),
	\label{eq:debye_response_gamma_alpha}
\end{equation}
which follows directly from the corresponding Wronskian definitions at \(r=R\).

Using Eqs.~\eqref{eq:debye_response_R22}--\eqref{eq:debye_response_R11}, together with Eq.~\eqref{eq:debye_response_gamma_alpha}, one obtains
\begin{subequations}
	\label{eq:debye_response_D_terms_coeffs}
	\begin{align}
		\mathcal{D}^{(0)}_\ell(\omega)
		&=
		P^{(-)}_\ell(\omega)
		+
		\frac{\beta^{\rm out}_\ell(\omega)}
		{\alpha^{\rm in}_\ell(\omega)}
		P^{(+)}_\ell(\omega),
		\label{eq:debye_response_D_terms_coeffs_a}
		\\
		\mathcal{D}^{(p)}_\ell(\omega)
		&=
		\left(
		\frac{k_{\rm int}(\omega,\ell)}{\omega}
		\right)^p
		\frac{
			\left[\delta^{\rm in}_\ell(\omega)\right]^{p-1}
			\xi_\ell(\omega)^p
		}
		{
			\left[\alpha^{\rm in}_\ell(\omega)\right]^{p+1}
		}
		P^{(+)}_\ell(\omega)
		\nonumber\\
		&\quad
		+
		\left(
		\frac{k_{\rm int}(\omega,\ell)}{\omega}
		\right)^{p-1}
		\frac{
			\left[\delta^{\rm in}_\ell(\omega)\right]^{p-1}
			\xi_\ell(\omega)^{p-1}
		}
		{
			\left[\alpha^{\rm in}_\ell(\omega)\right]^p
		}
		Q^{\rm in}_\ell(\omega)
		\nonumber\\
		&\quad
		+
		\left(
		\frac{k_{\rm int}(\omega,\ell)}{\omega}
		\right)^{p-1}
		\frac{
			\left[\delta^{\rm in}_\ell(\omega)\right]^{p-1}
			\xi_\ell(\omega)^p
		}
		{
			\left[\alpha^{\rm in}_\ell(\omega)\right]^p
		}
		Q^{\rm out}_\ell(\omega),
		\label{eq:debye_response_D_terms_coeffs_b}
	\end{align}
\end{subequations}

Equation~\eqref{eq:debye_response_D_terms_coeffs} makes explicit the singular structure of the individual Debye contributions. These singularities must be distinguished from the ordinary quasinormal frequencies of the complete response, which are defined by the global condition \(A^{(-)}_\ell(\omega)=0\). We shall therefore refer to the poles of the individual Debye terms as Debye quasinormal modes, or D-QNMs. The same qualitative labels---curvature, trapped, and interface---will be used below, but they should not be understood as implying a one-to-one correspondence with the QNM families of \(A^{(-)}_\ell(\omega)\).

This distinction is already apparent from Eq.~\eqref{eq:Aminus_Debye_reduced}. The global coefficient \(A^{(-)}_\ell(\omega)\) is not determined by \(\alpha^{\rm in}_\ell(\omega)\) and \(c^{\rm in}_\ell(\omega)\) separately, but by their Fabry--P\'erot-like combination with the factor describing repeated internal reflections. Thus, the ordinary QNMs are collective poles of the resummed response, whereas the D-QNMs are poles of the separate Debye building blocks.

More explicitly, the D-QNFs are the complex frequencies at which one of the Wronskians appearing in the denominators of the elementary Debye amplitudes vanishes. They can therefore be interpreted, in direct analogy with ordinary resonant modes, as solutions satisfying two homogeneous boundary conditions associated with a given Debye channel.

The zeroth-order Debye contribution \(\mathcal{D}^{(0)}_\ell\) contains poles associated with the zeros of \(\alpha^{\rm in}_\ell(\omega)\), namely
\begin{equation}
	\alpha^{\rm in}_\ell(\omega_{\ell n}^{(\alpha)})=0 .
\end{equation}
Using the Wronskian definition of \(\alpha^{\rm in}_\ell\) Eq.~\eqref{eq:connection_coeffs_Debye_alpha}, this condition implies that the corresponding Debye modes are purely down-going at the stellar surface and purely outgoing at spatial infinity. These modes will be referred to as interface D-QNMs.

For \(p\geq1\), the Debye terms have a richer singular structure. Besides the poles associated with \(\alpha^{\rm in}_\ell(\omega)\), they also contain the singularities carried by the interior phase factor \(\xi_\ell(\omega)\). Since \(\xi_\ell(\omega)\) involves the interior coefficient \(c^{\rm in}_\ell(\omega)\), additional poles occur at the zeros
\begin{equation}
	c^{\rm in}_\ell(\omega_{\ell n}^{(c)})=0 .
\end{equation}
These poles define the interior D-QNM sector. Using the Wronskian definition of \(c^{\rm in}_\ell\) Eq.~\eqref{eq:cin_cout_Debye_cin}, this condition implies that the corresponding modes are regular at the stellar centre and outgoing at the stellar surface. The labels curvature-type and trapped will be used below to characterize the different sequences appearing in this sector.

\subsection{Debye quasinormal mode spectrum}
\label{sec:sec_2_3}

To construct the Debye quasinormal-frequency spectra, we use a numerical procedure closely related to that introduced in Sec.~IV.A in Ref.~\cite{OuldElHadj:2026ewh} for the computation of the Regge--Debye pole spectrum. In the present case, however, the roles of the spectral variables are exchanged. Rather than fixing the frequency $\omega$ and determining the Regge-Debye poles in the complex angular momentum plane, we fix the multipolar index $\ell$ and search, in the complex-frequency plane, for the Debye quasinormal frequencies associated with the zeros of $\alpha_\ell^{\rm in}(\omega)$ and $c_\ell^{\rm in}(\omega)$.

It should be noted that, in order to construct the Debye quasinormal-frequency spectrum in the fourth quadrant, we must fix the analytic continuation of the interior wavenumber defined in Eq.~\eqref{eq:kint_def_Debye}. We therefore use
\begin{equation}
	k_{\rm int}^{-}(\omega,\ell)
	=
	-i\sqrt{V_\ell(R^-)-\omega^2},
	\label{eq:kint_lower_branch_DQNF}
\end{equation}
which selects the Riemann sheet appropriate to the fourth-quadrant D-QNF search.

\begin{figure}[!htb]
	\centering
	\includegraphics[scale=0.55]{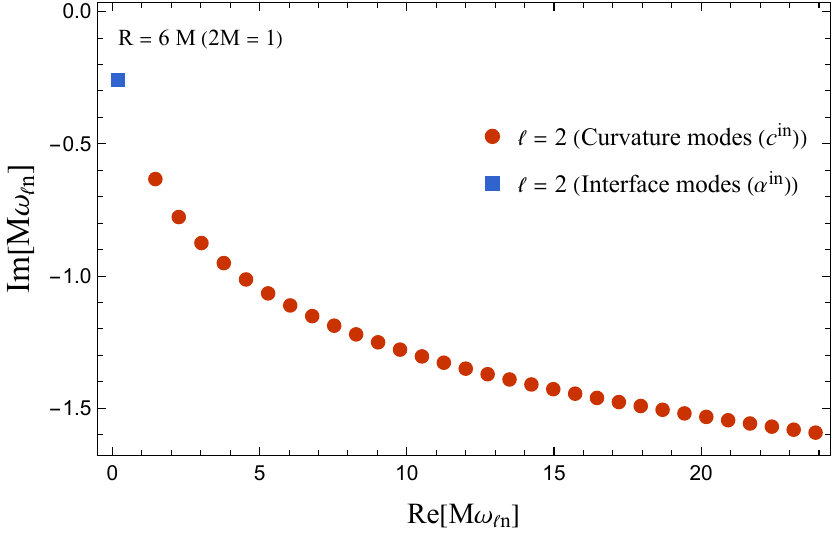}
\caption{Debye quasinormal-frequency spectrum in the complex-frequency plane for the \(\ell=2\) perturbation of a neutron-star-like compact object with radius \(R=6M\), in units where \(2M=1\). 
	The red circles denote the curvature-type D-QNMs associated with the zeros of \(c^{\rm in}_\ell(\omega)\), while the blue square denotes the interface D-QNM associated with the zero of \(\alpha^{\rm in}_\ell(\omega)\).}
	\label{fig:Debye-QNFs_NS_R6M_l_2}
\end{figure}

\begin{figure}[!htb]
	\centering
	\includegraphics[scale=0.52]{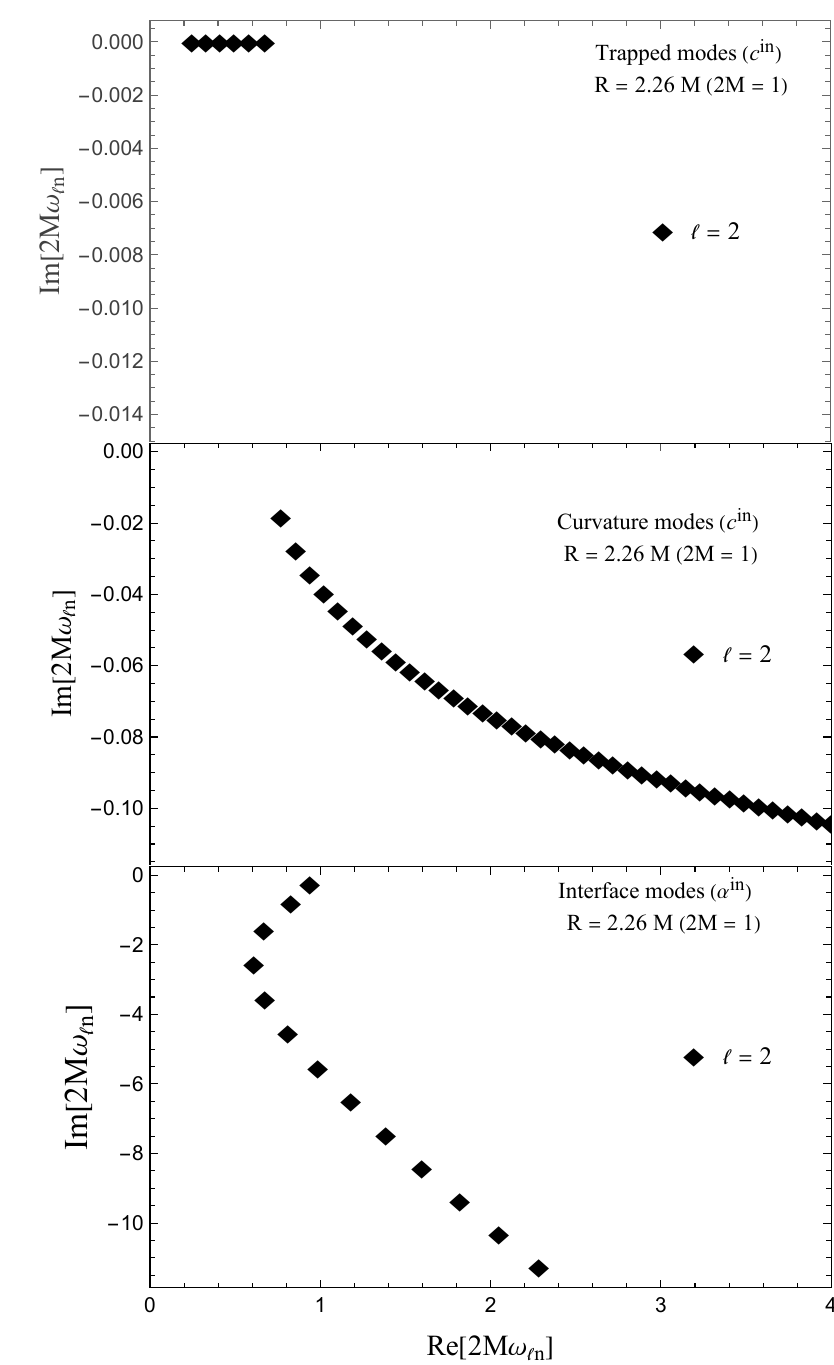}
	\caption{Debye quasinormal mode spectrum for the $\ell=2$ perturbation of an ultracompact object with radius $R=2.26M$ (in units where $2M=1$). The top, middle, and bottom panels display respectively the trapped, curvature, and interface-mode families.}
	\label{fig:Debye-QNFs_UCO_R226M_l_2}
\end{figure}

\begin{figure}[!htb]
	\centering
	\includegraphics[scale=0.55]{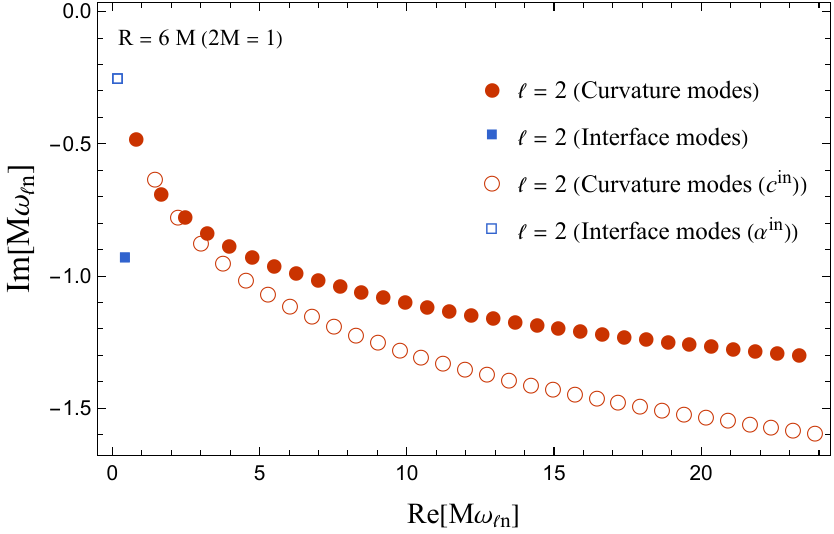}
		\caption{ Comparison between ordinary QNMs and D-QNMs for the \(\ell=2\) perturbation of a neutron-star-like compact object with radius \(R=6M\), in units where \(2M=1\). 
			Filled symbols correspond to the ordinary QNM spectrum, while open symbols correspond to the Debye quasinormal-frequency spectrum. 
			Red circles denote curvature-type modes, and blue squares denote interface modes.
		}
	\label{fig:QNFs_Vs_DQNFs_NS_R6M_l_2}
\end{figure}

\begin{figure}[!htb]
	\centering
	\includegraphics[scale=0.52]{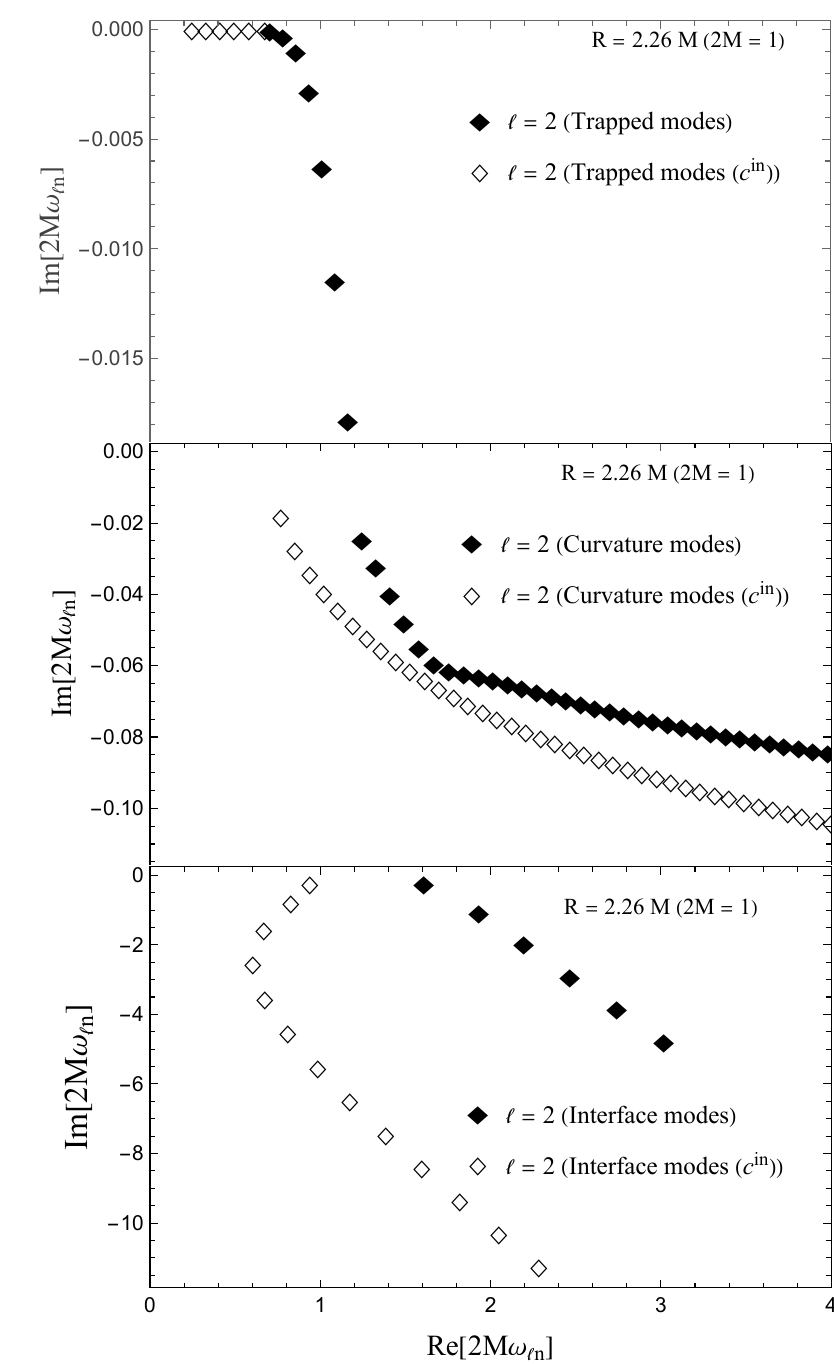}
		\caption{Comparison between ordinary QNMs and D-QNMs for the \(\ell=2\) perturbation of an ultracompact object with radius \(R=2.26M\), in units where \(2M=1\). 
			The three panels show, from top to bottom, the trapped, curvature, and interface families in the complex-frequency plane. 
			Filled diamonds correspond to the ordinary QNM spectrum of the fully resummed response, while open diamonds correspond to the Debye quasinormal-frequency spectrum of the individual Debye contributions.}
	\label{fig:QNFs_Vs_DQNFs_UCO_R226M_l_2}
\end{figure}

We now present numerical results for the Debye quasinormal-frequency spectrum of a massless scalar field in the same two representative configurations considered above: a neutron-star-like model with \(R=6M\), and an ultracompact object with \(R=2.26M\), close to the Buchdahl limit. In both cases, we focus on the \(\ell=2\) perturbation. The corresponding Debye spectra are displayed in Figs.~\ref{fig:Debye-QNFs_NS_R6M_l_2} and \ref{fig:Debye-QNFs_UCO_R226M_l_2}, while the lowest Debye quasinormal frequencies are listed in Tables~\ref{tab:Debye_qnm_modes_NS} and \ref{tab:Debye_qnm_modes_UCO}.

The neutron-star-like configuration exhibits a simpler Debye spectrum. As shown in Fig.~\ref{fig:Debye-QNFs_NS_R6M_l_2}, only two branches are present: a curvature branch associated with the zeros of $c^{\rm in}_\ell(\omega)$ and a low-lying interface branch associated with the zeros of $\alpha^{\rm in}_\ell(\omega)$. No trapped Debye branch is found in this case, consistently with the absence of a cavity structure in the corresponding effective potential. The values reported in Table~\ref{tab:Debye_qnm_modes_NS} show that the curvature Debye frequencies are significantly more damped than the trapped Debye modes of the ultracompact configuration, while the interface branch reduces essentially to a single low-lying mode.

For the ultracompact configuration, the Debye quasinormal spectrum separates into distinct structures, as shown in Fig.~\ref{fig:Debye-QNFs_UCO_R226M_l_2}. The curvature branch is associated with fourth-quadrant zeros of \(c^{\rm in}_\ell(\omega)\), while the interface branch is associated with the zeros of \(\alpha^{\rm in}_\ell(\omega)\). In addition, the \(c^{\rm in}_\ell\) sector contains sub-threshold trapped structures attached to the cut. These trapped structures, listed in Table~\ref{tab:Debye_qnm_modes_UCO}, should not be interpreted as ordinary weakly damped fourth-quadrant D-QNMs. They are instead treated as cut-supported singularities and enter the waveform through the regularized cut contribution. The curvature branch is more strongly damped than the cut-supported trapped structures, while the interface branch is the most rapidly damped.

The relation between the ordinary QNM spectrum and the Debye quasinormal-frequency spectrum is made more explicit in Figs.~\ref{fig:QNFs_Vs_DQNFs_NS_R6M_l_2} and \ref{fig:QNFs_Vs_DQNFs_UCO_R226M_l_2}. In these figures, filled symbols denote the ordinary QNMs of the fully resummed response, i.e. the zeros of \(A^{(-)}_\ell(\omega)\), while open symbols denote the D-QNMs associated with the individual Debye building blocks. For the neutron-star-like configuration, the comparison shows that the Debye spectrum separates into curvature and interface branches, as does the ordinary QNM spectrum, although the frequencies do not coincide exactly. Moreover, the relative damping is not the same in the two descriptions: the interface modes of the ordinary QNM spectrum are more strongly damped than the curvature modes, while the interface D-QNM is less damped than the curvature-type D-QNMs. 

For the ultracompact configuration, the comparison is shown in Fig.~\ref{fig:QNFs_Vs_DQNFs_UCO_R226M_l_2}. The ordinary QNM spectrum contains the familiar weakly damped trapped branch of the
fully resummed response. In the Debye representation, by contrast, the trapped structures associated with the zeros of \(c^{\rm in}_\ell(\omega)\) are attached to the two lips of the sub-threshold cut and are therefore
treated through the regularized cut contribution. The fourth-quadrant Debye residues are instead dominated by the curvature and interfac sectors. Thus the Debye spectral decomposition does not reproduce the
ordinary QNM families one by one; rather, it reorganizes the same physics according to the propagation channels resolved before the multiple-reflection series is resummed.

It is also worth emphasizing that the Debye quasinormal frequencies do not coincide with the exact quasinormal frequencies of the full problem. Rather, they encode the singular structure of the individual Debye terms entering the decomposition of the waveform response. In this sense, the Debye spectrum provides a partial resonant description, separating the contributions controlled by the interior quantity $c^{\rm in}_\ell(\omega)$ from those governed by the surface coefficient $\alpha^{\rm in}_\ell(\omega)$.

\begin{table}[!t]
	\centering
	\caption{The lowest Debye quasinormal frequencies for the massless scalar field. The radius of the compact body is $R=6M$.}
	\label{tab:Debye_qnm_modes_NS}
	\small
	\setlength{\tabcolsep}{6pt}
	\renewcommand{\arraystretch}{1.15}
	\begin{tabular}{c c l l l}
		\hline\hline
		$\ell$ & $n$ & $2M\omega_{\ell n}^{(c^{\rm in})}(\rm Curvature)$ & $2M\omega_{\ell n}^{(I, \alpha^{\rm in})}(\rm Interface)$ \\
		\hline
		2 & 0  & $1.459775 - 0.633220i$ & $0.201556 - 0.259562i$ \\
		& 1  & $2.254431 - 0.776885i$ & \\
		& 2  & $3.024166 - 0.875156i$ & \\
		& 3  & $3.783956 - 0.950950i$ & \\
		& 4  & $4.538499 - 1.012937i$ & \\
		& 5  & $5.289861 - 1.065481i$ & \\
		& 6  & $6.039120 - 1.111127i$ & \\
		& 7  & $6.786907 - 1.151499i$ & \\
		& 8  & $7.533616 - 1.187703i$ & \\
		& 9  & $8.279509 - 1.220525i$ & \\
		& 10 & $9.024768 - 1.250549i$ & \\
		\hline\hline
	\end{tabular}
\end{table}

\begin{table*}[!t]
	\centering
	\caption{The lowest Debye quasinormal frequencies for the massless scalar field. The radius of the compact body is $R=2.26M$.}
	\label{tab:Debye_qnm_modes_UCO}
	\small
	\setlength{\tabcolsep}{6pt}
	\renewcommand{\arraystretch}{1.15}
	\begin{tabular}{c c c l l}
		\hline\hline
		$\ell$ & $n$ & $2M\omega_{\ell n}^{(c^{\rm in})}(\rm Trapped)$ & $2M\omega_{\ell n}^{(c^{\rm in})}(\rm Curvature)$ & $2M\omega_{\ell n}^{(\alpha^{\rm in})}(\rm Interface)$ \\
		\hline
		2 & 0  & $0.248656$ & $0.771814 - 0.018150i$ & $0.940827 - 0.235419i$ \\
		& 1  & $0.328261  $ & $0.855127 - 0.027350i$ & $0.830280 - 0.800510i$ \\
		& 2  & $0.410534  $ & $0.938780 - 0.034080i$ & $0.668393 - 1.574199i$ \\
		& 3  & $0.495852  $ & $1.022701 - 0.039572i$ & $0.608762 - 2.538303i$ \\
		& 4  & $0.584527  $ & $1.106838 - 0.044266i$ & $0.677445 - 3.542707i$ \\
		& 5  & $0.679607  $ & $1.191151 - 0.048387i$ & $0.813834 - 4.536984i$ \\
		& 6  &                                  & $1.275610 - 0.052073i$ & $0.985588 - 5.518096i$ \\
		& 7  &                                  & $1.360190 - 0.055411i$ & $1.178193 - 6.488885i$ \\
		& 8  &                                  & $1.444872 - 0.058465i$ & $1.384369 - 7.451960i$ \\
		& 9  &                                  & $1.529641 - 0.061283i$ & $1.600030 - 8.409202i$ \\
		& 10 &                                  & $1.614484 - 0.063900i$ & $1.822677 - 9.361935i$ \\
		\hline\hline
	\end{tabular}
\end{table*}


\subsection{Extraction of the D-QNM contribution}
\label{sec:sec_2_4}

In this subsection, we extract the D-QNM contribution associated with the individual Debye terms entering the decomposition of the waveform response $\phi^{(p)}_\ell(t,r)$ [see Eq.~\eqref{eq:debye_response_time_p}], namely the contributions of $\phi^{(0)}_\ell(t,r)$ and of $\phi^{(p)}_\ell(t,r)$ for $p\geq 1$.

\subsubsection{Ringdown of the \(p = 0\) Debye terms: D-QNM and cut contributions}
\label{sec:sec_2_4_1}

By inserting Eq.~\eqref{eq:debye_response_D_terms_coeffs_a} into Eq.~\eqref{eq:debye_response_time_p}, the purely exterior contribution is given by
\begin{align}
	\phi_\ell^{(0)}(t,r)
	&=
	\frac{1}{2\pi}\,\Re\!\biggl[
	\int_{0+ic}^{+\infty+ic} \hspace{-5pt}
	e^{-i \omega(t-r_\ast)} P^{(-)}_{\ell}(\omega)\, d\omega
	\nonumber\\
	&\qquad\qquad +
	\int_{0+ic}^{+\infty+ic} \hspace{-5pt}
	e^{-i \omega(t-r_\ast)}
	\frac{\beta^{\rm out}_\ell(\omega)}{\alpha^{\rm in}_\ell(\omega)}\,
	P^{(+)}_\ell(\omega)\, d\omega
	\biggr].
	\label{eq:debye_response_0}
\end{align}

Since the term \(P^{(-)}_\ell(\omega)\) is regular, the pole contribution to \(\phi_\ell^{(0)}(t,r)\) is entirely governed by the zeros of \(\alpha^{\rm in}_\ell(\omega)\), namely by the Debye quasinormal frequencies \(\omega_{\ell n}^{(\alpha)}\), which are associated with the direct reflection at the stellar surface. In the Debye representation, the contour deformation also encounters the sub-threshold branch cut generated by the interior wave number \eqref{eq:kint_def_Debye}.  

This cut should be distinguished from the imaginary-axis branch cuts discussed in Ref.~\cite{Su:2026fvj}, whose contributions become non-oscillatory after the continuation \(\omega=-i\sigma\), with \(\sigma>0\), and are associated with the direct part and the late-time tail. The sub-threshold cut considered here instead lies on the real interval \(0<\omega<\omega_c\).  It therefore keeps the
oscillatory factor \(e^{-i\omega(t-r_\ast)}\), and its discontinuity has to be included in the ringing sector, together with the D-QNM residues.
\begin{figure}[!htb]
	\centering
	\includegraphics[scale=0.55]{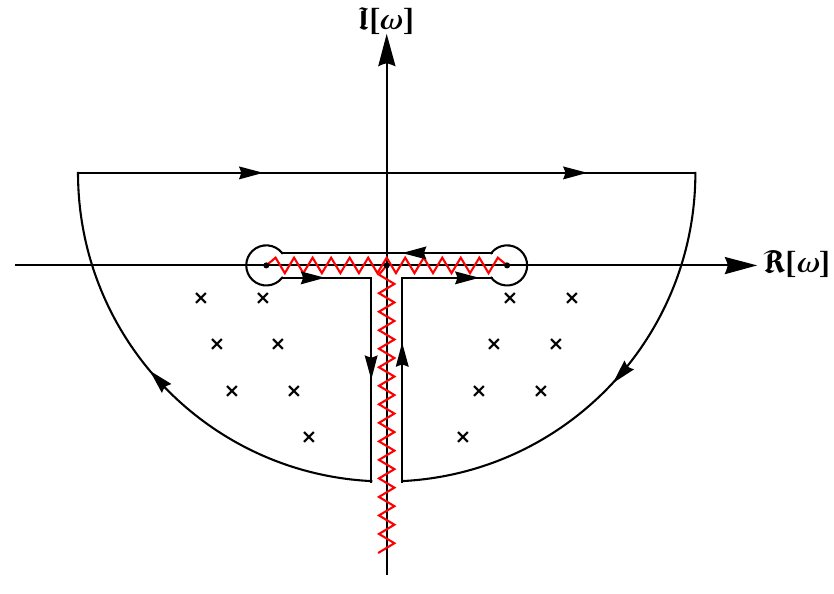}
	\caption{Contour deformation in the complex-frequency plane for the Debye contribution \(D^{(p)}(\omega)\).
		The original Bromwich contour is shifted into the lower half-plane, enclosing the D-QNM poles and the branch cuts associated with the interior momentum \(\kappa_{\rm int}\) and the Schwarzschild asymptotic branch point.
		The real-axis cut \(|\omega|<\omega_c\) gives the interior-cut contribution, while the negative-imaginary-axis cut gives the non-modal background/tail contribution.
		Small indentation circles around the branch points are shown explicitly; their contributions vanish when \(D^{(p)}(\omega)\) is locally integrable.}
	\label{fig:Contour_DQNMs_plan}
\end{figure}

Accordingly, after deforming the contour of integration in Eq.~\eqref{eq:debye_response_0} and using Cauchy's theorem (see Fig.~\ref{fig:Contour_DQNMs_plan}), we define the ringing part of the \(p=0\) Debye contribution as
\begin{equation}
	\phi_\ell^{(0),\mathrm{ring}}(t,r)
	=
	\phi_\ell^{(0),\mathrm{D\text{-}QNM}}(t,r)
	+
	\phi_\ell^{(0),\mathrm{cut}}(t,r).
	\label{eq:TimeEvolution_D-QNM_infinity_waveform_p0_with_cut}
\end{equation}
Here, \(\phi_\ell^{(0),\mathrm{D\text{-}QNM}}(t,r)\) denotes the residue contribution associated with the poles \(\omega_{\ell n}^{(\alpha)}\), whereas \(\phi_\ell^{(0),\mathrm{cut}}(t,r)\) is the contribution coming from the discontinuity across the sub-threshold cut \(0<\omega<\omega_c\). 

The D-QNM contribution is given by
\begin{equation}
	\phi_\ell^{(0),\mathrm{D\text{-}QNM}}(t,r)
	=
	2\,\Re\!\left[
	\sum_n
	-i\,\omega_{\ell n}^{(\alpha)}\,
	\mathcal{C}_{\ell n}^{(0)}\,
	e^{-i\omega_{\ell n}^{(\alpha)}(t-r_\ast)}
	\right],
	\label{eq:TimeEvolution_D-QNM_infinity_waveform_p0}
\end{equation}
while the corresponding cut contribution reads
\begin{align}
	&\phi_\ell^{(0),\mathrm{cut}}(t,r)
	= \nonumber \\
	&\frac{1}{2\pi}\,\Re\!\left[
	\int_{0}^{\omega_c}
	e^{-i \omega(t-r_\ast)}
	\left(
	\mathcal{F}^{(0)+}_\ell(\omega)
	-
	\mathcal{F}^{(0)-}_\ell(\omega)
	\right)\,d\omega
	\right],
	\label{eq:TimeEvolution_D-QNM_infinity_waveform_p0_cut}
\end{align}
with
\begin{equation}
	\mathcal{F}^{(0)\pm}_\ell(\omega)
	=
	\left[
	\frac{\beta^{\rm out}_\ell(\omega)}{\alpha^{\rm in}_\ell(\omega)}\,
	P^{(+)}_\ell(\omega)
	\right]^{\pm},
\end{equation}
where $0<\omega<\omega_c$ and  the superscripts \(+\) and \(-\) denote the values taken on the two lips of the cut.

The corresponding intrinsic Debye excitation factor is
\begin{equation}
	\mathcal{B}_{\ell n}^{(0)}
	=
	\left[
	\frac{1}{2\omega}\,
	\frac{\beta^{\rm out}_\ell(\omega)}
	{\displaystyle \frac{d\alpha^{\rm in}_\ell(\omega)}{d\omega}}
	\right]_{\omega=\omega_{\ell n}^{(\alpha)}},
	\label{eq:Debye_ExcitationFactor_waveform}
\end{equation}
while the associated Debye excitation coefficient, which incorporates the chosen initial data, is
\begin{equation}
	\mathcal{C}_{\ell n}^{(0)}
	=
	\mathcal{B}_{\ell n}^{(0)}\,
	P^{(+)}_\ell(\omega)\Big|_{\omega=\omega_{\ell n}^{(\alpha)}}.
	\label{eq:Debye_ExcitationCoefficient_waveform}
\end{equation}

\subsubsection{Ringdown of the \(p\geq1\) Debye terms: D-QNM and cut contributions}
\label{sec:sec_2_4_2}

For $p\geq1$, the Debye contribution $\phi_\ell^{(p)}(t,r)$ has a richer singular structure than the leading term. It receives pole contributions from the zeros of $\alpha_\ell^{\rm in}(\omega)$, denoted by $\omega_{\ell n}^{(\alpha)}$, which define the interface D-QNMs. It also receives contributions from the zeros of $c_\ell^{\rm in}(\omega)$, denoted by $\omega_{\ell n}^{(c)}$. In the residue sum below, we include only those $c_\ell^{\rm in}$-zeros which are strictly enclosed by the deformed contour in the fourth quadrant; these give the curvature-type D-QNMs, as illustrated in Fig.~\ref{fig:Contour_DQNMs_plan}.

In the ultracompact case, as discussed in Sec.~\ref{sec:sec_2_3}, the zeros of $c_\ell^{\rm in}(\omega)$ contain, in addition to the curvature modes, trapped modes which are attached to the lower side of the sub-threshold cut; see Fig.~\ref{fig:Debye-QNFs_UCO_R226M_l_2} and Tab.~\ref{tab:Debye_qnm_modes_UCO}. These trapped modes, together with those associated with the upper-lip branch of the cut, $k_{\rm int}^{+}$, are included in the cut contribution rather than in the fourth-quadrant residue sum.

\paragraph*{(i) Poles at $\alpha^{\rm in}_\ell(\omega)=0$.}

Substituting Eq.~\eqref{eq:debye_response_D_terms_coeffs_b} into Eq.~\eqref{eq:debye_response_time_p}, the Debye contribution of order $p\geq1$ takes the form
\begin{align}
	&\hspace{-10pt}\phi_\ell^{(p)}(t,r)
	= \nonumber \\
	&\hspace{5pt}\frac{1}{2\pi}\,
	\Re\!\biggl[
	\sum_{j\in\{P,\mathrm{in},\mathrm{out}\}}
	\int_{0+ic}^{+\infty+ic} \hspace{-5pt}
	e^{-i \omega(t-r_\ast)} \mathcal{D}^{(p)}_{\ell, j}(\omega)\, d\omega
	\biggr],
	\label{eq:debye_response_p}
\end{align}
with
\begin{align}
	\mathcal{D}^{(p)}_{\ell, P}(\omega)
	&=
	\frac{\mathcal{K}^{(p)}_{\ell,P}(\omega)}
	{\bigl(\alpha^{\rm in}_\ell(\omega)\bigr)^{p+1}}\,
	P^{(+)}_\ell(\omega),
	\\
	\mathcal{D}^{(p)}_{\ell, \mathrm{in}}(\omega)
	&=
	\frac{\mathcal{K}^{(p)}_{\ell,\mathrm{in}}(\omega)}
	{\bigl(\alpha^{\rm in}_\ell(\omega)\bigr)^p}\,
	Q^{\rm in}_\ell(\omega),
	\\
	\mathcal{D}^{(p)}_{\ell, \mathrm{out}}(\omega)
	&=
	\frac{\mathcal{K}^{(p)}_{\ell,\mathrm{out}}(\omega)}
	{\bigl(\alpha^{\rm in}_\ell(\omega)\bigr)^p}\,
	Q^{\rm out}_\ell(\omega),
	\label{eq:debye_Dp_defs}
\end{align}
where
\begin{subequations}\label{eq:Debye_Kp_defs}
	\begin{align}
		\mathcal{K}^{(p)}_{\ell,P}(\omega)
		&=
		\left(\frac{k_{\rm int}(\omega,\ell)}{\omega}\right)^p
		\bigl(\delta^{\rm in}_\ell(\omega)\bigr)^{p-1}
		\bigl(\xi_\ell(\omega)\bigr)^p,
		\label{eq:Debye_Kp_P}
		\\
		\mathcal{K}^{(p)}_{\ell,\mathrm{in}}(\omega)
		&=
		\left(\frac{k_{\rm int}(\omega,\ell)}{\omega}\right)^{p-1}
		\bigl(\delta^{\rm in}_\ell(\omega)\bigr)^{p-1}
		\bigl(\xi_\ell(\omega)\bigr)^{p-1},
		\label{eq:Debye_Kp_in}
		\\
		\mathcal{K}^{(p)}_{\ell,\mathrm{out}}(\omega)
		&=
		\left(\frac{k_{\rm int}(\omega,\ell)}{\omega}\right)^{p-1}
		\bigl(\delta^{\rm in}_\ell(\omega)\bigr)^{p-1}
		\bigl(\xi_\ell(\omega)\bigr)^p.
		\label{eq:Debye_Kp_out}
	\end{align}
\end{subequations}
In Eq.~\eqref{eq:debye_response_p}, $j\in\{P,\mathrm{in},\mathrm{out}\}$ labels the three pieces of the Debye integrand: the exterior overlap contribution, and the two contributions associated with the interior basis functions \(u^{\rm in}_{\omega\ell}\) and \(u^{\rm out}_{\omega\ell}\).

To introduce the associated Debye excitation factors, we define
\begin{align}
	h_{\ell n}^{(\alpha)}(\omega)
	&\equiv
	\frac{\alpha^{\rm in}_\ell(\omega)}
	{\omega-\omega_{\ell n}^{(\alpha)}},
	\\
	h_{\ell n}^{(\alpha)}\!\left(\omega_{\ell n}^{(\alpha)}\right)
	&=
	\left.\frac{d}{d\omega}\alpha^{\rm in}_\ell(\omega)\right|_{\omega=\omega_{\ell n}^{(\alpha)}} .
\end{align}
More generally, the Taylor expansion of $\alpha^{\rm in}_\ell(\omega)$ around $\omega=\omega_{\ell n}^{(\alpha)}$ yields
\begin{equation}
	\left.
	\frac{d^k}{d\omega^k}h_{\ell n}^{(\alpha)}(\omega)
	\right|_{\omega=\omega_{\ell n}^{(\alpha)}}
	=
	\frac{1}{k+1}
	\left.
	\frac{d^{k+1}}{d\omega^{k+1}}\alpha^{\rm in}_\ell(\omega)
	\right|_{\omega=\omega_{\ell n}^{(\alpha)}},
\end{equation}
with $k=0,1,2,\dots$.

The pole order associated with each channel is
\begin{equation}
	m_P^{(\alpha)}=p+1,
	\qquad
	m_{\mathrm{in}}^{(\alpha)}=m_{\mathrm{out}}^{(\alpha)}=p.
\end{equation}

The intrinsic Debye excitation factors for the channel $j$ are defined by
\begin{align}
	\mathcal{B}_{\ell n, j,q}^{(p,\alpha)}
	&=
	\left[
	\frac{1}{2\omega}\,
	\frac{1}{q!}
	\frac{d^{\,q}}{d\omega^{\,q}}
	\left(
	\frac{\mathcal{K}^{(p)}_{\ell,j}(\omega)}
	{\bigl(h_{\ell n}^{(\alpha)}(\omega)\bigr)^{m_j^{(\alpha)}}}
	\right)
	\right]_{\omega=\omega_{\ell n}^{(\alpha)}},
	\label{eq:Debye_ExcitationFactor_waveform_alpha_compact}
\end{align}
for
\begin{equation}
	q=0,1,\dots,m_j^{(\alpha)}-1.
\end{equation}

The corresponding Debye excitation coefficients, which encode the chosen initial data, are given by
\begin{align}
	&\mathcal{C}_{\ell n, j,s}^{(p,\alpha)}
	= \nonumber\\
	&\sum_{k=0}^{m_j^{(\alpha)}-1-s} \hspace{-5pt}
	\frac{1}{k!}
	\left[
	\frac{d^{\,k}}{d\omega^k}\mathcal{S}_{\ell,j}(\omega)
	\right]_{\omega=\omega_{\ell n}^{(\alpha)}}
	\mathcal{B}_{\ell n, j,\,m_j^{(\alpha)}-1-s-k}^{(p,\alpha)},
	\label{eq:Debye_ExcitationCoefficient_waveform_alpha_compact}
\end{align}
with
\begin{equation}
	s=0,1,\dots,m_j^{(\alpha)}-1.
\end{equation}
The source amplitudes read
\begin{align}
	\mathcal{S}_{\ell,P}(\omega)&=P^{(+)}_\ell(\omega),
	\nonumber\\
	\mathcal{S}_{\ell,\mathrm{in}}(\omega)&=Q^{\rm in}_\ell(\omega),
	\nonumber\\
	\mathcal{S}_{\ell,\mathrm{out}}(\omega)&=Q^{\rm out}_\ell(\omega).
	\label{eq:source_terms}
\end{align}
For the three channels, one obtains
\begin{align}
	\mathcal{C}_{\ell n, P,s}^{(p,\alpha)}
	&=
	\sum_{k=0}^{p-s}
	\frac{1}{k!}
	\left[
	\frac{d^{\,k}}{d\omega^k}P^{(+)}_\ell(\omega)
	\right]_{\omega=\omega_{\ell n}^{(\alpha)}}
	\mathcal{B}_{\ell n, P,\,p-s-k}^{(p,\alpha)},
	\\
	\mathcal{C}_{\ell n, \mathrm{in},s}^{(p,\alpha)}
	&=
	\sum_{k=0}^{p-1-s}
	\frac{1}{k!}
	\left[
	\frac{d^{\,k}}{d\omega^k}Q^{\rm in}_\ell(\omega)
	\right]_{\omega=\omega_{\ell n}^{(\alpha)}}
	\mathcal{B}_{\ell n, \mathrm{in},\,p-1-s-k}^{(p,\alpha)},
	\\
	\mathcal{C}_{\ell n, \mathrm{out},s}^{(p,\alpha)}
	&=
	\sum_{k=0}^{p-1-s}
	\frac{1}{k!}
	\left[
	\frac{d^{\,k}}{d\omega^k}Q^{\rm out}_\ell(\omega)
	\right]_{\omega=\omega_{\ell n}^{(\alpha)}}
	\mathcal{B}_{\ell n, \mathrm{out},\,p-1-s-k}^{(p,\alpha)},
\end{align}
where \(s=0,\dots,p\) for the \(P\)-channel and \(s=0,\dots,p-1\) for the \(\mathrm{in}\) and \(\mathrm{out}\) channels.

By deforming the contour of integration and using Cauchy's theorem, one isolates the pole contribution associated with the zeros $\omega_{\ell n}^{(\alpha)}$ of $\alpha^{\rm in}_\ell(\omega)$. Taking into account that the derivatives associated with the higher-order poles act both on the non-exponential part of the integrand and on the phase factor \(e^{-i\omega(t-r_\ast)}\), and using the definitions of the Debye excitation factors \eqref{eq:Debye_ExcitationFactor_waveform_alpha_compact} and of the Debye excitation coefficients \eqref{eq:Debye_ExcitationCoefficient_waveform_alpha_compact}, one finds
\begin{align}
	&\phi_{\ell}^{(p,\alpha),\mathrm{D\text{-}QNM}}(t,r)
	=\nonumber \\
	&\hspace{20pt}2\,\Re\!\biggl[
	\sum_{j\in\{P,\mathrm{in},\mathrm{out}\}}\sum_n
	-i\,\omega_{\ell n}^{(\alpha)}\,
	e^{-i\omega_{\ell n}^{(\alpha)}(t-r_\ast)}
	\nonumber\\
	&\hspace{20pt}\qquad\qquad\qquad \times
	\sum_{s=0}^{m_j^{(\alpha)}-1}
	\frac{\bigl[-i(t-r_\ast)\bigr]^s}{s!}\,
	\mathcal{C}_{\ell n, j,s}^{(p,\alpha)}
	\biggr].
	\label{eq:TimeEvolution_D-QNM_infinity_waveform_p_alpha}
\end{align}
This expression gives the residue contribution associated with the poles \(\omega_{\ell n}^{(\alpha)}\), and makes explicit the polynomial factors in \((t-r_\ast)\) generated by the higher-order poles.

\paragraph*{(ii) Poles at $c^{\rm in}_\ell(\omega)=0$.}

Using Eq.~\eqref{eq:debye_response_xi}, the three channel contributions entering Eq.~\eqref{eq:debye_response_p} can be rewritten as
\begin{align}
	\mathcal{D}^{(p)}_{\ell, P}(\omega)
	&=
	\frac{\widetilde{\mathcal{K}}^{(p)}_{\ell,P}(\omega)}
	{\bigl(c^{\rm in}_\ell(\omega)\bigr)^p}\,
	P^{(+)}_\ell(\omega),
	\\
	\mathcal{D}^{(p)}_{\ell, \mathrm{in}}(\omega)
	&=
	\frac{\widetilde{\mathcal{K}}^{(p)}_{\ell,\mathrm{in}}(\omega)}
	{\bigl(c^{\rm in}_\ell(\omega)\bigr)^{p-1}}\,
	Q^{\rm in}_\ell(\omega),
	\\
	\mathcal{D}^{(p)}_{\ell, \mathrm{out}}(\omega)
	&=
	\frac{\widetilde{\mathcal{K}}^{(p)}_{\ell,\mathrm{out}}(\omega)}
	{\bigl(c^{\rm in}_\ell(\omega)\bigr)^p}\,
	Q^{\rm out}_\ell(\omega),
\end{align}
where
\begin{align}
	\widetilde{\mathcal{K}}^{(p)}_{\ell,P}(\omega)
	&=
	\left(\frac{k_{\rm int}(\omega,\ell)}{\omega}\right)^p
	\frac{\bigl(c^{\rm out}_\ell(\omega)\bigr)^p
		\bigl(\delta^{\rm in}_\ell(\omega)\bigr)^{p-1}}
	{\bigl(\alpha^{\rm in}_\ell(\omega)\bigr)^{p+1}},
	\\
	\widetilde{\mathcal{K}}^{(p)}_{\ell,\mathrm{in}}(\omega)
	&=
	\left(\frac{k_{\rm int}(\omega,\ell)}{\omega}\right)^{p-1}
	\frac{\bigl(c^{\rm out}_\ell(\omega)\bigr)^{p-1}
		\bigl(\delta^{\rm in}_\ell(\omega)\bigr)^{p-1}}
	{\bigl(\alpha^{\rm in}_\ell(\omega)\bigr)^p},
	\\
	\widetilde{\mathcal{K}}^{(p)}_{\ell,\mathrm{out}}(\omega)
	&=
	\left(\frac{k_{\rm int}(\omega,\ell)}{\omega}\right)^{p-1}
	\frac{\bigl(c^{\rm out}_\ell(\omega)\bigr)^p
		\bigl(\delta^{\rm in}_\ell(\omega)\bigr)^{p-1}}
	{\bigl(\alpha^{\rm in}_\ell(\omega)\bigr)^p}.
\end{align}

It is convenient to introduce
\begin{equation}
	J_p^{(c)}=
	\begin{cases}
		\{P,\mathrm{out}\}, & p=1,\\
		\{P,\mathrm{in},\mathrm{out}\}, & p\ge2.
	\end{cases}
\end{equation}

To construct the corresponding Debye excitation factors, we define
\begin{align}
	h_{\ell n}^{(c)}(\omega)
	&\equiv
	\frac{c^{\rm in}_\ell(\omega)}
	{\omega-\omega_{\ell n}^{(c)}},
	\\
	h_{\ell n}^{(c)}\!\left(\omega_{\ell n}^{(c)}\right)
	&=
	\left.\frac{d}{d\omega}c^{\rm in}_\ell(\omega)\right|_{\omega=\omega_{\ell n}^{(c)}} .
\end{align}
More generally, the Taylor expansion of \(c^{\rm in}_\ell(\omega)\) around \(\omega=\omega_{\ell n}^{(c)}\) gives
\begin{equation}
	\left.
	\frac{d^k}{d\omega^k}h_{\ell n}^{(c)}(\omega)
	\right|_{\omega=\omega_{\ell n}^{(c)}}
	=
	\frac{1}{k+1}
	\left.
	\frac{d^{k+1}}{d\omega^{k+1}}c^{\rm in}_\ell(\omega)
	\right|_{\omega=\omega_{\ell n}^{(c)}},
\end{equation}
with \(k=0,1,2,\dots\).

The pole order associated with each channel is
\begin{equation}
	m_P^{(c)}=p,
	\qquad
	m_{\mathrm{in}}^{(c)}=p-1,
	\qquad
	m_{\mathrm{out}}^{(c)}=p.
\end{equation}
In particular, the \(\mathrm{in}\)-channel is regular for \(p=1\) and therefore does not contribute in that case.

The intrinsic Debye excitation factors for channel \(j\) are defined by
\begin{align}
	\mathcal{B}_{\ell n, j,q}^{(p,c)}
	&=
	\left[
	\frac{1}{2\omega}\,
	\frac{1}{q!}
	\frac{d^{\,q}}{d\omega^{\,q}}
	\left(
	\frac{\widetilde{\mathcal{K}}^{(p)}_{\ell,j}(\omega)}
	{\bigl(h_{\ell n}^{(c)}(\omega)\bigr)^{m_j^{(c)}}}
	\right)
	\right]_{\omega=\omega_{\ell n}^{(c)}},
	\label{eq:Debye_ExcitationFactor_waveform_c_compact}
\end{align}
for
\begin{equation}
	q=0,1,\dots,m_j^{(c)}-1.
\end{equation}

The associated Debye excitation coefficients, which encode the chosen initial data, are then
\begin{align}
	&\mathcal{C}_{\ell n,j,s}^{(p,c)}
	= \nonumber \\
	&\sum_{k=0}^{m_j^{(c)}-1-s} \hspace{-5pt}
	\frac{1}{k!}
	\left[
	\frac{d^{\,k}}{d\omega^k}\mathcal{S}_{\ell,j}(\omega)
	\right]_{\omega=\omega_{\ell n}^{(c)}}
	\mathcal{B}_{\ell n,j,\,m_j^{(c)}-1-s-k}^{(p,c)},
	\label{eq:Debye_ExcitationCoefficient_waveform_c_compact}
\end{align}
with
\begin{equation}
	s=0,1,\dots,m_j^{(c)}-1.
\end{equation}
and the source terms are given by \eqref{eq:source_terms}.

For the contributing channels, one obtains
\begin{align}
	\mathcal{C}_{\ell n, P,s}^{(p,c)}
	&=
	\sum_{k=0}^{p-1-s}
	\frac{1}{k!}
	\left[
	\frac{d^{\,k}}{d\omega^k}P^{(+)}_\ell(\omega)
	\right]_{\omega=\omega_{\ell n}^{(c)}}
	\mathcal{B}_{\ell n, P,\,p-1-s-k}^{(p,c)},
	\\
	\mathcal{C}_{\ell n, \mathrm{in},s}^{(p,c)}
	&=
	\sum_{k=0}^{p-2-s}
	\frac{1}{k!}
	\left[
	\frac{d^{\,k}}{d\omega^k}Q^{\rm in}_\ell(\omega)
	\right]_{\omega=\omega_{\ell n}^{(c)}}
	\mathcal{B}_{\ell n, \mathrm{in},\,p-2-s-k}^{(p,c)},
	\\
	\mathcal{C}_{\ell n, \mathrm{out},s}^{(p,c)}
	&=
	\sum_{k=0}^{p-1-s}
	\frac{1}{k!}
	\left[
	\frac{d^{\,k}}{d\omega^k}Q^{\rm out}_\ell(\omega)
	\right]_{\omega=\omega_{\ell n}^{(c)}}
	\mathcal{B}_{\ell n, \mathrm{out},\,p-1-s-k}^{(p,c)},
\end{align}
where \(s=0,\dots,p-1\) for the \(P\)- and \(\mathrm{out}\)-channels, and \(s=0,\dots,p-2\) for the \(\mathrm{in}\)-channel, with \(p\ge2\).

By deforming the contour of integration and applying Cauchy's theorem, one isolates the pole contribution associated with the zeros \(\omega_{\ell n}^{(c)}\) of \(c^{\rm in}_\ell(\omega)\). Taking into account that the derivatives generated by the higher-order poles act both on the non-exponential part of the integrand and on the phase factor \(e^{-i\omega(t-r_\ast)}\), and using the definitions \eqref{eq:Debye_ExcitationFactor_waveform_c_compact} and \eqref{eq:Debye_ExcitationCoefficient_waveform_c_compact}, one arrives at
\begin{align}
	&\phi_{\ell}^{(p,c),\mathrm{D\text{-}QNM}}(t,r)
	= \nonumber\\
	&\hspace{25pt} 2\,\Re\!\biggl[
	\sum_{j\in J_p^{(c)}}\sum_n
	-i\,\omega_{\ell n}^{(c)}\,
	e^{-i\omega_{\ell n}^{(c)}(t-r_\ast)} \nonumber \\
	&\hspace{15pt}\qquad\qquad\qquad \times
	\sum_{s=0}^{m_j^{(c)}-1}
	\frac{\bigl[-i(t-r_\ast)\bigr]^s}{s!}\,
	\mathcal{C}_{\ell n, j,s}^{(p,c)}
	\biggr].
		\label{eq:TimeEvolution_D-QNM_infinity_waveform_p_c}
\end{align}

This expression represents the residue contribution associated with the subset of zeros $\omega_{\ell n}^{(c)}$. In the ultracompact case, this subset corresponds to the curvature-type D-QNMs.

\paragraph*{(iii) Cut contribution of the Debye term of order $p$.}

Besides the pole contributions associated with the zeros of $\alpha^{\rm in}_\ell(\omega)$ and with the curvature-type zeros of $c^{\rm in}_\ell(\omega)$, the contour deformation of the Debye term of order $p$ also encounters the sub-threshold branch cut generated by the interior wavenumber $k_{\rm int}$, Eq.~\eqref{eq:kint_def_Debye} (see Fig.~\ref{fig:Contour_DQNMs_plan}).

This contribution is associated with the discontinuity of the Debye integrand across the two lips of the cut. We therefore define
\begin{equation}
	\mathcal{F}^{(p)\pm}_\ell(\omega)
	=
	\left[
	\sum_{j\in\{P,\mathrm{in},\mathrm{out}\}}
	\mathcal{D}^{(p)}_{\ell,j}(\omega)
	\right]^{\pm},
\end{equation}
where the superscripts \(+\) and \(-\) denote the values taken on the two sides of the cut. The corresponding cut contribution is
\begin{align}
	\phi_\ell^{(p),\mathrm{cut}}(t,r)
	&=
	\frac{1}{2\pi}\,
	\Re\!\biggl[
	\int_0^{\omega_c}
	e^{-i\omega(t-r_\ast)} \nonumber \\
	&\quad\times
	\Bigl(
	\mathcal{F}^{(p)+}_\ell(\omega)
	-
	\mathcal{F}^{(p)-}_\ell(\omega)
	\Bigr)\,d\omega
	\biggr].
	\label{eq:TimeEvolution_D-QNM_infinity_waveform_p_cut}
\end{align}

It is important to note that Eq.~\eqref{eq:TimeEvolution_D-QNM_infinity_waveform_p_cut} is written in a compact form. When no pole lies on the cut, the integral is understood in the usual sense. In the ultracompact case, however, trapped poles lie on the two lips of the sub-threshold cut, associated with the two branch values of \(k_{\rm int}\). The cut contribution must then be understood in a regularized sense: for the first Debye order \(p=1\), the integral is evaluated as a Cauchy principal value, whereas for higher orders \(p\geq2\), the higher-order singularities are treated by Hadamard finite parts. The details of this prescription and its numerical implementation are given in Appendix~\ref{app:cut_regularization}.

Combining the two D-QNM sectors with the discontinuity across the sub-threshold cut, the ringing part of the Debye term of order \(p\geq1\) is written as
\begin{align}
	\phi_\ell^{(p),\mathrm{ring}}(t,r)
	&=
	\phi_{\ell}^{(p,\alpha),\mathrm{D\text{-}QNM}}(t,r)
	+
	\phi_{\ell}^{(p,c),\mathrm{D\text{-}QNM}}(t,r)
	\nonumber\\
	&\quad
	+
	\phi_\ell^{(p),\mathrm{cut}}(t,r).
	\label{eq:TimeEvolution_D-QNM_infinity_waveform_p_ring}
\end{align}

\section{The Debye reconstruction: D-QNMs and echoes}
\label{sec:sec_3}

\subsection{Computational methods}
\label{sec:sec_3_1}

To construct the waveforms obtained from the closed-form Debye response, Eq.~\eqref{eq:debye_response_phi_inf} [see also Eqs.~\eqref{eq:debye_response_D_def} and \eqref{eq:debye_response_D_exact}], we first compute the effective scattering coefficients appearing in Eq.~\eqref{eq:debye_response_RijTij} [cf. Eq.~\eqref{eq:connection_coeffs_Debye}]. This is done by solving the radial equation~\eqref{eq:RW_freq_waveform} both inside and outside the compact object. In the interior region, we integrate the equation with the potential~\eqref{eq:V_general_scalar_waveform}, starting close to the center, typically at \(r=10^{-6}\), with initial data supplied by the regular Frobenius expansion (see Ref.~\cite{OuldElHadj:2026ewh} for more details). The solution is then propagated up to the surface \(r=R\) and matched there to the interior basis \(\{u^{\rm in}_{\omega\ell},u^{\rm out}_{\omega\ell}\}\). In the exterior region, we integrate Eq.~\eqref{eq:RW_freq_waveform} with the potential~\eqref{eq:Vext_waveform}, starting from a large radius where the solution is initialized by its asymptotic expansion at spatial infinity, and propagate it down to the surface. The comparison with the same surface basis then provides the effective coefficients used in the Debye representation.

 Once these coefficients have been obtained, the frequency-domain kernel is reconstructed from Eq.~\eqref{eq:debye_response_D_exact}. In the applications considered here, the initial Cauchy data are supported outside the compact object. As a result, the interior overlap integrals \(Q^{\rm in}_\ell\) and \(Q^{\rm out}_\ell\) vanish, and only the exterior overlaps \(P^{(-)}_\ell\) and \(P^{(+)}_\ell\) contribute. The time-domain waveform is then obtained from the real-frequency inverse transform associated with Eq.~\eqref{eq:debye_response_phi_inf}. The real-axis kernel is taken as the boundary value of the retarded response approached from the upper half-plane, \(D_\ell(\omega+i0)\). For the regular part of the kernel, this reduces to a Fourier-type real-frequency quadrature. When the sub-threshold sector \(0<\omega<\omega_c\) contributes, the two lips of the cut associated with the analytic continuation of \(k_{\rm int}(\omega,\ell)\) must be distinguished. In the ultracompact case, possible trapped-pole singularities attached to these lips are treated with the principal-value and Hadamard finite-part prescriptions described in Appendix~\ref{app:cut_regularization}.

The same procedure is used to construct the individual Debye contributions defined in Eq.~\eqref{eq:debye_response_time_p}. The leading term is obtained from Eq.~\eqref{eq:debye_response_D0}, while the higher-order terms are obtained from Eq.~\eqref{eq:debye_response_Dp_exact} [see also Eqs.~\eqref{eq:debye_response_D_terms_coeffs_a}~and~\eqref{eq:debye_response_D_terms_coeffs_b}]. As for the closed-form Debye response, only the \(P^{(-)}_\ell\) and \(P^{(+)}_\ell\) channels are retained. This construction is applied in the same way to the two configurations considered below, namely the neutron-star-like model with \(R=6M\) and the ultracompact model with \(R=2.26M\).

We next extract the ringing content of each Debye order. For the leading term \(p=0\), the D-QNM contribution is obtained from Eq.~\eqref{eq:TimeEvolution_D-QNM_infinity_waveform_p0}. To this end, we first compute the intrinsic excitation factors \(\mathcal B_{\ell n}^{(0)}\), Eq.~\eqref{eq:Debye_ExcitationFactor_waveform}, and the corresponding excitation coefficients \(\mathcal C_{\ell n}^{(0)}\), Eq.~\eqref{eq:Debye_ExcitationCoefficient_waveform}. These quantities are evaluated at the Debye quasinormal frequencies \(\omega_{\ell n}^{(\alpha)}\), i.e., at the zeros of \(\alpha^{\rm in}_\ell(\omega)\). The coefficients \(\alpha^{\rm in}_\ell\) and \(\beta^{\rm out}_\ell\) appearing in the excitation factors are obtained from the same exterior integration and surface matching procedure described above. In addition to this residue contribution, we also compute the contribution of the sub-threshold branch cut \(0<\omega<\omega_c\). The two lips of the cut are constructed according to the branch convention chosen for the interior wavenumber \(k_{\rm int}(\omega,\ell)\), Eq.~\eqref{eq:kint_def_Debye}, using the same numerical ingredients as in the construction of the full Debye waveform.

For \(p\geq1\), the ringing part contains two possible D-QNM sectors. The first one is associated with the same zeros \(\omega_{\ell n}^{(\alpha)}\) of \(\alpha^{\rm in}_\ell(\omega)\). Its contribution is obtained by computing the excitation factors of Eq.~\eqref{eq:Debye_ExcitationFactor_waveform_alpha_compact} and the corresponding excitation coefficients of Eq.~\eqref{eq:Debye_ExcitationCoefficient_waveform_alpha_compact}, restricted here to the exterior \(P\)-channel since the interior \(Q^{\rm in/out}\) overlaps vanish for the Cauchy data used in this work. Substitution of these coefficients into Eq.~\eqref{eq:TimeEvolution_D-QNM_infinity_waveform_p_alpha} gives the \(\alpha^{\rm in}\)-D-QNM contribution at order \(p\).

The second D-QNM sector is associated with the zeros \(\omega_{\ell n}^{(c)}\) of \(c^{\rm in}_\ell(\omega)\). In this case, we compute the excitation factors from Eq.~\eqref{eq:Debye_ExcitationFactor_waveform_c_compact} and the excitation coefficients from Eq.~\eqref{eq:Debye_ExcitationCoefficient_waveform_c_compact}, again retaining only the exterior \(P\)-channel. Inserting these coefficients into Eq.~\eqref{eq:TimeEvolution_D-QNM_infinity_waveform_p_c} gives the corresponding \(c^{\rm in}\)-D-QNM contribution. The physical content of this sector depends on the compactness. For the neutron-star-like configuration \(R=6M\), the relevant \(c^{\rm in}\)-D-QNMs form the curvature-mode family. For the ultracompact configuration \(R=2.26M\), this sector also contains trapped modes. Finally, the branch-cut contribution on \(0<\omega<\omega_c\) is included for all Debye orders in the \(R=6M\) case, where it gives a non-negligible contribution. For the ultracompact case \(R=2.26M\), the cut is included for \(p=0\), while for \(p\geq1\), the main echo/ringdown packets are dominated by the curvature-type D-QNM residues. The regularized sub-threshold cut, however, is required to account consistently for the trapped poles attached to the two lips of the cut and it becomes visible mainly in the low-amplitude late-time tail. These ingredients finally give the ringing part of the Debye term of order \(p\geq1\), \(\phi_\ell^{(p),\mathrm{ring}}(t,r)\), through Eq.~\eqref{eq:TimeEvolution_D-QNM_infinity_waveform_p_ring}. All numerical results presented below were obtained using Mathematica~\cite{Mathematica}.

\subsection{Debye reconstruction of the waveform, ringdown content, and echoes}
\label{sec:sec_3_2}

\begin{figure*}[!htb]
	\centering
	\includegraphics[scale=0.55]{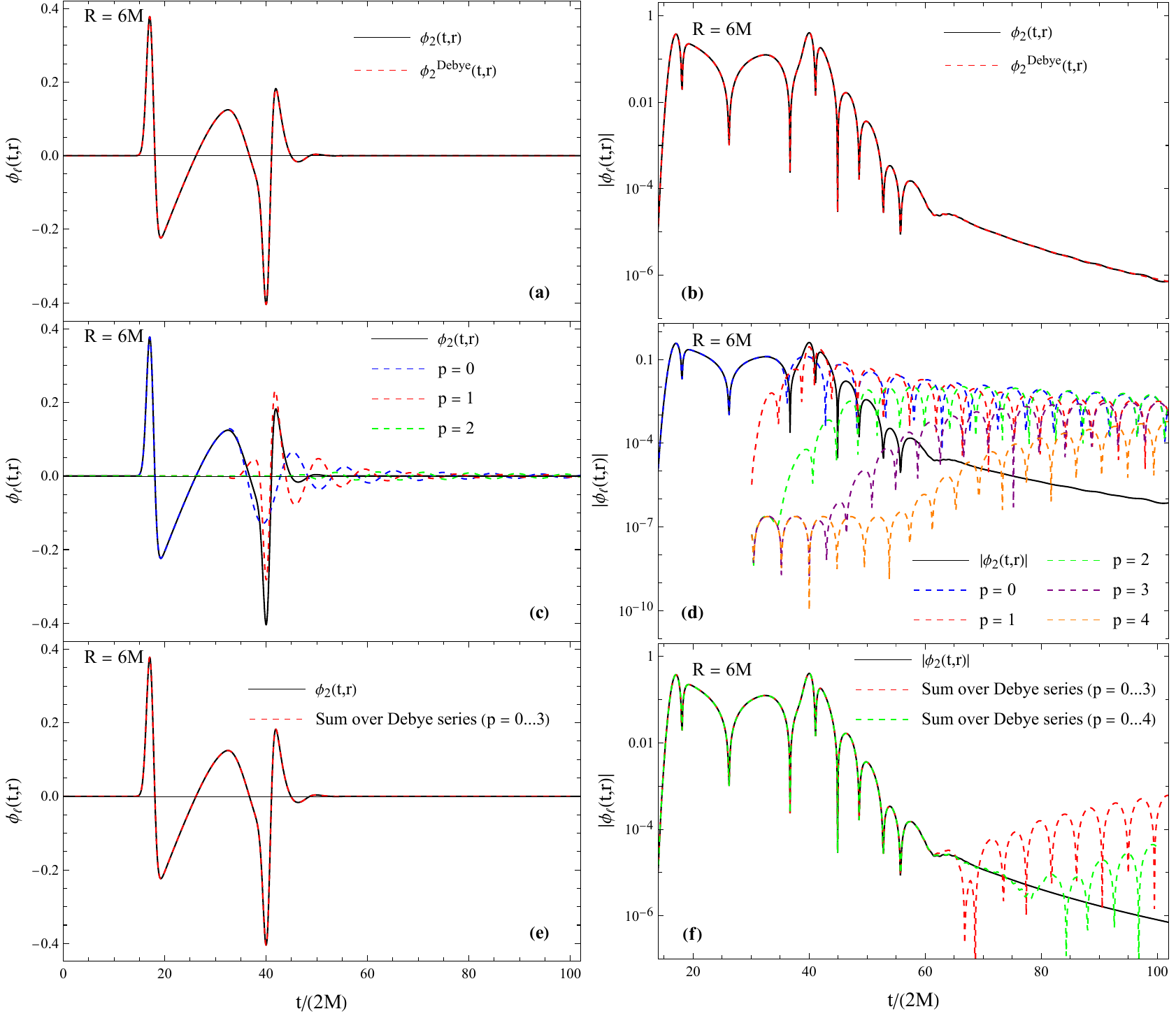}
	\caption{
		Debye reconstruction of the \(\ell=2\) waveform for the neutron-star-like configuration \(R=6M\).  The observer is located at \(r_{\rm obs}=50M\). The Gaussian initial data in Eq.~\eqref{eq:Cauchy_data_waveform} are specified by \(\phi_0=1\), \(a=1\), and \(r_0=17.84M\).
		Panel (a) compares the exact frequency-domain waveform \(\phi_2(t,r)\), computed from the full response kernel in Eq.~\eqref{eq:response_infinity_waveform}, with the Debye-reconstructed waveform \(\phi^{\rm Debye}_2(t,r)\), obtained by inserting the Debye expansion of the closed cavity-form kernel, Eq.~\eqref{eq:debye_response_D_exact}, into Eq.~\eqref{eq:response_infinity_waveform}. 
		Panel (b) shows the corresponding logarithmic representation of the absolute value. 
		Panel (c) displays the exact waveform together with the first individual Debye contributions, \(p=0\), \(p=1\), and \(p=2\). 
		Panel (d) shows, in logarithmic scale, the exact waveform together with the individual Debye contributions from \(p=0\) to \(p=4\). 
		Panel (e) compares the exact waveform with the partial Debye sum truncated at \(p=3\). 
		Panel (f) shows the corresponding logarithmic representation, together with the partial sums truncated at \(p=3\) and \(p=4\). 
		The Debye reconstruction obtained from the closed cavity form is almost indistinguishable from the exact waveform, while the low-order Debye sums reproduce the signal accurately on a linear scale but display residual late-time oscillations in logarithmic scale. 
		In this non-ultracompact configuration, the Debye terms do not form a sequence of well-separated echoes, consistently with the characteristic propagation times discussed in Sec.~\ref{sec:sec_1_2_1}.
	}
	\label{fig:Debye_Waveform_NS}
\end{figure*}
\begin{figure*}[!htb]
	\centering
	\includegraphics[scale=0.55]{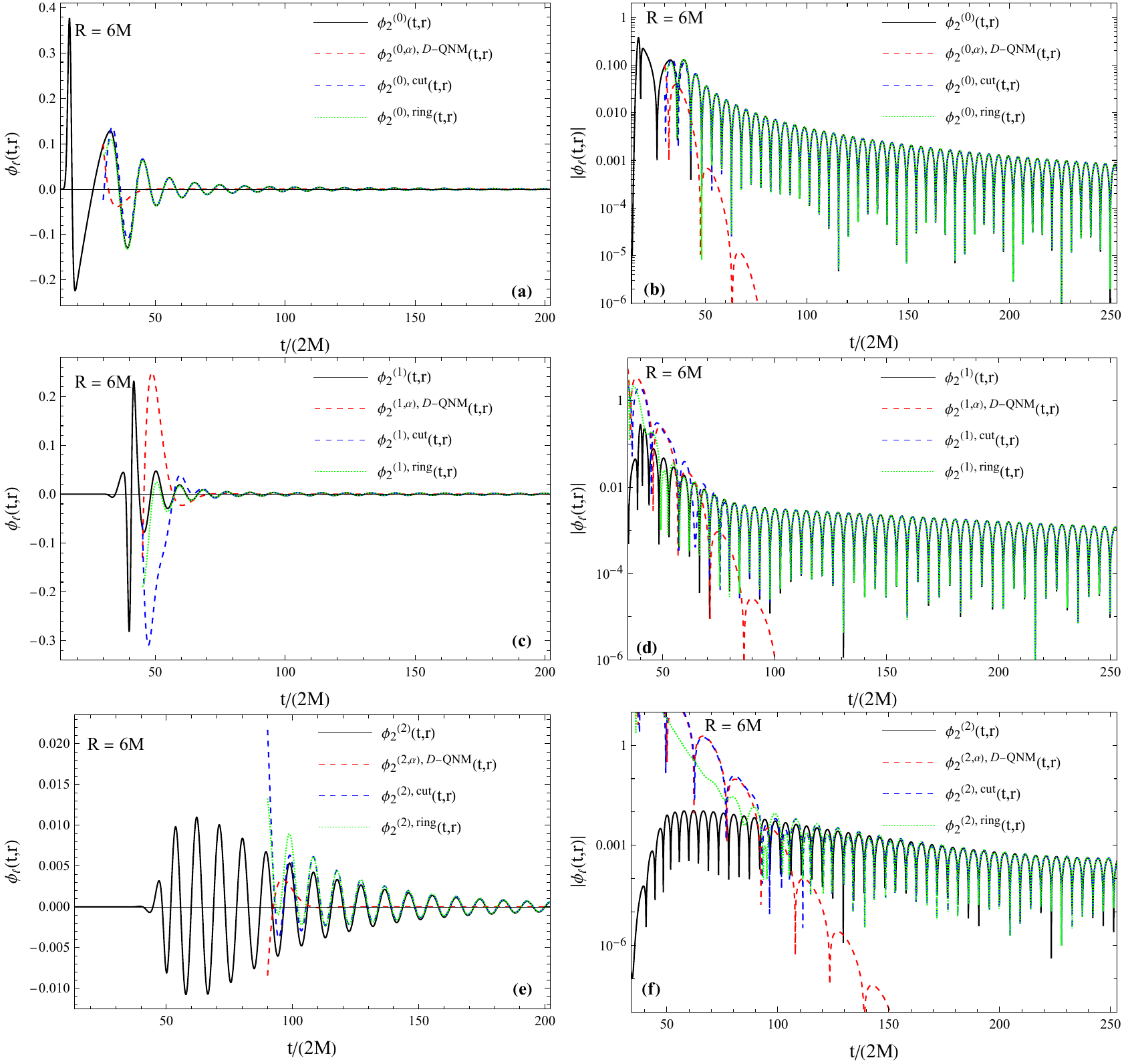}
	\caption{
		D-QNM and branch-cut content of the first  Debye-series terms of the $\ell=2$ waveform for the neutron-star-like configuration $R=6M$. The observer is located at \(r_{\rm obs}=50M\). The Gaussian initial data in Eq.~\eqref{eq:Cauchy_data_waveform} are specified by \(\phi_0=1\), \(a=1\), and \(r_0=17.84M\).  
		For each order $p$, the exact Debye contribution $\phi^{(p)}_2(t,r)$ is compared with its D-QNM contribution associated with the poles of $\alpha^{\rm in}_\ell(\omega)$, and with the branch-cut contribution associated with the sub-threshold cut $0<\omega<\omega_c$. 
		Panels (a), (c), and (e) correspond to $p=0$, $p=1$, and $p=2$, respectively, while panels (b), (d), and (f) show the same quantities in logarithmic scale. 
		For $p=0$, the branch-cut contribution provides the dominant part of both the ringdown and the late-time tail of the leading Debye term, while the $\alpha^{\rm in}_\ell$ D-QNM sector contributes mainly near the onset of the ringdown. 
		For $p\geq1$, the ringdown receives contributions from the interface-mode D-QNMs associated with the poles of $\alpha^{\rm in}_\ell(\omega)$, while the late-time tail remains dominated by the branch-cut contribution. Additional curvature-mode D-QNM contributions associated with the poles $\omega^{(c)}_{\ell n}$ of $c^{\rm in}_\ell(\omega)$ are also present in principle, but they are not shown here since they are strongly suppressed for the present configuration.
	}
	\label{fig:NS_Exact_DQNMS_Cut_p}
\end{figure*}
\begin{figure*}[!htb]
	\centering
	\includegraphics[scale=0.50]{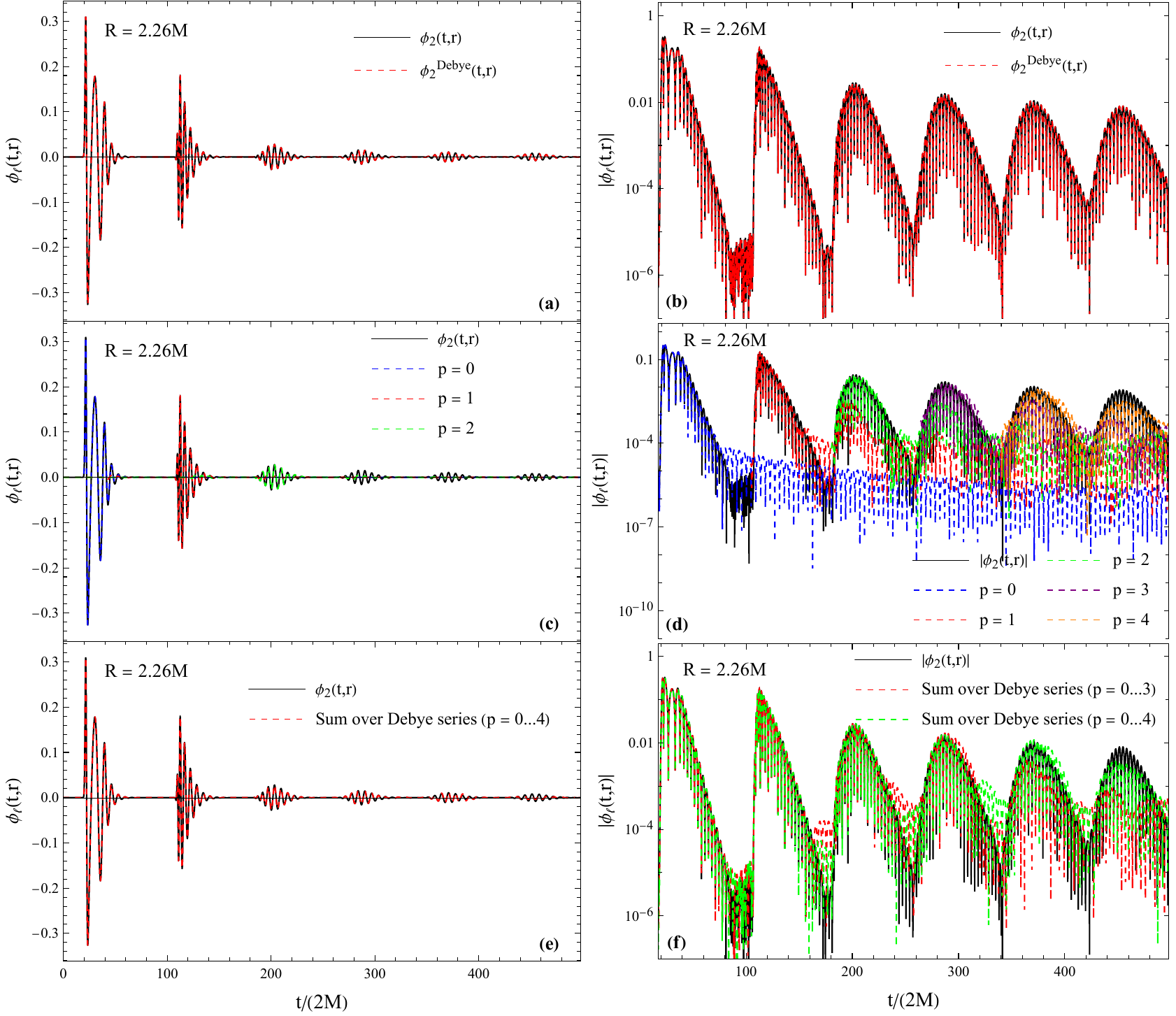}
	\caption{
		Debye reconstruction of the $\ell=2$ waveform for the ultracompact-object configuration $R=2.26M$.  The observer is located at \(r_{\rm obs}=50M\). The Gaussian initial data in Eq.~\eqref{eq:Cauchy_data_waveform} are specified by \(\phi_0=1\), \(a=1\), and \(r_0=10M\).
		Panel (a) compares the time-domain waveform $\phi_2(t,r)$, obtained from Eq.~\eqref{eq:response_infinity_waveform}, with the waveform reconstructed from the closed-form Debye expression, $\phi^{\rm Debye}_2(t,r)$, Eq.~\eqref{eq:debye_response_phi_inf} (cf. Eq.~\eqref{eq:debye_response_D_exact}). 
		Panel (b) shows the corresponding logarithmic representation of the absolute value. 
		Panel (c) displays the exact waveform together with the first individual Debye contributions, $p=0$, $p=1$, and $p=2$. 
		Panel (d) shows, in logarithmic scale, the exact waveform together with the individual Debye contributions from $p=0$ to $p=4$. 
		Panel (e) compares the exact waveform with the partial Debye sum truncated at $p=4$. 
		Panel (f) shows the corresponding logarithmic representation, together with the partial sums truncated at $p=3$ and $p=4$. 
		For this ultracompact configuration, the closed-form Debye reconstruction is almost indistinguishable from the exact waveform over the full time interval shown. 
		The individual Debye terms separate the signal into a prompt/ringdown part, dominated by the leading contribution, followed by a sequence of echo-like responses associated with successive internal reflections. The dominant packets occur near the geometric echo times estimated in Sec.~\ref{sec:sec_1_2_1}, with a separation controlled by the interior round-trip time. The partial sums reproduce the main echo structure, with the agreement improving as higher Debye orders are included.
	}
	\label{fig:Debye_Waveform_UCO}
\end{figure*}
\begin{figure}[!htb]
	\centering
	\includegraphics[scale=0.50]{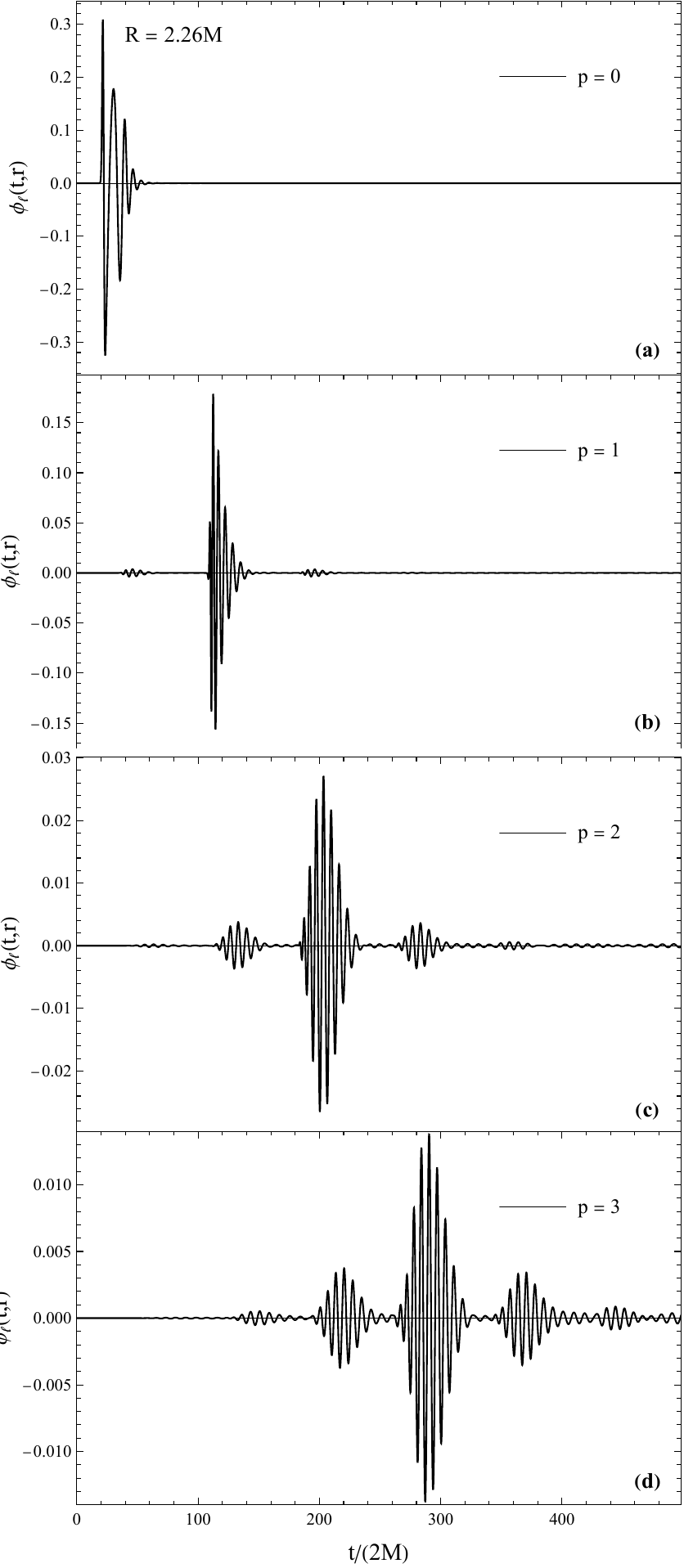}
	\caption{Individual Debye contributions to the \(\ell=2\) waveform for the ultracompact configuration \(R=2.26M\). The observer is located at \(r_{\rm obs}=50M\). The Gaussian initial data in Eq.~\eqref{eq:Cauchy_data_waveform} are specified by \(\phi_0=1\), \(a=1\), and \(r_0=10M\). Panels (a)--(d) show the first terms of the Debye expansion, from \(p=0\) to \(p=3\), respectively. The leading term \(p=0\) contains the prompt response followed by the first ringdown burst, while the higher-order terms are dominated by echo-like packets associated with successive interior traversals. The location of the dominant packet of each contribution is consistent with the geometric arrival times discussed in Sec.~\ref{sec:sec_1_2_1}. Each Debye term is nevertheless not strictly localized in time: weaker oscillatory components appear before and after the main arrival, reflecting the frequency-domain nature of the Debye paths and their overlap in the time domain.}
	\label{fig:Debye_Decomposition_p_UCO}
\end{figure}
\begin{figure*}[!htb]
	\centering
	\includegraphics[scale=0.48]{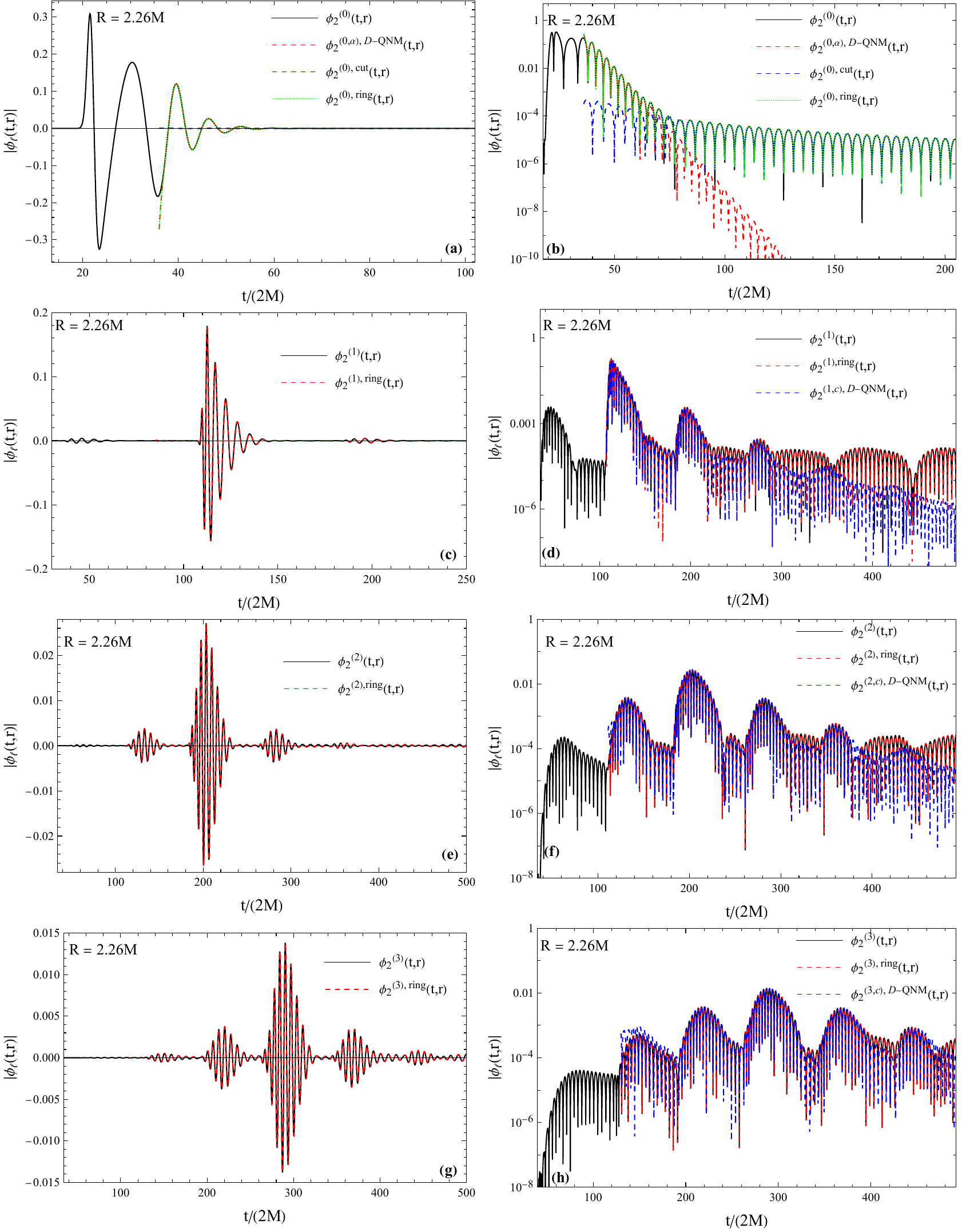}
\caption{
	D-QNM and branch-cut content of the first Debye-series terms of the \(\ell=2\) waveform for the ultracompact configuration \(R=2.26M\). The observer is located at \(r_{\rm obs}=50M\). The Gaussian initial data in Eq.~\eqref{eq:Cauchy_data_waveform} are specified by \(\phi_0=1\), \(a=1\), and \(r_0=10M\).
	Panels (a), (c), (e), and (g) show the exact Debye contributions \(\phi^{(p)}_2(t,r)\), for \(p=0,1,2,3\), together with their corresponding ringdown reconstructions.
	Panels (b), (d), (f), and (h) show the same quantities in logarithmic scale.
	For \(p=0\), the ringdown reconstruction is decomposed into the interface D-QNM contribution, associated with the poles of \(\alpha^{\rm in}_\ell(\omega)\), and the branch-cut contribution.
	For \(p\geq1\), the quantity denoted by \(\phi^{(p),\mathrm{ring}}_2\) contains the curvature-type \(c^{\rm in}\)-D-QNM contribution together with the regularized sub-threshold cut contribution.
	In the logarithmic panels, the curvature-type D-QNM contribution \(\phi^{(p,c),\mathrm{D\text{-}QNM}}_2\) is shown separately in order to display the additional contribution carried by the cut.
	The comparison is made in the ringdown and echo windows selected from the characteristic propagation times introduced in Sec.~\ref{sec:sec_1_2_1}.
}
	\label{fig:UCO_Exact_DQNMS_Cut_p}
\end{figure*}
\begin{figure}[!htb]
	\centering
	\includegraphics[scale=0.55]{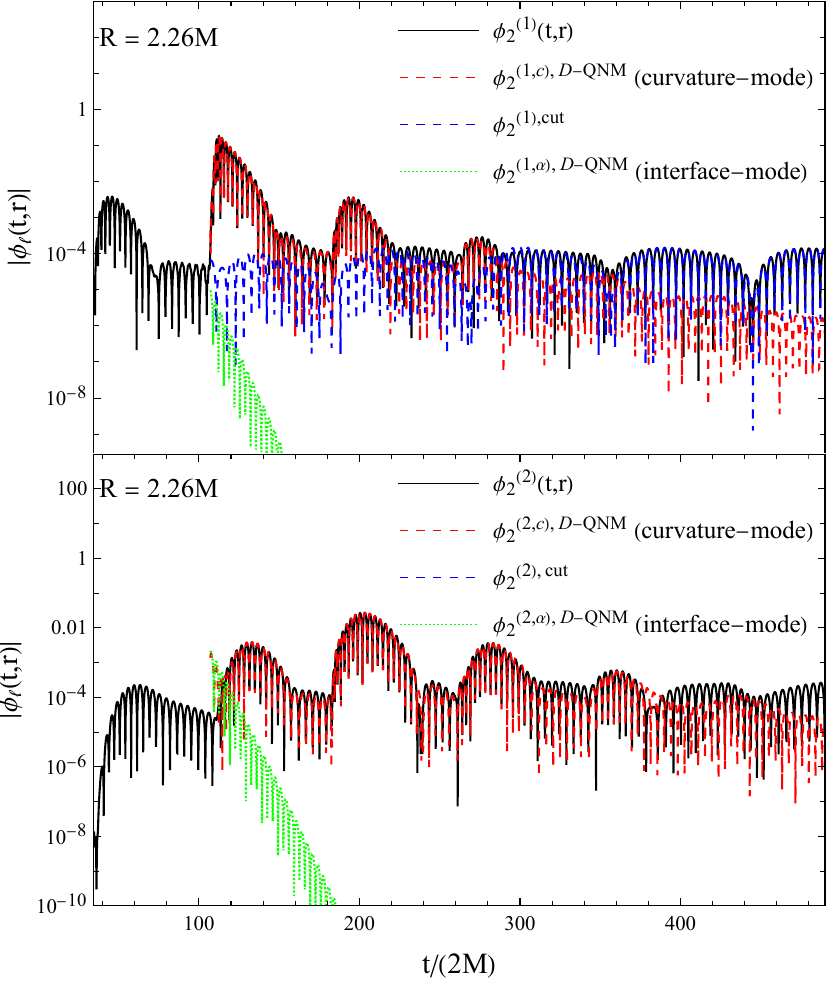}
\caption{
	Logarithmic decomposition of the \(\ell=2\) Debye contributions \(p=1\) and \(p=2\) for the ultracompact configuration \(R=2.26M\). The observer is located at \(r_{\rm obs}=50M\). The Gaussian initial data in Eq.~\eqref{eq:Cauchy_data_waveform} are specified by \(\phi_0=1\), \(a=1\), and \(r_0=10M\).
	The exact contributions are compared with the curvature-type D-QNM sector, the regularized sub-threshold cut contribution, and the interface D-QNM sector. 
	The curvature-type D-QNMs dominate the main echo/ringdown packets, while the cut contribution, which contains the trapped poles attached to the two lips of the cut, mainly affects the late-time signal. 
	The interface contribution is strongly suppressed in this time window.
}
	\label{fig:UCO_p_1_2_DQNMs}
\end{figure}

We now discuss the Debye-series reconstruction of the time-domain waveforms for the two representative configurations considered here: the neutron-star-like model with $R=6M$ and the ultracompact configuration with $R=2.26M$. In both cases we focus on the quadrupolar sector $\ell=2$. The reference waveform is obtained from the frequency-domain representation \eqref{eq:response_infinity_waveform}, whereas the Debye reconstruction is computed from the closed-form cavity expression \eqref{eq:debye_response_phi_inf} [cf. Eq.~\eqref{eq:debye_response_D_exact}]. We also compare the exact Debye result with the individual Debye contributions $\phi^{(p)}_2(t,r)$, defined through \eqref{eq:debye_response_time_p} [cf. \eqref{eq:debye_response_D_terms_coeffs_a}~and~\eqref{eq:debye_response_D_terms_coeffs_b}], and with the corresponding low-order partial Debye sums.

For the neutron-star-like configuration $R=6M$, Fig.~\ref{fig:Debye_Waveform_NS} shows that the waveform reconstructed from the closed-form Debye expression is essentially indistinguishable from the exact time-domain signal over the displayed time interval. This agreement provides a direct numerical check of the cavity representation \eqref{eq:debye_response_D_exact}. The individual Debye terms shown in panels (c) and (d) reveal that the leading contribution $p=0$ already captures the dominant part of the response, while the higher-order terms $p\geq1$ provide progressively smaller corrections.  In this configuration the Debye components do not organize themselves into well-separated echoes. This is consistent with the absence of an ultracompact cavity: the stellar surface lies outside the photon-sphere region, and repeated interior reflections do not generate clearly identifiable secondary pulses in the full waveform. In terms of the characteristic times introduced in Sec.~\ref{sec:sec_1_2_1}, the relevant reflected timescales are not sufficiently separated from the main ringdown stage. The response associated with the surface and with the first partial penetration into the stellar interior therefore remains mixed with the dominant ringdown, rather than appearing as a distinct echo packet.

The same figure also illustrates the convergence properties of the truncated Debye sums. Panels (e) and (f) show that the first few partial sums reproduce the waveform accurately on a linear scale, in particular during the main ringdown stage. However, the convergence is slower in the low-amplitude late-time regime, where residual oscillations remain visible in logarithmic scale. This behaviour is consistent with the geometric nature of the Debye expansion: near the interior threshold frequency $\omega_c$, the expansion parameter is less strongly suppressed, so that a low-order truncation is not uniformly accurate over the full frequency range. Increasing the number of Debye terms progressively reduces this residual structure. 

Figure~\ref{fig:NS_Exact_DQNMS_Cut_p} further decomposes the first Debye contributions into their D-QNM and sub-threshold branch-cut parts. For $p=0$, the cut contribution associated with $0<\omega<\omega_c$ reconstructs the dominant part of both the ringdown and the subsequent tail of the leading Debye term, while the interface D-QNMs, associated with the poles of $\alpha^{\rm in}_\ell(\omega)$, provide a correction near the onset of the ringing. For $p\geq1$, the interface-mode D-QNMs still contribute to the oscillatory part of the signal, but the branch cut remains essential for describing the low-amplitude nonmodal component. In the $R=6M$ case, the D-QNM sums displayed in Fig.~\ref{fig:NS_Exact_DQNMS_Cut_p} are therefore built from the single relevant pole of $\alpha^{\rm in}_\ell(\omega)$. Additional poles of $c^{\rm in}_\ell(\omega)$ are present in principle; in the present neutron-star-like configuration they correspond to curvature modes, but their contribution is strongly suppressed and is not displayed.

We now turn to the ultracompact configuration $R=2.26M$. The comparison shown in Fig.~\ref{fig:Debye_Waveform_UCO} again demonstrates that the closed-form Debye reconstruction \eqref{eq:debye_response_phi_inf} [cf. Eq.~\eqref{eq:debye_response_D_exact}] reproduces the exact waveform with high accuracy. In contrast with the $R=6M$ case, however, the individual Debye terms now acquire a clear time-domain interpretation. The leading term $p=0$ contains the prompt response and the first ringdown burst, while the higher-order terms $p\geq1$ are associated with successive echo-like packets generated by repeated propagation through the interior and re-emission to infinity. The dominant packet of each Debye contribution is delayed with respect to the previous one, in agreement with the optical-length interpretation of the cavity. More precisely, the leading contribution is associated with the prompt/ringdown window, whereas the dominant packets of the higher Debye orders occur around the echo times \(t_{{\rm echo},n}\) estimated in Sec.~\ref{sec:sec_1_2_1}. The separation between successive packets is therefore controlled by the interior round-trip time \(\Delta t_{\rm echo}\).

This separation is displayed more explicitly in Fig.~\ref{fig:Debye_Decomposition_p_UCO}, where the first Debye contributions are shown individually. The dominant burst of the \(p\)-th contribution is located close to the corresponding geometric arrival window discussed in Sec.~\ref{sec:sec_1_2_1}. Each order $p$ is dominated by a main burst, but it is not strictly localized in time: weaker oscillatory components appear both before and after the dominant arrival. These features reflect the fact that a Debye term is a frequency-domain propagation channel rather than a compactly supported time-domain pulse. Consequently, neighbouring Debye orders overlap and interfere in the intermediate-time regime, even though their dominant echo packets remain well separated.

The D-QNM and regularized branch-cut content of the first ultracompact Debye terms is shown in Fig.~\ref{fig:UCO_Exact_DQNMS_Cut_p}. For \(p=0\), the ringdown is mainly reconstructed by the interface D-QNMs associated with the poles of \(\alpha^{\rm in}_\ell(\omega)\), while the sub-threshold cut provides the nonmodal contribution visible after the main burst. This comparison is therefore made in the ringdown window identified by the characteristic time estimates. In this case, the interface-mode sum includes the first five poles of \(\alpha^{\rm in}_\ell(\omega)\).

For \(p\geq1\), the ringing part of each Debye contribution is reconstructed from the curvature-type D-QNMs associated with the zeros of \(c^{\rm in}_\ell(\omega)\), together with the regularized sub-threshold cut. The trapped poles attached to the two lips of the cut are not included as ordinary fourth-quadrant D-QNM residues. Instead, they are recovered through the regularized cut contribution, using the principal-value prescription for \(p=1\) and the Hadamard finite-part prescription for \(p\geq2\), as described in Appendix~\ref{app:cut_regularization}. Numerically, the main echo/ringdown packets in the displayed windows are dominated by the curvature-type \(c^{\rm in}\)-D-QNMs. The reconstructions shown in Fig.~\ref{fig:UCO_Exact_DQNMS_Cut_p} are therefore obtained by summing over the first 46 curvature-mode poles of \(c^{\rm in}_\ell(\omega)\), while the regularized cut accounts for the trapped poles attached to the lower and upper lips of the sub-threshold cut.

This behaviour is displayed more explicitly in Fig.~\ref{fig:UCO_p_1_2_DQNMs}, where the \(p=1\) and \(p=2\) Debye contributions are decomposed into the curvature-type D-QNM contribution, the regularized cut contribution, and the interface D-QNM contribution. The curvature-type sector reconstructs the dominant echo/ringdown packets, whereas the cut contribution mainly affects the lower-amplitude late-time part of the signal. This is precisely where the trapped poles attached to the two lips of the cut leave their imprint. The interface contribution associated with the zeros of \(\alpha^{\rm in}_\ell(\omega)\) remains strongly suppressed in the displayed time window.

This comparison also clarifies the relation between the ordinary QNM description of the ringing signal and the D-QNM description of the individual Debye terms. Although we use the same qualitative labels---curvature, trapped, and interface modes---the corresponding poles do not play the same role in the two representations. The ordinary QNMs are defined by the zeros of the global coefficient \(A^{(-)}_\ell(\omega)\), whereas the D-QNMs are associated with the singularities of the separate Debye building blocks, in particular with the zeros of \(\alpha^{\rm in}_\ell(\omega)\) and \(c^{\rm in}_\ell(\omega)\). The absence of a one-to-one correspondence is already visible from Eq.~\eqref{eq:Aminus_Debye_reduced}: the zeros of \(A^{(-)}_\ell(\omega)\) are determined by the full Fabry--P\'erot-like combination of the Debye building blocks, not by the zeros of \(\alpha^{\rm in}_\ell(\omega)\) and \(c^{\rm in}_\ell(\omega)\) separately.

This distinction is particularly important in the ultracompact configuration. In the ordinary QNM reconstruction, the late-time echo train is mainly described by the trapped QNMs of \(A^{(-)}_\ell(\omega)\), which are collective resonances of the fully resummed cavity response. In the Debye representation, by contrast, the signal is organized according to propagation histories: a fixed Debye order represents a definite propagation channel rather than the complete multiple-reflection dynamics. As a result, the dominant D-QNM contribution to the main echo packets at fixed \(p\geq1\) can come from the curvature-type zeros of \(c^{\rm in}_\ell(\omega)\), while the trapped poles attached to the two lips of the sub-threshold cut are recovered through the regularized branch-cut term and mainly affect the low-amplitude late-time part of the signal. This is not a contradiction: the ordinary QNM expansion identifies the collective poles of the complete ringing response, whereas the Debye-QNM expansion identifies the poles controlling each individual propagation channel before resummation.

\section{Conclusions and outlook}
\label{sec:conclusion}

We have developed a Debye-series description of the time-domain response of compact, horizonless objects. Starting from the frequency-domain Green-function representation, we rewrote the response kernel in a cavity form and expanded the corresponding Fabry--P\'erot-like denominator as a Debye series. This construction separates the waveform into contributions associated with direct exterior propagation, surface reflection, and successive transmissions through the stellar interior. It therefore provides a propagation-based organization of the signal, complementary to the ordinary quasinormal-mode expansion.

We applied this framework to scalar perturbations of a constant-density compact object with Schwarzschild exterior, focusing on two representative configurations: a neutron-star-like model with \(R=6M\), and an ultracompact model with \(R=2.26M\). For the neutron-star-like case, the Debye decomposition does not lead to a sequence of well-separated echoes. Instead, the low-order Debye terms contribute mainly to the ringdown and tail, and the sub-threshold branch cut \(0<\omega<\omega_c\) plays an important role in the non-modal part of the response. For the ultracompact configuration, by contrast, the Debye terms organize the signal into a prompt/ringdown contribution followed by a sequence of echo-like packets. The arrival times of these packets are consistent with the optical-length interpretation based on propagation through the stellar interior.

We also analyzed the spectral content of the Debye terms. The ordinary QNMs are defined by the zeros of the global coefficient \(A^{(-)}_\ell(\omega)\), whereas the Debye QNMs are associated with the poles of the individual Debye building blocks, in particular the zeros of \(\alpha^{\rm in}_\ell(\omega)\) and \(c^{\rm in}_\ell(\omega)\). The absence of a one-to-one correspondence between the two spectra follows from the factorized relation \eqref{eq:Aminus_Debye_reduced}, where the global coefficient \(A^{(-)}_\ell\) contains the Fabry--P\'erot-like resummation of repeated internal reflections. Thus, the QNMs of the full response describe collective resonances of the resummed cavity, while the D-QNMs describe the modal content of fixed propagation channels.

This distinction is particularly important in the ultracompact case. In the ordinary QNM reconstruction, the late-time echo train is mainly governed by the trapped QNMs of \(A^{(-)}_\ell(\omega)\), which arise as collective resonances of the fully resummed response. In the Debye representation, however, the signal is organized before this full resummation is performed. At fixed Debye order, the dominant modal contribution to the principal echo packets is carried by the curvature-type zeros of \(c^{\rm in}_\ell(\omega)\). The trapped structures attached to the two lips of the sub-threshold cut are instead recovered through the regularized cut contribution, where they mainly contribute to the lower-amplitude late-time part of the signal. The interface D-QNM contribution remains weakly excited in the time windows considered here. Thus, the two descriptions are not in conflict: the ordinary QNM expansion identifies collective poles of the complete response, whereas the Debye representation separates the propagation channels and their associated pole and cut contributions.

A useful feature of the closed-form Debye reconstruction is that it is constructed directly from the full frequency-domain response kernel and reorganizes it into physically interpretable propagation channels before any pole decomposition is performed. After inverse transformation, this closed-form representation describes the prompt response, the initial ringdown, and the subsequent echo-like packets at the level of the full reconstructed waveform. This statement should be distinguished from a pole reconstruction: a standard QNM pole expansion or a D-QNM pole expansion remains a residue expansion, and its practical use is therefore subject to the usual starting-time limitation. The prompt signal is captured by the closed-form Debye representation itself, not by extrapolating a pole sum to early times.

The cavity-like expansion of frequency-domain response kernels has precedents in the echo literature, in particular in effective-reflectivity and Green-function approaches~\cite{Mark:2017dnq,Hui:2019aox}. In the present work, the Debye expansion is used to go further by resolving the D-QNM residues and cut contributions associated with each propagation channel in a resolved compact-star model. The association between a Debye order and an echo-like packet is therefore a dominant-channel association, not a statement of compact temporal support. The numerical results show that each order \(p\) is dominated by a main burst near the corresponding geometric arrival window, while also having weaker temporal support before and after this burst. Neighbouring Debye orders can consequently overlap, with the overlap becoming more pronounced at higher order.

This points to complementary regimes of usefulness for the Debye and standard-QNM reconstructions. The Debye expansion is most transparent for the prompt response and the first echo-like packets, where a small number of low-order propagation channels dominates. At later times, more Debye orders may contribute within the same time window. In that regime, the standard QNM expansion can become more economical when the signal is governed by the long-lived collective resonances of the fully resummed response.

The present analysis also highlights the role of the branch-cut sector. For the neutron-star-like configuration, the sub-threshold cut generated by the interior wavenumber contributes significantly to the ringing and tail of the leading Debye terms. For the ultracompact model considered here, the \(p=0\) Debye term also receives an important non-modal contribution from the sub-threshold cut. For \(p\geq1\), however, the cut requires a regularized treatment because of the trapped structures attached to its two lips. This regularized cut contribution is subdominant in the main echo packets, where the curvature-type D-QNM residues dominate, but it becomes visible in the low-amplitude late-time tail. This behavior shows that the relative importance of pole and cut contributions is sensitive both to the compactness of the object and to the propagation channel selected by the Debye order.

In the stellar model considered here, the resolved interior response entering the Debye construction provides a concrete realization of the effective-reflectivity picture. In a reduced \((K,x_0)\) model, the complicated interior dynamics is compressed into a prescribed reflectivity and an effective propagation phase. By contrast, in the resolved stellar problem, the internal response is computed from the regular interior solution itself. Up to convention-dependent phase factors, the role of an effective reflectivity is played by \(\xi_\ell(\omega)=c_\ell^{\rm out}(\omega)/c_\ell^{\rm in}(\omega)\). This quantity carries the pole and cut structure inherited from the interior problem. Consequently, in the present setting, the Debye expansion does more than reproduce the approximate echo spacing: it resolves the fine structure of the response and separates the contributions associated with direct propagation, surface reflection, repeated interior traversals, D-QNM residues, and the regularized sub-threshold cut.

Several extensions are natural. First, this study could be extended to perturbations of the gravitational field, governed by the \(s=2\) Regge--Wheeler--Zerilli equations. Second, the large-order and high-frequency behavior of the Debye-QNM spectrum should be studied analytically, for example by WKB methods, in order to relate the D-QNM frequencies to the optical lengths and phase shifts of the corresponding propagation paths. Relatedly, it should be possible to obtain exact results for \(c\)-D-QNM frequencies for conformally coupled fields by exploiting a conformal transformation in the interior region \cite{Seenivasan:2025ysy}. Third, it would be useful to develop a complex-angular-momentum version of the present time-domain construction, connecting the Debye-QNM spectrum with the Regge--Debye pole spectrum obtained in the scattering problem. Finally, the method could be extended to massive fields or to other compact-object models. In such cases, additional branch cuts and multi-sheet structures may appear, and the Debye representation may provide a useful way of separating pole, cut, and propagation-path contributions to the waveform.

\begin{acknowledgments}
	S.D.~acknowledges financial support from the Science and Technology Facilities Council (STFC) under Grant No.~ST/X000621/1. The authors are grateful to the anonymous referee for careful and constructive reports, which helped improve the clarity and presentation of the manuscript.
\end{acknowledgments}

\appendix
\section{Regularization of the sub-threshold contribution: UCO-case}
\label{app:cut_regularization}

In this appendix, we describe the numerical prescription used to evaluate the sub-threshold part of the Debye contributions. This issue arises because the interior wavenumber \(k_{\rm int}\) has a branch point at \(\omega=\omega_c\). For \(0<\omega<\omega_c\), the two sides of the cut correspond to different analytic continuations of \(k_{\rm int}\). In the ultracompact case, trapped poles may lie on, or exponentially close to, these lips. They must therefore be treated by an appropriate principal-value or finite-part prescription.

\subsection{Direct reconstruction of the Debye contribution}
\label{app:direct_upper_lip}

\begin{figure}[!htb]
	\centering
	\includegraphics[scale=0.48]{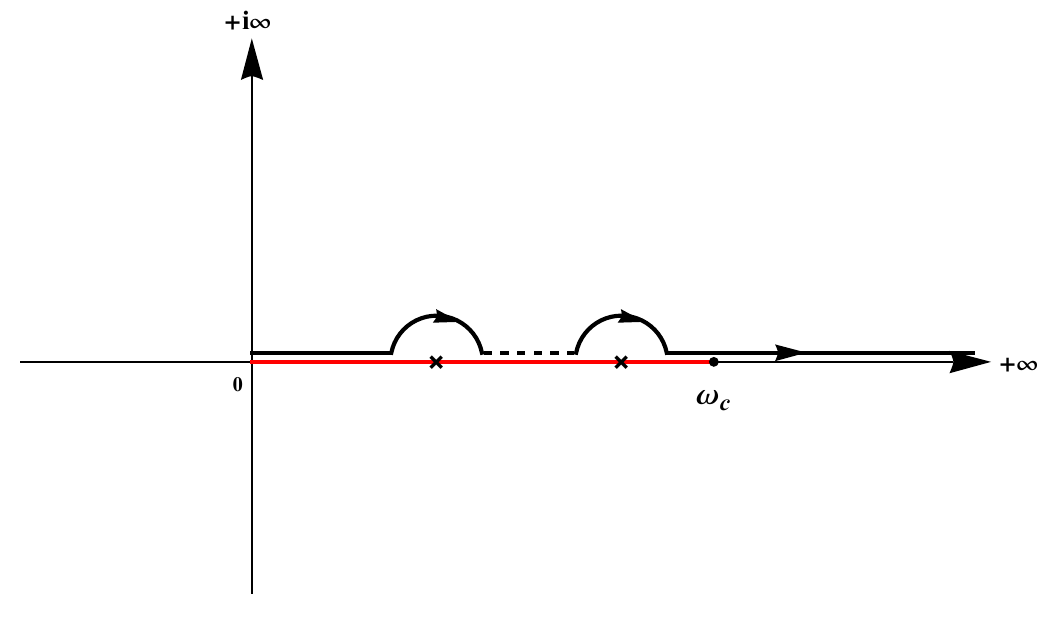}
		\caption{Upper-lip contour used to evaluate the sub-threshold part of the direct Debye contribution. 
			The interval \(0<\omega<\omega_c\) is taken on the upper lip of the cut, and the trapped poles on this path are bypassed by small semicircles in the upper half-plane. 
			This yields the principal-value prescription for \(p=1\) and the Hadamard finite-part prescription for \(p\geq2\).}
		\label{fig:Contour_Lip_Sup}
\end{figure}

The Debye contribution of order \(p\geq1\) is reconstructed from the frequency-domain integral (cf. Eq.~\eqref{eq:debye_response_time_p})
\begin{equation}
	\phi^{(p)}_\ell(t,r)
	=
	\frac{1}{2\pi}\,
	\Re\!\left[
	\int_{0+ic}^{+\infty+ic}
	e^{-i\omega u}
	\mathcal{D}^{(p)}_\ell(\omega)
	d\omega
	\right],
	\label{eq:app_debye_response_time_p}
\end{equation}
where $u=t-r_\ast$. 

When the contour is brought down to the real-frequency axis, the branch point at \(\omega=\omega_c\) naturally splits the integral into two pieces,
\begin{align}
	\int_{0+ic}^{+\infty+ic}
	&e^{-i\omega u}
	\mathcal{D}^{(p)}_\ell(\omega)
	d\omega
	=
	\int_{0+i0}^{\omega_c+i0}
	e^{-i\omega u}
	\mathcal{D}^{(p)+}_\ell(\omega)
	d\omega
	\nonumber\\
	&\quad
	+
	\int_{\omega_c}^{+\infty}
	e^{-i\omega u}
	\mathcal{D}^{(p)}_\ell(\omega)
	d\omega .
	\label{eq:app_split_direct_integral}
\end{align}
The first integral is evaluated on the upper lip of the sub-threshold cut. The second integral is regular and is computed directly on the real axis.

In the ultracompact case, the first integral may contain trapped poles associated with the upper-lip branch; see Tab.~\ref{tab:trapped_poles_cut_UCO}. The integration path cannot then be identified with the real segment without specifying how these singularities are avoided. We deform the path into the upper half-plane by small semicircles around the trapped poles, as illustrated in Fig.~\ref{fig:Contour_Lip_Sup}. This prescription gives a Cauchy principal value for simple poles and a Hadamard finite part for higher-order poles.

Let \(\omega_n^+\) denote the positions of the trapped poles on the upper lip. On the sub-threshold interval, the singular part of the Debye integrand of order \(p\) is written as
\begin{equation}
	\mathcal{D}^{(p)+}_\ell(\omega)
	=
	\mathcal{D}^{(p)+}_{\ell,\mathrm{reg}}(\omega)
	+
	\sum_{n}
	\sum_{m=1}^{p}
	\frac{A^{(p)+}_{m,n}}{(\omega-\omega_n^+)^m}.
	\label{eq:app_laurent_upper_direct}
\end{equation}
The coefficients \(A^{(p)+}_{m,n}\) are the Laurent coefficients of the full Debye integrand on the upper lip. They are obtained from the excitation coefficients defined in Eq.~\eqref{eq:Debye_ExcitationCoefficient_waveform_c_compact}, evaluated on the \(k_{\rm int}^{+}\) branch. More precisely,
\begin{equation}
	A_{m,n}^{(p)+}
	=
	2\omega_n^+
	\sum_j
	\mathcal{C}_{\ell n,j,m-1}^{(p,+)},
	\qquad
	m=1,\ldots,p .
	\label{eq:app_A_coefficients_from_C}
\end{equation}
Here the sum over \(j\) runs over the Debye channels contributing to the pole of order \(m\)  (although, in practice, only the \(P\)-channel is retained, since the  \(\mathrm{in}\) and \(\mathrm{out}\) channels are negligible).  The factor \(2\omega_n^+\) restores the normalization convention used in  Eq.~\eqref{eq:Debye_ExcitationCoefficient_waveform_c_compact}, where the factor  \(1/(2\omega)\) is included in the definition of the excitation factors in  Eq.~\eqref{eq:Debye_ExcitationFactor_waveform_c_compact}.

Using the local expansion \eqref{eq:app_laurent_upper_direct}, the first integral in Eq.~\eqref{eq:app_split_direct_integral} is understood as
\begin{align}
	\int_{0+i0}^{\omega_c+i0}
	e^{-i\omega u}
	\mathcal{D}^{(p)+}_{\ell}(\omega)
	d\omega
	&=
	\operatorname{F.p.}
	\int_0^{\omega_c}
	e^{-i\omega u}
	\mathcal{D}^{(p)+}_{\ell}(\omega)
	d\omega
	\nonumber\\
	&\quad
	-i\pi
	\sum_{n}
	{\cal R}^{(p)+}_{n}(u).
	\label{eq:app_upper_lip_regularized}
\end{align}
Here \(\operatorname{F.p.}\) denotes a Cauchy principal value for \(p=1\), and a Hadamard finite part for \(p\geq2\). The quantity entering the indentation term is the effective residue of the full integrand,
\begin{equation}
	{\cal R}^{(p)+}_{n}(u)
	=
	\operatorname*{Res}_{\omega=\omega_n^+}
	\left[
	e^{-i\omega u}
	\mathcal{D}^{(p)+}_{\ell}(\omega)
	\right].
\end{equation}
Using the Laurent expansion \eqref{eq:app_laurent_upper_direct}, it is given by
\begin{equation}
	{\cal R}^{(p)+}_{n}(u)
	=
	e^{-i\omega_n^+u}
	\sum_{m=1}^{p}
	A^{(p)+}_{m,n}
	\frac{(-iu)^{m-1}}{(m-1)!}.
	\label{eq:app_indentation_upper_direct}
\end{equation}
The sign \(-i\pi\) follows from the fact that, on the upper lip, the pole is bypassed by a small semicircle in the upper half-plane while the integration runs from left to right.

In practice, we subtract the singular part from the numerical data and interpolate only the regular remainder,
\begin{equation}
	\mathcal{D}^{(p)+}_{\ell,\mathrm{reg}}(\omega)
	=
	\mathcal{D}^{(p)+}_{\ell}(\omega)
	-
	\sum_{n}
	\sum_{m=1}^{p}
	\frac{A^{(p)+}_{m,n}}{(\omega-\omega_n^+)^m}.
	\label{eq:app_regular_part_direct}
\end{equation}
The finite part of the integral is then reconstructed by adding back the singular contributions analytically,
\begin{align}
	\operatorname{F.p.}
	\int_{a}^{b}
	&e^{-i\omega u}
	\mathcal{D}^{(p)+}_{\ell}(\omega)d\omega
	=
	\int_{a}^{b}
	e^{-i\omega u}
	\mathcal{D}^{(p)+}_{\ell,\mathrm{reg}}(\omega)d\omega
	\nonumber\\
	&\quad
	+
	\sum_{n}
	\sum_{m=1}^{p}
	A^{(p)+}_{m,n}
	J_m[e^{-i\omega u};\omega_n^+].
	\label{eq:app_reconstructed_finite_part_direct}
\end{align}
Here \(a\) and \(b\) denote the endpoints of the sub-threshold interval. Formally, \(a=0\) and \(b=\omega_c\). The elementary finite-part integrals are defined by
\begin{equation}
	J_m[\phi;\omega_n^+]
	=
	\operatorname{F.p.}
	\int_a^b
	\frac{\phi(\omega)}{(\omega-\omega_n^+)^m}
	d\omega .
	\label{eq:app_Jm_definition}
\end{equation}

For a simple pole, \(m=1\), the elementary singular integral is evaluated as a Cauchy principal value,
\begin{equation}
	J_1[\phi;\omega_n^+]
	=
	\operatorname{PV}
	\int_a^b
	\frac{\phi(\omega)}{\omega-\omega_n^+}
	d\omega .
	\label{eq:app_pv_simple_pole}
\end{equation}
where
\[
\phi(\omega)=e^{-i\omega u}.
\]
For \(m\geq2\), the Hadamard finite part is evaluated recursively as
\begin{align}
	&J_m[\phi;\omega_n^+]
	=
	\frac{1}{m-1}
	J_{m-1}[\phi';\omega_n^+]\nonumber\\
	&-
	\frac{\phi(b)}{(m-1)(b-\omega_n^+)^{m-1}}
	+
	\frac{\phi(a)}{(m-1)(a-\omega_n^+)^{m-1}} .
	\label{eq:app_hadamard_recursion}
\end{align}

The effective residue associated with the indentation term is obtained from the same Laurent coefficients. For example, for \(p=2\),
\begin{equation}
	{\cal R}^{(2)+}_{n}(u)
	=
	e^{-i\omega_n^+u}
	\biggl[
	A^{(2)+}_{1,n}
	+
	(-iu)A^{(2)+}_{2,n}
	\biggr],
\end{equation}
while for \(p=3\),
\begin{align}
	{\cal R}^{(3)+}_{n}(u)
	=
	e^{-i\omega_n^+u}
	\biggl[
	&A^{(3)+}_{1,n}
	+
	(-iu)A^{(3)+}_{2,n}
	\nonumber\\
	&\hspace{45pt}+
	\frac{(-iu)^2}{2}
	A^{(3)+}_{3,n}
	\biggr].
\end{align}

\subsection{Regularized cut contribution}
\label{app:cut_two_lips}

\begin{figure}[!htb]
	\centering
	\includegraphics[scale=0.55]{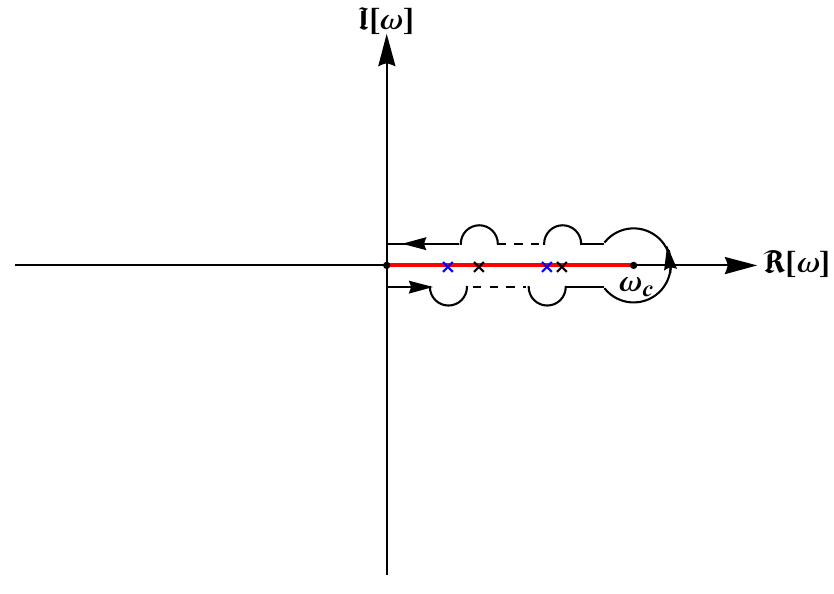}
	\caption{
		Regularization of the sub-threshold cut \(0<\omega<\omega_c\). The cut contribution is obtained from the discontinuity between the upper and lower lips. In the ultracompact case, trapped poles may lie on the two lips of the cut.  The poles associated with the upper-lip branch are bypassed by small semicircles in the upper half-plane, whereas those associated with the lower-lip branch are bypassed by small semicircles in the lower half-plane. This prescription gives the indentation terms appearing in the regularized cut contribution.
	}
	\label{fig:Contour_Cut}
\end{figure}
We now consider the contribution of the sub-threshold cut which appears after the contour deformation. This contribution is defined as the discontinuity between the two lips of the cut, as in Eq.~\eqref{eq:TimeEvolution_D-QNM_infinity_waveform_p_cut}
\begin{align}
	&\phi_\ell^{(p),\mathrm{cut}}(t,r)
	= \nonumber \\
	&\hspace{10	pt}\frac{1}{2\pi}\,
	\Re\!\biggl[
	\int_0^{\omega_c}
	e^{-i\omega u}
	\Bigl(
	\mathcal{F}^{(p)+}_\ell(\omega)
	-
	\mathcal{F}^{(p)-}_\ell(\omega)
	\Bigr)
	d\omega
	\biggr].
	\label{eq:TimeEvolution_D-QNM_infinity_waveform_p_cut_bis}
\end{align}
When no pole lies on the cut, this expression is understood in the ordinary sense. In the ultracompact case, however, the two lips may contain trapped poles and the integral has to be regularized. This situation is illustrated in Fig.~\ref{fig:Contour_Cut}, while the trapped poles associated with the lower- and upper-lip branches are listed in Tab.~\ref{tab:trapped_poles_cut_UCO}.

Near a trapped pole \(\omega_n^+\) on the upper lip, we write
\begin{equation}
	\mathcal{F}^{(p)+}_\ell(\omega)
	=
	\mathcal{F}^{(p)+}_{\ell,\mathrm{reg}}(\omega)
	+
	\sum_n
	\sum_{m=1}^{p}
	\frac{A^{(p)+}_{m,n}}
	{(\omega-\omega_n^+)^m}.
	\label{eq:app_laurent_cut_upper}
\end{equation}
Similarly, near a trapped pole \(\omega_n^-\) on the lower lip,
\begin{equation}
	\mathcal{F}^{(p)-}_\ell(\omega)
	=
	\mathcal{F}^{(p)-}_{\ell,\mathrm{reg}}(\omega)
	+
	\sum_n
	\sum_{m=1}^{p}
	\frac{A^{(p)-}_{m,n}}
	{(\omega-\omega_n^-)^m}.
	\label{eq:app_laurent_cut_lower}
\end{equation}
The coefficients \(A^{(p)\pm}_{m,n}\) are the Laurent coefficients of the cut integrands on the corresponding lips.

For the upper lip, the indentation is taken in the upper half-plane. Therefore,
\begin{align}
	&\int_{0+i0}^{\omega_c+i0}
	e^{-i\omega u}
	\mathcal{F}^{(p)+}_\ell(\omega)
	d\omega
	= 	\nonumber\\
	&\quad\operatorname{F.p.}
	\int_0^{\omega_c}
	e^{-i\omega u}
	\mathcal{F}^{(p)+}_\ell(\omega)
	d\omega
	-i\pi
	\sum_n
	\mathcal{R}^{(p)+}_n(u).
	\label{eq:app_cut_upper_lip_regularized}
\end{align}
For the lower lip, the indentation is taken in the lower half-plane, giving
\begin{align}
	&\int_{0-i0}^{\omega_c-i0}
	e^{-i\omega u}
	\mathcal{F}^{(p)-}_\ell(\omega)
	d\omega
	= \nonumber\\
	&\quad \operatorname{F.p.}
	\int_0^{\omega_c}
	e^{-i\omega u}
	\mathcal{F}^{(p)-}_\ell(\omega)
	d\omega
	+i\pi
	\sum_n
	\mathcal{R}^{(p)-}_n(u).
	\label{eq:app_cut_lower_lip_regularized}
\end{align}
Here \(\operatorname{F.p.}\) denotes a Cauchy principal value for \(p=1\), and a Hadamard finite part for \(p\geq2\).

The quantities \(\mathcal{R}^{(p)\pm}_n(u)\) are the effective residues of the full cut integrands,
\begin{equation}
	\mathcal{R}^{(p)\pm}_n(u)
	=
	\operatorname*{Res}_{\omega=\omega_n^\pm}
	\left[
	e^{-i\omega u}
	\mathcal{F}^{(p)\pm}_\ell(\omega)
	\right].
	\label{eq:app_cut_effective_residue_def}
\end{equation}
Using the Laurent expansions \eqref{eq:app_laurent_cut_upper} and \eqref{eq:app_laurent_cut_lower}, this gives
\begin{equation}
	\mathcal{R}^{(p)\pm}_n(u)
	=
	e^{-i\omega_n^\pm u}
	\sum_{m=1}^{p}
	A^{(p)\pm}_{m,n}
	\frac{(-iu)^{m-1}}{(m-1)!}.
	\label{eq:app_cut_effective_residue}
\end{equation}

Substituting Eqs.~\eqref{eq:app_cut_upper_lip_regularized} and \eqref{eq:app_cut_lower_lip_regularized} into the discontinuity formula, the regularized cut contribution becomes
\begin{align}
	&\phi_\ell^{(p),\mathrm{cut}}(t,r)
	= \nonumber \\
	&\frac{1}{2\pi}
	\Re\!\biggl[
	\operatorname{F.p.}
	\int_0^{\omega_c}
	e^{-i\omega u}
	\Bigl(
	\mathcal{F}^{(p)+}_\ell(\omega)
	-
	\mathcal{F}^{(p)-}_\ell(\omega)
	\Bigr)
	d\omega
	\nonumber\\
	&\qquad
	-i\pi
	\sum_n
	\mathcal{R}^{(p)+}_n(u)
	-i\pi
	\sum_n
	\mathcal{R}^{(p)-}_n(u)
	\biggr].
	\label{eq:app_regularized_cut_final}
\end{align}

\begin{table}[htp]
	\centering
	\caption{Trapped poles attached to the sub-threshold cut for the massless scalar field. The radius of the ultracompact body is $R=2.26M$. The first trapped sector is associated with the lower-side branch $k_{\rm int}^{-}$, while the second one is associated with the upper-lip branch $k_{\rm int}^{+}$.}
	\label{tab:trapped_poles_cut_UCO}
	\small
	\setlength{\tabcolsep}{5pt}
	\renewcommand{\arraystretch}{1.15}
	\begin{tabular}{c c c c}
		\hline\hline
		$\ell$ & $n$ 
		& $2M\omega_{\ell n}^{(c^{\rm in},-)}(\rm Trapped)$ 
		& $2M\omega_{\ell n}^{(c^{\rm in},+)}(\rm Trapped)$ \\
		\hline
		2 & 0  
		& $0.248656 $ 
		& $0.248083 $ \\
		& 1  
		& $0.328261 $ 
		& $0.325250 $ \\
		& 2  
		& $0.410534 $ 
		& $0.402100 $ \\
		& 3  
		& $0.495852 $ 
		& $0.478544 $ \\
		& 4  
		& $0.584527 $ 
		& $0.554326 $ \\
		& 5  
		& $0.679607 $ 
		& $0.628739 $ \\
		& 6  
		&                                  
		& $0.698267 $ \\
		\hline\hline
	\end{tabular}
\end{table}

In practice, the numerical construction of the regularized cut follows the same steps as those described in Eqs.~\eqref{eq:app_regular_part_direct}--\eqref{eq:app_hadamard_recursion}, applied separately to the upper and lower lips. Namely, on each lip we first subtract the singular Laurent part, interpolate the regular remainder, and then reconstruct the principal-value or finite-part contribution by adding back the analytic singular integrals. For \(p=1\), this gives a Cauchy principal value prescription, while for \(p\geq2\) the higher-order pole terms are evaluated as Hadamard finite parts. The two regularized lip contributions are then combined according to Eq.~\eqref{eq:app_regularized_cut_final} to obtain the full cut contribution.

\section{Relation with effective-reflectivity echo models}
\label{app:effective_reflectivity}

This appendix clarifies the relation between the Debye representation developed in this work and the effective-reflectivity models commonly used in echo phenomenology. In many phenomenological descriptions, the near-surface or near-horizon structure of a compact object is compressed into an effective boundary condition at some tortoise-coordinate location \(x=x_0\), characterized by a frequency-dependent reflectivity \(K(\omega)\)~\cite{Mark:2017dnq,Wang:2018gin,Testa:2018bzd,Maggio:2019zyv}. Schematically, one writes
\begin{equation}
	\psi \sim e^{-i\omega(x-x_0)}
	+
	K(\omega)e^{+i\omega(x-x_0)} ,
	\label{eq:app_effective_reflectivity_bc}
\end{equation}
so that \(K(\omega)\) and \(x_0\) encode, in a reduced way, the response of the unresolved inner region. Such models provide a compact description of the existence and approximate spacing of echoes, and of the repeated filtering of the signal between the inner boundary and the exterior curvature barrier.

In the present stellar problem, this replacement is not made. The wave equation is solved through the regular interior of the compact object, and the interior response is obtained by matching the regular solution at the stellar surface to the local basis
\(\{u^{\rm in}_{\omega\ell},u^{\rm out}_{\omega\ell}\}\). With the conventions of the main text,
\begin{equation}
	\phi^{\rm reg}_{\omega\ell}
	=
	c_\ell^{\rm in}(\omega) u^{\rm in}_{\omega\ell}
	+
	c_\ell^{\rm out}(\omega) u^{\rm out}_{\omega\ell}.
	\label{eq:app_regular_solution_surface_decomp}
\end{equation}
Equivalently,
\begin{equation}
	\phi^{\rm reg}_{\omega\ell}
	=
	c_\ell^{\rm in}(\omega)
	\left[
	u^{\rm in}_{\omega\ell}
	+
	\xi_\ell(\omega)u^{\rm out}_{\omega\ell}
	\right],
	\qquad
	\xi_\ell(\omega)
	=
	\frac{c_\ell^{\rm out}(\omega)}
	{c_\ell^{\rm in}(\omega)} .
	\label{eq:app_xi_def_reflectivity}
\end{equation}
The quantity \(\xi_\ell(\omega)\) therefore plays the role of a resolved, dynamically generated internal reflection factor. It is not prescribed externally, but follows from the solution of the interior wave equation and from the regularity condition at the center.

Up to convention-dependent phase factors, a reduced effective-reflectivity model may be viewed as replacing this resolved quantity by a prescribed function, for example
\begin{equation}
	\xi_\ell(\omega)
	\longrightarrow
	K(\omega)e^{2i\omega x_0}.
	\label{eq:app_xi_to_K}
\end{equation}
This removes part of the spectral information contained in the stellar problem. Indeed, \(\xi_\ell(\omega)\) has poles at the zeros of \(c_\ell^{\rm in}(\omega)\). These poles generate the \(c^{\rm in}\)-D-QNM sector discussed in the main text, while the branch structure associated with the interior wavenumber gives rise to the cut-supported trapped structures. These features are absent from the simplest effective-reflectivity models unless they are inserted by hand into \(K(\omega)\).

The same distinction appears in the Fabry--P\'erot denominator of the full response. The ordinary QNMs are collective resonances of the resummed cavity response, whereas the Debye expansion resolves the successive propagation channels before this resummation is performed. A reduced \((K,x_0)\) model is therefore well suited to describing the existence and approximate spacing of echoes, while the Debye representation gives access to their fine structure. In this sense, within the present resolved stellar model, the Debye decomposition turns echoes into a probe of the compact-object interior: it makes explicit how the internal response, encoded in \(\xi_\ell(\omega)\) and in its associated pole and cut structure, enters the successive propagation channels and carries a spectral fingerprint of the object's internal structure.

\bibliography{Debye_Echoes_Compact_Objects}

\end{document}